\documentclass[preprint]{aastex}
\usepackage{natbib}

\begin{document}                                                                     

\title{Binary Properties from  {\it C}epheid {\it Ra}dial {\it V}elocities (CRaV)   }


\author{Nancy Remage Evans}
\affil{Smithsonian Astrophysical Observatory,    
MS 4, 60 Garden St., Cambridge, MA 02138, USA; nevans@cfa.harvard.edu}
\author{Leonid Berdnikov }  
\affil{Astronomy and Astrophysics Research Division, Entoto Observatory and Research Center, P.O.Box 8412, Addis Ababa, Ethiopia}
\affil{Sternberg Astronomical Institute of the Moscow State University,
13  Universitetskii Prospect, Moscow 119992, Russia and Isaac Newton Institute of
  Chile, Moscow Branch, 13 Universitetskij Prospect Moscow 119992,
  Russia  }
\author{Jennifer Lauer}
\affil{Smithsonian Astrophysical Observatory,    
MS 34, 60 Garden St., Cambridge, MA 02138, USA}
\author{Douglas Morgan}
\affil{Smithsonian Astrophysical Observatory,    
MS 34, 60 Garden St., Cambridge, MA 02138, USA}
\author{Joy Nichols}
\affil{Smithsonian Astrophysical Observatory,    
MS 34, 60 Garden St., Cambridge, MA 02138, USA}
\author{H. Moritz G\"unther}
\affil{Smithsonian Astrophysical Observatory,    
MS 6, 60 Garden St., Cambridge, MA 02138, USA}
\affil{Massachusetts Institute of Technology,
Kavli Institute for Astrophysics and Space Research,
77 Massachusetts Avenue, Cambridge, MA 02139, USA}
\author{Natalya Gorynya} 
\affil{Institute of Astronomy, Russian Academy of Sciences, 48 Pyatnitskaya Str., 
Moscow, Russia and
 Lomonosov Moscow State University, Sternberg State Astronomical Institute, 
13 Universitetskii prosp., Moscow, Russia}
\author{Alexey Rastorguev } 
\affil{Lomonosov Moscow State University, Sternberg State Astronomical Institute, 
13 Universitetskii prosp., Moscow, Russia}
\author{Pawel Moskalik } 
\affil{Copernicus Astronomical Center, Warsaw, Poland}









\begin{abstract}
We have examined high accuracy radial velocities of Cepheids  
to determine the binary
frequency. The data  are largely from the CORAVEL spectrophotometer and 
the Moscow version, with a typical uncertainty of $\leq1$~km~s$^{-1}$,
and a time span from 1 to 20 years.   A 
systemic velocity was obtained by removing the pulsation component using a high
order Fourier series.  From this data we have developed a list of stars showing
no orbital velocity larger than $\pm1$~km~s$^{-1}$.   
The binary fraction was analyzed as a function of magnitude, 
and yields an apparent decrease in this fraction for fainter stars. 
We interpret this as 
 incompleteness at fainter magnitudes, 
and derive the preferred binary fraction of $29\pm8$\% 
( $20\pm6$\% per decade of orbital period)
from the brightest
40 stars.  Comparison of this fraction in this  period range 
(1-20 years) implies a large
fraction for the full period range.   This is
reasonable in that the high accuracy velocities are sensitive to the longer periods 
and smaller orbital velocity amplitudes in the period range sampled here. 
Thus the Cepheid velocity sample provides a sensitive detection in the period range 
between short period spectroscopic binaries and resolved companions.    
The recent identification of $\delta$ Cep as a binary with very low amplitude
and high eccentricity underscores the fact that the binary fractions we derive 
are lower limits, to which other low amplitude systems will probably be added. 
The mass ratio (q) 
distribution derived from ultraviolet observations of the secondary is 
consistent with a flat distribution for the 
applicable period range (1 to 20 years). 
 
\end{abstract}




\section{Introduction}


Star formation is a very active area of research both observationally
and theoretically, with many unanswered questions. One such area is 
binary and multiple systems.  They are very common and their formation 
involves both disk   fragmentation
and accretion.  Binaries are particularly important for a number of reasons. 
They result in
redistribution of angular momentum during formation and have an important effect on the
distribution of masses, the Initial Mass Function (IMF). 
Furthermore,  stars on the Zero Age
Main Sequence (ZAMS) are at their minimum radius, and hence as they evolve and expand,
some binary stars will diverge in their evolution from single stars  if
they become large enough to undergo Roche lobe overflow.  For the most
massive stars (O stars) it has recently been estimated that  more
than 70\% will exchange  mass  through this process (Sana et al.
2012), 
and a third of those will merge.  This has a dramatic effect on the predictions for post-main 
sequence evolution.  It is  important
to determine what fraction of less massive B stars meets this
fate. Furthermore, high mass stars  are destined to
become compact objects.  Some of those in binary systems will ultimately
provide a zoo of exotic end-stage objects:  symbiotic
stars, novae, cataclysmic variable stars, supernovae.   For all these
reasons, we need to determine the properties of the population of
binaries in order to link these stages together.  (In this paper we will use
``binaries'' as short hand for binaries or higher order multiple systems.
Indeed, in many contexts  the tightest subsystem in a
multiple functions as a binary.)

In order to untangle and understand  all these processes, binary properties are
needed as a function of stellar mass, binary mass ratio, and separation.  
Because star formation covers many  decades of separation (with very
different properties of the ``raw material'') from the
beginning of the collapse to the ZAMS, it is probable that  binary properties 
also  differ depending on, for instance, the separation. 
We begin by summarizing what is known about binary properties of stars with 
different masses. 

{\bf Solar Mass Stars:}
Binary properties of low mass stars are reasonably well known, notably
through the seminal study of Duquennoy and Mayor (1991).
The combination of CORAVEL velocities (below) with visual binary results
and common proper motion pairs produced a distribution of binary periods which 
has become the cornerstone of the discussion of binary systems.    
The recent  study of Raghavan et al. (2010) expanded the results 
considerably, and 
Tokovinin (2014) added further discussion of hierarchical multiples.

{\bf O Stars:}
Properties of massive and intermediate mass stars (O and B) stars are
less well determined.  They are rarer, and hence typically more
distant.  Their spectral lines are broad, and hence velocities cannot
be as precisely determined as for cooler stars.  Finally,
particularly for the most massive stars (O stars) the interpretation is
complicated by high rotation and continuous mass loss.  
 O star binary properties have been discussed by Sana et al. (2013)
 and Kimiki and Kobulnicky (2012) and references therein, as well as Kobulnicky, et 
al. (2014),  Caballero-Nieves et al. (2014) and Aldoretta et al. (2015).

{\bf B Stars:}
B stars  provide a valuable step in tracing the
progression of binary properties from low mass to high mass, and
largely avoid the complications of mass loss in interpretation. 
Radial velocity studies of B stars  date back many years.  
Wolff (1978) 
found that approximately 24\% of late B stars are
binaries with periods $<$ 100 days and mass ratios M$_2$/M$_1$ $>$
0.1.  Abt, and coworkers (e.g. Abt,  Gomez, and Levy [1987]) combined
radial velocities with visual binaries and common proper motion stars
for B stars 
and discussed the implications on star formation.  Similar work was
done by Levato et al. (1987).  
 
Binary studies of B stars have been enhanced greatly by interferometry
and high resolution imaging techniques.  For instance, using speckle 
interferometry, Mason et al.
(2009) found that 64\% of the B stars in their OB star sample 
had companions with $\Delta$ V $<$ 3
mag between 0.03'' and 5'' separation.   However, many stars were included on the 
target list because they already 
showed some indication of being binary (Mason 2014, private
communication), so the sample is not unbiased.


Intermediate mass stars binaries in Sco OB2 were studied by
Kouwenhoven et al. (2007), who found a binary fraction of $>$70\%.
Shatsky and Tokovinin (2002) made an adaptive optics survey of the B
stars in Sco OB2, identifying essentially all companions with
separations  between 45 and 900 AU.  They found a companion star
fraction of 0.20 $\pm$ 0.04 per decade of separation. 

{\bf A Stars:}
Binary properties of A stars  (only slightly less massive than B stars) 
 are discussed by De Rosa et al. (2014). They find a different
distribution of mass ratios for separations smaller and larger than
125 AU (a period of 570 years).  They find 44\% of their volume limited sample is binary or
multiple.

Stellar multiplicity over all masses and
separations was recently summarized  by Duchene and Krauss (2013), who
find an increasing multiplicity fraction with increasing mass.

{\bf Cepheids}:  Studies to identify and characterize binary Cepheids have 
been numerous, particularly in the effort to measure masses to provide 
information to resolve ``the Cepheid mass problem''.  
Cepheids, which began on the main sequence 
as B stars, provide several ways to improve our
knowledge of binary properties of intermediate mass stars.  It is the aim
of the present study to use Cepheid velocities to contribute to our 
understanding of binary properties of intermediate and massive stars.  
A variety of techniques 
have been used to determine the distributions of periods and mass ratios
of Cepheids.  
Two recent summaries are provided by Szabados\footnote{http://www.konkoly.hu/CEP/orbit.html}
and by Evans et al. (2013).  

We have undertaken three related studies to determine the properties
of intermediate mass stars, in particular Cepheids, and their
progenitor B stars.  This is both to determine whether the
consequences of binarity are as severe as for O stars, and also
because  some binary properties  can be particularly
well determined making use of the characteristics of these stars.  
In this study we focus on a property of Cepheids which sets
them apart from their main sequence counterparts, namely their sharp
spectral lines and accurate velocities.  This classic approach to
binary studies lets us probe both systems with periods longer than a
few days and also systems with low mass secondaries.  
In the second approach, we  use  X-ray studies of late B stars to determine
the fraction which have low mass companions (Evans et al. 2011). 
Late B stars themselves do not in general produce X-rays, but
companions later than mid-F spectral type are copious X-ray producers
at the young age of the system.  Thus low mass companions can be 
identified through X-ray observations of late B stars in a cluster.
Third,  we have  made a survey with the 
Hubble Space Telescope (HST) Wide Field Camera 3 (WFC3) to
determine the properties of resolved companions of Cepheids (Evans, et
al. 2013). This study demonstrates another contribution
Cepheids can make to the understanding of binaries.  Since many
companions dominate in the satellite ultraviolet, the mass of the
secondary, and hence the mass ratio, can be determined from an
uncontaminated spectrum.  

Ultimately, this combination of techniques will enable us to provide a
much improved description of the binary properties of these
intermediate mass stars.   
There is one feature of Cepheids, however, which makes 
their binary properties more difficult
to interpret (or possibly will help us get a handle on another aspect
of binarity).
Since Cepheids are post red-giant branch stars, short period binary
systems have undergone interactions, presumably resulting in mergers
in a number of cases.  Indeed, the shortest period system in the Milky
Way containing a Cepheid has an orbital period of a year (Sugars and
Evans 1996). An additional complication to the interpretation is 
that  multiple systems may undergo dynamical
evolution  until they reach a stable hierarchical state. A
component (typically the smallest) may be ejected from the system in the 
process.  

The advent of correlation spectrometers vastly increased the quantity
of accurate radial velocities available for Cepheids.  The CORAVEL instrument at
the Geneva observatory 
(Baranne et al. 1979) observed 
a large  number of Cepheids for
studies of galactic structure and to obtain Cepheid distances via a
variant of the Baade Wesselink technique.   
The accuracy of the velocities is typically less than 1~km~s$^{-1}$ in the
magnitude range of the stars in this study. A similar instrument was
built at Moscow University (Tokovinin 1987).  Between these
instruments (including an extension to the Southern hemisphere), a large
number of Cepheids has been observed since 1978.  A major purpose of
these observations was to discover binaries which can then be
used to measure Cepheid masses.


\subsection{The  Scope of this Project}

While radial velocity spectrometers have produced a huge amount of
velocity data for Cepheids, the data have never been analyzed to determine
the fraction of stars which show no orbital motion over the more than
three  decades they cover.  That is the goal of this project, CRaV
 ({\bf C}epheid  {\bf Ra}dial  {\bf V}elocities).  
  Many Cepheids have been identified as
 binary systems.  However, the  binary frequency is only
obtained if we know how many single stars are also included in the sample.  
The main aim of this project is  to examine the sample of stars 
for orbital motion, or in its absence, to produce a
well characterized sample of stars without orbital motion.  
We have deliberately  limited the data examined 
to observations with a typical accuracy of 1 km s$^{-1}$ per
observation and studies which included many Cepheids to allow
cross-checks on the accuracy.  In this first paper, we include data
from 1978 to 2000 of ``Northern stars'' down to -20.9$^0$ (as discussed
below). We plan two subsequent papers on Southern stars and
observations since 2000.  As discussed in the detection limits
section, with this sample, we will detect almost all binaries with
orbital periods of 1 to 20 years down to mass ratios q = M$_2$/ M$_1$
= 0.1 (where M$_1$ is $\ge$ M$_2$), that is companions as cool as K stars.

We begin with the 38 stars in Table 2 which meet these criteria, omitting
stars known to be binary.  To complete the analysis, we examine all 
stars brighter than 9$^{th}$ magnitude (Table 6, 62 stars), ultimately
omitting stars deemed to have too little information. 

The purpose of CRaV is to investigate the annual mean systemic
velocities of Cepheids with accurate velocities.  
The observed Cepheid velocity, of course, is a combination of the
systemic velocity of the star and the pulsation velocity curve.  To
correct for the pulsation velocity, we use a Fourier representation of
the curve with up to 20 terms.   The velocity curves have
to be aligned over decades by means of a  pulsation period.  Both
these steps will be discussed in sections below.  Finally,
cross-checking the results of the annual means between studies
identifies and removes small systematic differences.

\section{The Sample}

The accuracy with
which a Cepheid velocity can be measured (1 km s$^{-1}$) is easily the highest for
high and intermediate mass stars.
While this accuracy in the annual means does not come
close to the highest accuracy possible today (e.g. Anderson 2014), the long sequence of
data at this level is valuable.   The details of the sample which is available
at this level of accuracy are discussed in this section.     

The Moscow velocities (Gorynya et al. 1992, 1996, 1998, referred to as 
Gorynya et al. below) are a large dataset for
which many stars are covered annually in the 1990's.  For this reason,
we are defining our ``Northern Sample'' to be stars with Dec
$>$ -20$^\circ$.9,  the region they covered.


For the most southern stars in the sample, data from a few studies have
been deferred to
the next paper (Southern stars) so that we can assess 
 the zeropoints of the whole study.
Datasets in this category are Petterson et al. (2004), Coulson and
Caldwell   (1985), Coulson, Caldwell, and Gieren,
(1985), Gieren (1981), and  Caldwell et al. (2001).
On the other hand, CORAVEL data from both  Observatoire de
Haute-Provence and ESO La Silla 
 (Bersier et al. 1994 and Bersier 2002) have been included since they
were taken and analyzed with the N hemisphere stars by the Geneva group.

\begin{deluxetable}{llll}
\footnotesize
\tablecaption{Sources \label{sources}}
\tablewidth{0pt}
\tablehead{
\colhead{Id} & \colhead{Symbol}& \colhead{Symbol}  & \colhead{Source} \\ 
 \colhead{} & \colhead{Table}& \colhead{Plot*}  & \colhead{} \\  
}
\startdata


 1 & gg & r circle &  Gorynya et al.  1992, 1996, 1998   \\
 2 & b9 & r diamond & Bersier et al. 1994  \\
3 & im & c \^ & Imbert  1999 \\
4 & ba & g x &  Barnes et al. 2005 \\
 5 & kk & m x & Kiss and Vinko 2000  \\
 6 & b0 & y diamond & Bersier  2002 \\
 7 & s4 & c square  & Storm et al. 2004 \\

\enddata

* r = red; c = cyan; g = green; y = yellow; m = magenta

\vskip .1truein


\noindent

\end{deluxetable}


\begin{deluxetable}{lllrlrllrll}
\footnotesize
\tablecaption{Data Sources  \qquad\qquad\qquad\qquad  \label{master.list}}
\tablewidth{0pt}
\tablehead{
\colhead{Star} & \colhead{1*}  
  & \colhead{2*} & \colhead{3*} &
\colhead{4*}  & \colhead{5*} & \colhead{6*}  &  \colhead{7*} 
  & \colhead{P} & \colhead{$<$V$>$} & \colhead{Mode} \\
\colhead{} & \colhead{}  
  & \colhead{} & \colhead{} &
\colhead{}  & \colhead{} & \colhead{}  &  \colhead{} 
  & \colhead{$^d$} & \colhead{mag} &  \colhead{} \\
}

\startdata


$\eta$ Aql    & gg &  &    & ba   &   kk & b0  & s4  &  7.17  & 3.90 &    \\
SZ Aql &  &  &  & ba     &  &   b0   & &   17.14 &  8.60  &    \\
TT Aql    & gg & b9 & im & ba   &    & b0  &   &   13.75  &   7.14   &   \\
FM Aql    & gg &    &    & ba   &    &   &   &   6.11 &   8.27   &    \\
FN Aql    & gg &    &    & ba   &    &   &  &   9.48  &  8.38  &     \\
V1162 Aql & gg &    &    &     &    &   &   &   5.38  &  7.80  &   \\
RT Aur    & gg &    &    &     &   kk &   &  &  3.73   &   5.45 &    \\
RX Aur    & gg &    & im  &    &    &   &   &   11.62  & 7.66   &   \\
CK Cam    & gg &    &    &     &   kk &   &   & 3.29  & 7.54  &   \\
SU Cas    & gg & b9 &    &     &   kk &   & s4  &   2.74  &   5.99   & o  \\
V379 Cas  & gg &    &    &     &      &   &   &   4.31 &  9.05   &  o  \\
V636 Cas  & gg & b9 &    &     &      &   &  &  8.38  &    7.20   &   \\
$\delta$ Cep    &  & b9 &     & ba   &   kk &   & s4    &   5.37   &    3.99  &  \\
IR Cep    & gg &    &    &     &      &   &   &  2.98   &   7.78  &  o  \\
X Cyg     & gg & b9 &    & ba  &    kk &   & s4  &   16.39 &    6.39  &   \\
CD Cyg    & gg &    & im &     &       &   &   &  17.07 &  8.95 &   \\
DT Cyg    & gg & b9 &    &     &    kk &   &  &   3.53  &    5.77  & o   \\
V1726 Cyg & gg &    &    &     &      &   &   &    4.24  &  9.01 &  o  \\
$\zeta$ Gem  & gg & b9   &     &     & kk &   &    &  10.15   &   3.92 &   \\
W Gem     & gg &    & im  &    &      &   &  &   7.91   &   6.95  &   \\
V Lac     & gg &    &    &     &      &   &   &   4.98  &  8.94 &   \\
X Lac     & gg & b9 &    & ba  &      &   &   &   5.44  &  8.41  & o? \\
RR Lac    & gg & b9 & im &     &      &   &   &   6.42  &  8.85  &  \\
BG Lac    &  &  &  im    & ba   &       &   &  & 5.33  &  8.88  &  \\
SV Mon    & gg &    & im &     &      &   &   &  15.23 &  8.22  &  \\
Y Oph     & gg &    &    &     &      &   &   &   17.13  &   6.17   &   \\ 
RS Ori    & gg &    & im &     &      &   &    &  7.57  &  8.41  &   \\
V440 Per  & gg & b9 &    &     &      &   &   & 10.94  &    6.28    & o \\
U Sgr     & gg & b9 &    &     &      & b0 & s4  &   6.75  &    6.70  &   \\
WZ Sgr    & gg &    &    &     &      & b0 &    &   21.85 &  8.03  & \\
BB Sgr    & gg &    &    &     &       &   &   &  6.64  &    6.95  &   \\ 
ST Tau    & gg & b9 & im &     &      &   &   &   4.03  &  8.22    &    \\
SZ Tau    & gg & b9 &    &    &    kk &   &   &  4.47  &    6.53  &  o \\
EU Tau    & gg & b9 &    &     &      &   &   &  2.97  &  8.09   & o  \\
S Vul     & gg &  &      &     &      &   &    &  68.46  & 8.96  &  \\  
T Vul    &  & b9 &       & ba   &   kk &   &   &   4.43  &    5.76  &   \\
X Vul     & gg & b9 &    &     &      &   &    &   6.32   &   8.85  &  \\
SV Vul    & gg & b9 & im & ba   &   kk   &   & s4  &  44.99  &  7.22  &   \\   
  &  &  &  &  &  &  &  &   &     \\


\enddata    

*:  data sources identified in Table~\ref{sources}

\end{deluxetable}


The sample consists of Cepheids brighter than 9.0 mag north of
declination -20$^\circ$.9, not already known to be binaries.  
There are a few stars which otherwise fall within these criteria which
we have not included.  
We have not used double mode pulsators (CO Aur, TU Cas, and EW Sct)
because  of the complexity of their light and velocity curves.  We have also
omitted V473 Lyr, the only Cepheid know to have Blazhko-like
variations in amplitude.  RY CMa did not have a well covered velocity curve 
during a single season,  making a Fourier fit unreliable.   


Table~\ref{sources} lists the sources of data used in the study and
 Table~\ref{master.list}
summarizes the data used for each of the stars. 
Period and $<$V$>$ are
taken from Fernie et al. (1995) (except for CK Cam, which is from
Berdnikov et al. 2000).  The farthest right column identifies stars
pulsating in an overtone mode, based on the discussion of Evans et al.
(2015).  IR Cep is also classified as an overtone (Groenewegen and Oudmaijer 2000).

During the course of this project, relevant data, means, and other
parameters such as Fourier coefficients were stored in a database
(created by D. Morgan).  This proved invaluable for tracking the
steps, and occasional updating as needed.




\section{Data Treatment}

\subsection{Fourier Curves}

In order to remove the pulsation velocity from the observed velocity,
we (LB) fit the pulsation curve with a Fourier series.
In general,  all the data from Gorynya et al. sources were fit in order to
get well determined curves tightly constrained by the
data. Occasionally other data were included to obtain a well covered
pulsation cycle. (Exceptions were also sometimes made for 
stars with more difficult periods as discussed in the next section.) 
With this
data up to 20 Fourier coefficients are needed from the fit.
Note that one reason for restricting  the sample to stars with velocity 
curves which are well covered  
by high accuracy data is that this quality and quantity of data is required 
for a Fourier fit which will provide a pulsation velocity for any phase with the 
necessary accuracy (1 km s$^{-1}$).  

The pulsation curve is well represented by a  Fourier  series:




$$  V_R = c(0) + \sum_{i=1}^{10} (a_isin(2\pi i\phi) + b_icos(2\pi i\phi)) $$


\noindent where V$_R$ is the radial velocity, c(0) is the systemic velocity, $\phi$ is the phase, 
$a_i$ and $b_i$ are the Fourier coefficients, and $i$ runs from 1 to 10 (as needed).  

\begin{figure}
 \includegraphics[width=\columnwidth]{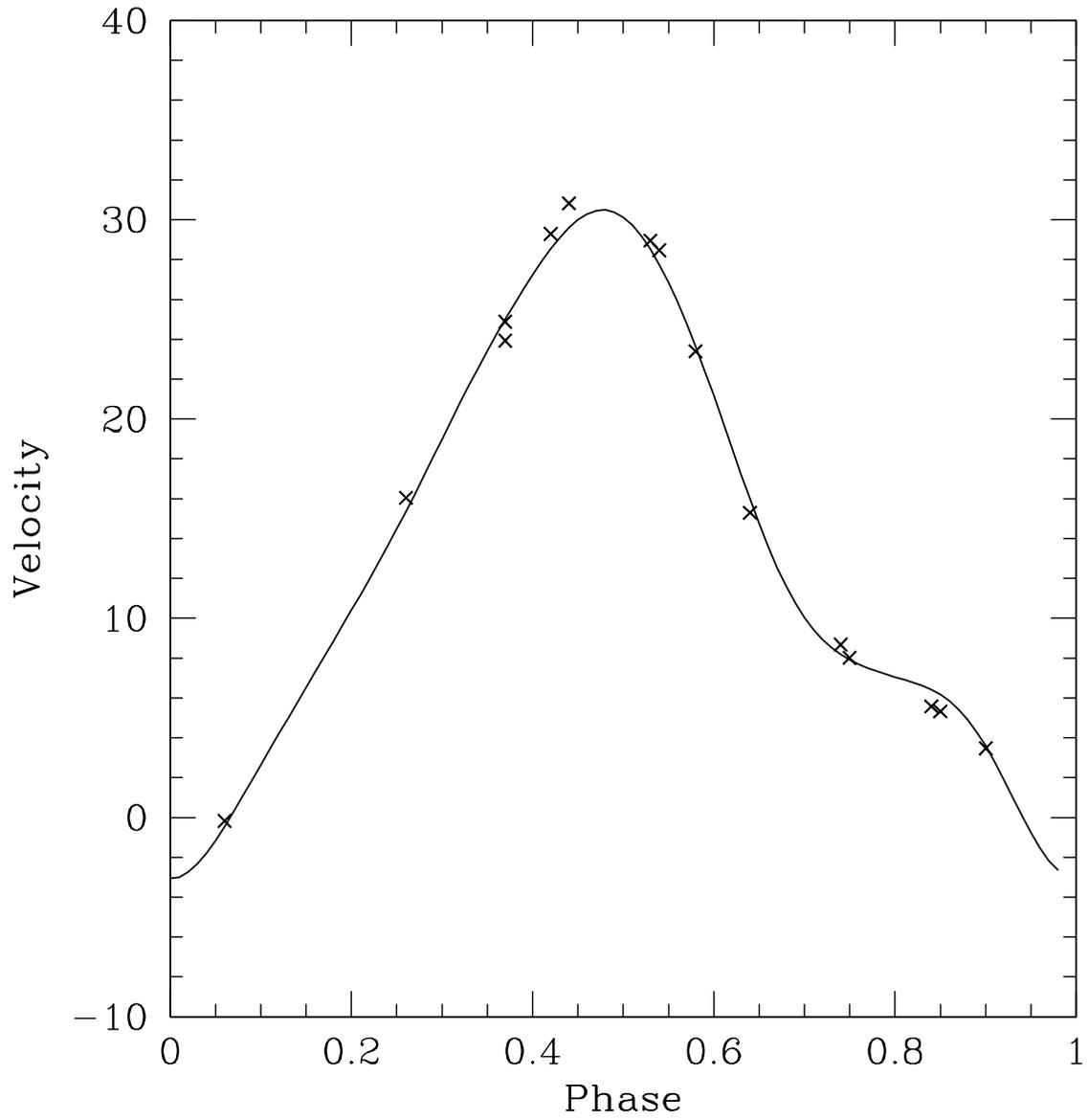}
\caption{Sample velocity curve for FN Aql.  Solid line is the Fourier
  fit;  the data (shown with x) are from Gorynya et al. for the year {\bf 1989}. 
Velocities are in km s$^{-1}$.
As is typical,
the Fourier fit was made using the Gorynya et al. data for all seasons (79 points in 
total). 
\label{fnaql.four.eps}}
\end{figure}

A sample curve is provided in Fig.~\ref{fnaql.four.eps}
showing the data for FN Aql
from Gorynya et al. for 1989, and the Fourier representation.  The
Fourier coefficients for each star are listed at http://hea-www.cfa.harvard.edu/\~\/evans/.

 The fitting
process also determined a period and epoch of maximum light which is
included in the table on the website.  While this period should be used to generate
the Fourier plot, it should {\bf not} be used when
data taken over many years are to be phased together. As needed, the
phase shift between the two periods was incorporated in the analysis. 
(Fourier coefficients for S Vul and SV Vul are included, although 
ultimately the means used were from  independent year by year solutions  as
discussed in Appendix A.)

\subsection{Pulsation Periods}

Cepheids have famously repetitive light variations. For most stars we
have been able to identify a constant period or variation
in period which has been
parametrized as a parabola (changing period).  Using these well
determined periods, largely derived from photometry, phases were
computed for the velocity measures.  Combining these velocities with
the Fourier representation resulted in the mean velocity difference for
the observations for each year, as discussed in the next section.  
With this approach assuming the phase
from the periods in the Fourier fit 
(http://hea-www.cfa.harvard.edu/\~\/evans/), years in which comparatively few
observations were obtained still provide a mean velocity.


In a few cases pulsation periods are too   unstable for long term
projection.   Long period high luminosity stars sometimes fall in this
category  (S Vul, and SV Vul). Stars pulsating in an overtone also
seem to have unusually large period variations (Szabados 1983;
Berdnikov et al. 1997; Evans et al. 2015).  Some
stars in these groups have period fluctuations on longer
timescales than the 20 years discussed here, so a single period was
adequate to phase the data.  
Since it is desirable to have
binary information on both high mass (long period stars) and overtone
pulsators, in some cases we
have adopted a more complicated approach.  
Specifically, for four stars (EU Tau, SV Mon, ST
Tau, and X Lac) we had to develop an individual approach to the
phasing of the curves, typically based on photometric curves.  Details
are provided in Appendix A. 
Working through the sample,  we became aware that the 5$^d$ Cepheid
X Lac has an unusually erratic period.  Because of this
we suggest that it may be pulsating in the
first overtone. Various diagnostics of overtone pulsation are
considered in the case of X Lac and discussed in Appendix A.    

Where the periods are erratic (SV Vul, and  S Vul)
fits to the velocity curve for each year have
been made.  The fits produce the
amplitude of the curve, the phase of maximum light, and also the mean
velocity without assuming a predicted period.
 For these  cases  we  are restricted to seasons with a
significant amount of data to obtain the
 systemic ($\gamma$)  velocity corrected for pulsation. We do not
include  these seasonal
means in the final cross check on annual means of individual data
sets (see Sect 5).
All these stars for which we have used a special approach are discussed in
Appendix A.









\subsection{Cepheid Velocities}

This section discusses the way in which the Fourier curves, velocity
data, and periods are used to generate annual mean systemic ($\gamma$)
velocities for the Cepheids.


When the period varied a little (or was a little different from the 
one in the Fourier fits), we made an adjustment with  small phase shifts.  
This may sound arbitrary, but in cases of stars with  erratic variations or
even a constantly changing period, the single best group of data
(typically the Gorynya et al. velocities) was required to tightly
constrain the Fourier fit.  Only high quality data providing a well
covered pulsation cycle will constrain the Fourier fit to the level of
accuracy needed in the project ($<1$~km~s$^{-1}$).   
To emphasize, 
the period in the Fourier coefficient table is the one that 
should be used to generate the Fourier curve.  Any deviation from 
that period  during the whole time span of the data
was included with a subsequent phase shift. 
The next step is to compare data from each season 
with the Fourier curve.    The average difference between the data and
the  curve is computed, creating the annual mean velocity. 
As discussed in the next section, a small instrumental
correction was computed for each instrument for 
each season using  all Cepheids (Table \ref{ann.corr}). 
The annual mean velocity from the data and the Fourier curves
is listed  in Table \ref{ceph.ann}.   
 Columns in Table \ref{ceph.ann} list the star, the data
source, the year, the mean and standard deviation $\sigma$ 
and the number of observations.    The corrections in
 Table \ref{ann.corr} have been included.  

Since Cepheids occur close to the galactic plane instead of randomly
over the sky, observations are typically confined to a season. Hence,
the division of
velocities by calender year is appropriate.  For Cepheids in the
winter sky, some observations from a given source might be taken in
December, with related observations in the following January.  We have
investigated whether division of these stars by calendar year affects
our results.  For Cepheid periods the optimal
cadence is typically at least a day. Many of the observations were taken
during observing runs of a month, often with more than one run a year.  
We have examined the few instances where a star had a December--January
combination of observations and found that only a very few points
would be assigned to a different year in a more complicated scheme, so
adding complexity does not seem warranted in deriving the annual mean
velocities.



\subsection{Annual Corrections}


In this project we  examine the annual means of the
velocity data for each star (corrected for pulsation velocity) to identify  any
long term variation due to orbital motion.  
Even though the datasets we have used were taken with high dispersion
and well controlled instruments, in order to make comparisons to the
desired accuracy over  two decades, we have made the
following check.   
To test for small instrumental zero point variations, we
have formed the mean and standard deviation ($\sigma$) of all stars observed in
each season for each instrument.  From this we have created small
annual corrections to be incorporated in the analysis, listed in 
Table \ref{ann.corr}.  
Columns in Table \ref{ann.corr} are the data source (Table 2), the year, mean and standard
deviation, and the number of stars.
Corrections $>$0.3 km s$^{-1}$ (absolute value) from at least 4 stars (indicated with
*)  have been added to
Table \ref{ceph.ann} and Fig \ref{vels}. 
Using corrections derived from a smaller number of stars puts
too large a
weight on individual stars, i.e. corrects a possible orbital variation
to 0.0.
In Table \ref{ann.corr}, Column 5, an entry for only 1 star lists 
s.d. of the fit of that data to the Fourier curve as the error.
A sense of the overall velocity comparison between instruments can also
be obtained from Table \ref{ann.corr}. 
 The full table of annual corrections is in the
electronic version;  only a partial section is included here to
provide information about the form and content.

\begin{deluxetable}{llclr}
\footnotesize
\tablecaption{Annual Corrections   \qquad\qquad\qquad\qquad \label{ann.corr}}
\tablewidth{0pt}
\tablehead{
\colhead{Source} & \colhead{Year} & \colhead{Mean} & \colhead{s.d.}
& \colhead{N } \\
\colhead{} & \colhead{} & \colhead{km s$^{-1}$} & \colhead{km s$^{-1}$}
& \colhead{} \\
}

\startdata


 & & & & \\
   gg &    1986 &   0.16 &   0.19 &    1 \\ 
   gg &    1987 &   0.71* &   0.17 &    6 \\ 
   gg &    1989 &   0.25 &   0.07 &    3 \\ 
   gg &    1990 &   0.41* &   0.10 &    6 \\ 
   gg &    1991 &   0.35* &   0.04 &   20 \\ 
   gg &    1992 &   0.25 &   0.07 &   10 \\ 
   gg &    1993 &  -0.39* &   0.09 &   16 \\ 
   gg &    1994 &  -0.34* &   0.05 &   22 \\ 
   gg &    1995 &  -0.42* &   0.04 &   23 \\ 
   gg &    1996 &   0.08 &   0.04 &   22 \\ 
   gg &    1997 &   0.17 &   0.06 &   17 \\ 
   gg &    1998 &   0.18 &   0.10 &   12 \\ 
 & & & & \\
   b0 &    1978 &  -0.16 &   0.06 &    4 \\ 
   b0 &    1979 &  -0.38* &   0.07 &   4 \\ 
   b0 &    1980 &  -0.05 &   0.03 &    5 \\ 
   b0 &    1981 &   0.02 &   0.12 &    4 \\ 
   b0 &    1982 &  -0.69 &   0.39 &    1 \\ 
   b0 &    1987 &  -0.22 &   0.17 &    3 \\ 
   b0 &    1988 &  -0.85 &   0.37 &    3 \\

\enddata

\vskip .1truein


\noindent


*: corrections incorporated in  Table~\ref{ceph.ann}

\end{deluxetable}

\begin{deluxetable}{lllrrr}
\footnotesize
\tablecaption{Cepheid Annual Mean Velocities \label{ceph.ann}}
\tablewidth{0pt}
\tablehead{
\colhead{Star} & \colhead{Source} & \colhead{Year} & \colhead{Mean} &
\colhead{$\sigma$}
 & \colhead{N } \\
\colhead{} & \colhead{} & \colhead{} & \colhead{km s$^{-1}$} &
\colhead{km s$^{-1}$ }
 & \colhead{ }
}
\startdata



 & & & & & \\


  Eta  Aql &    b0 &    1983 &   0.74 &   0.48 &    5 \\ 
  Eta  Aql &    gg &    1986 &   0.16 &   0.19 &   25 \\ 
  Eta  Aql &    s4 &    1986 &   0.12 &   0.28 &   25 \\ 
  Eta  Aql &    b0 &    1989 &   0.37 &   0.36 &   33 \\ 
  Eta  Aql &    ba &    1995 &   2.71 &   ---  &    1 \\ 
  Eta  Aql &    kk &    1996 &   0.58 &   0.33 &    8 \\ 
  Eta  Aql &    ba &    1996 &   1.14 &   0.51 &   13 \\ 
  Eta  Aql &    ba &    1997 &  -0.05 &   0.51 &   16 \\ 
  Eta  Aql &    kk &    1997 &   0.71 &   0.30 &    6 \\ 
 & & & & & \\
   SZ  Aql &    b0 &    1996 &  -0.38 &   0.24 &   21 \\ 
   SZ  Aql &    ba &    1996 &  -0.16 &   0.23 &   13 \\ 
   SZ  Aql &    ba &    1997 &  -0.15 &   0.27 &   19 \\ 
   SZ  Aql &    b0 &    1997 &  -0.29 &   0.57 &   10 \\ 
 & & & & & \\
   TT  Aql &    im &    1989 &  -0.07 &   0.18 &   14 \\ 
   TT  Aql &    im &    1991 &   0.28 &   0.12 &    5 \\ 
   TT  Aql &    im &    1993 &  -0.14 &   0.45 &    5 \\ 
\enddata
\end{deluxetable}



\section{Results}

\begin{figure}
\includegraphics[totalheight=1.5in]{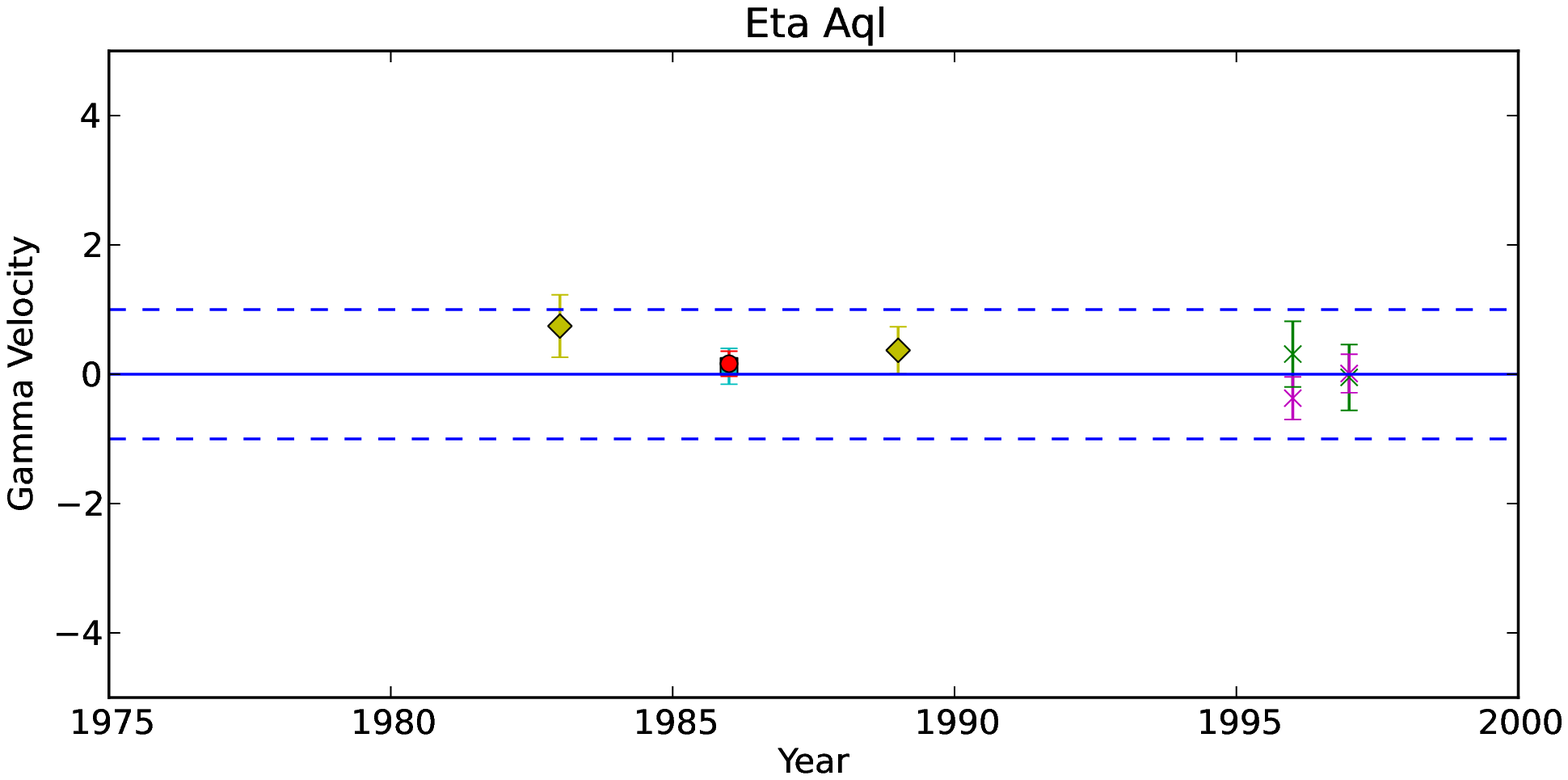} 
\includegraphics[totalheight=1.5in]{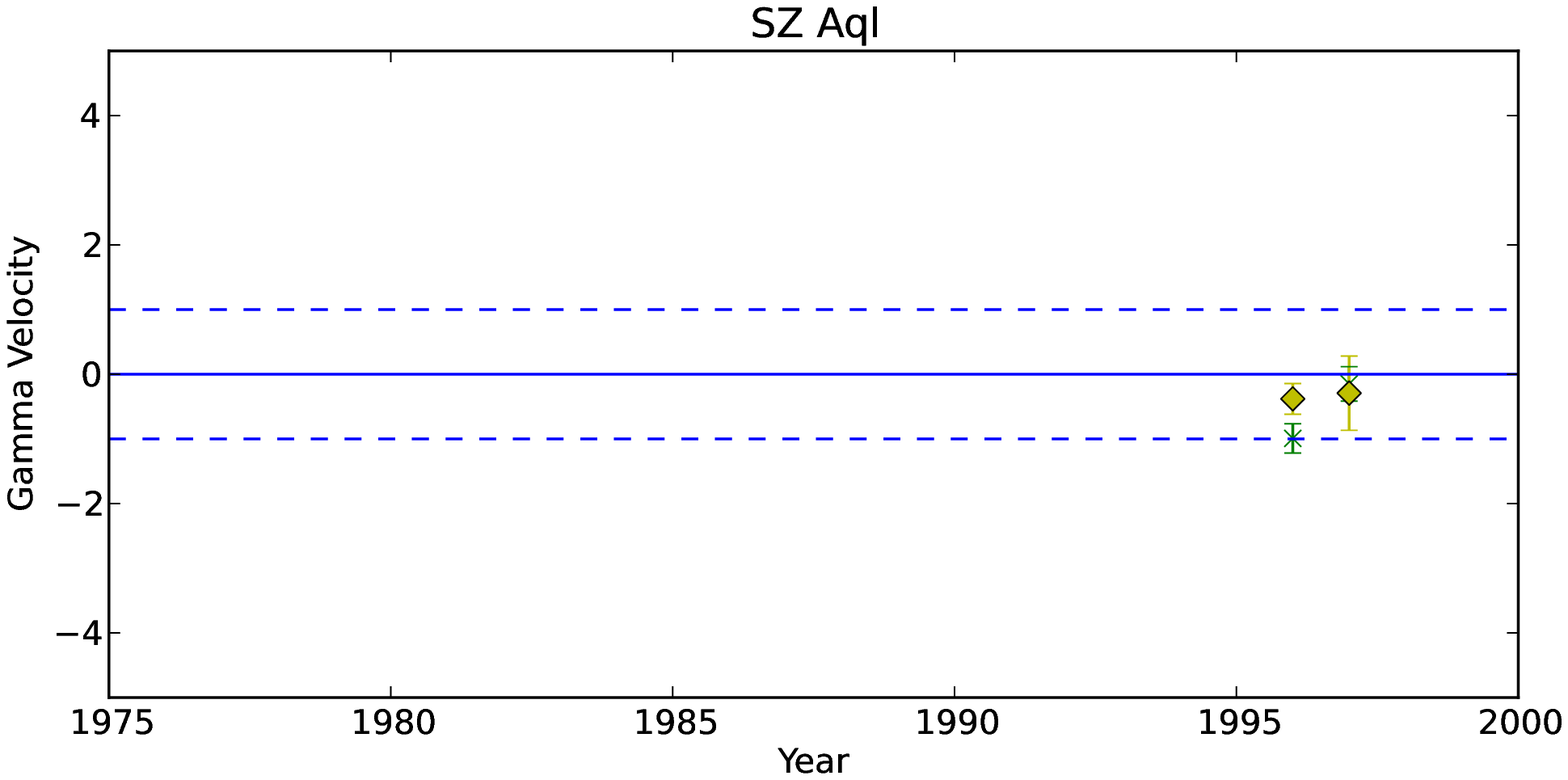} \\
\includegraphics[totalheight=1.5in]{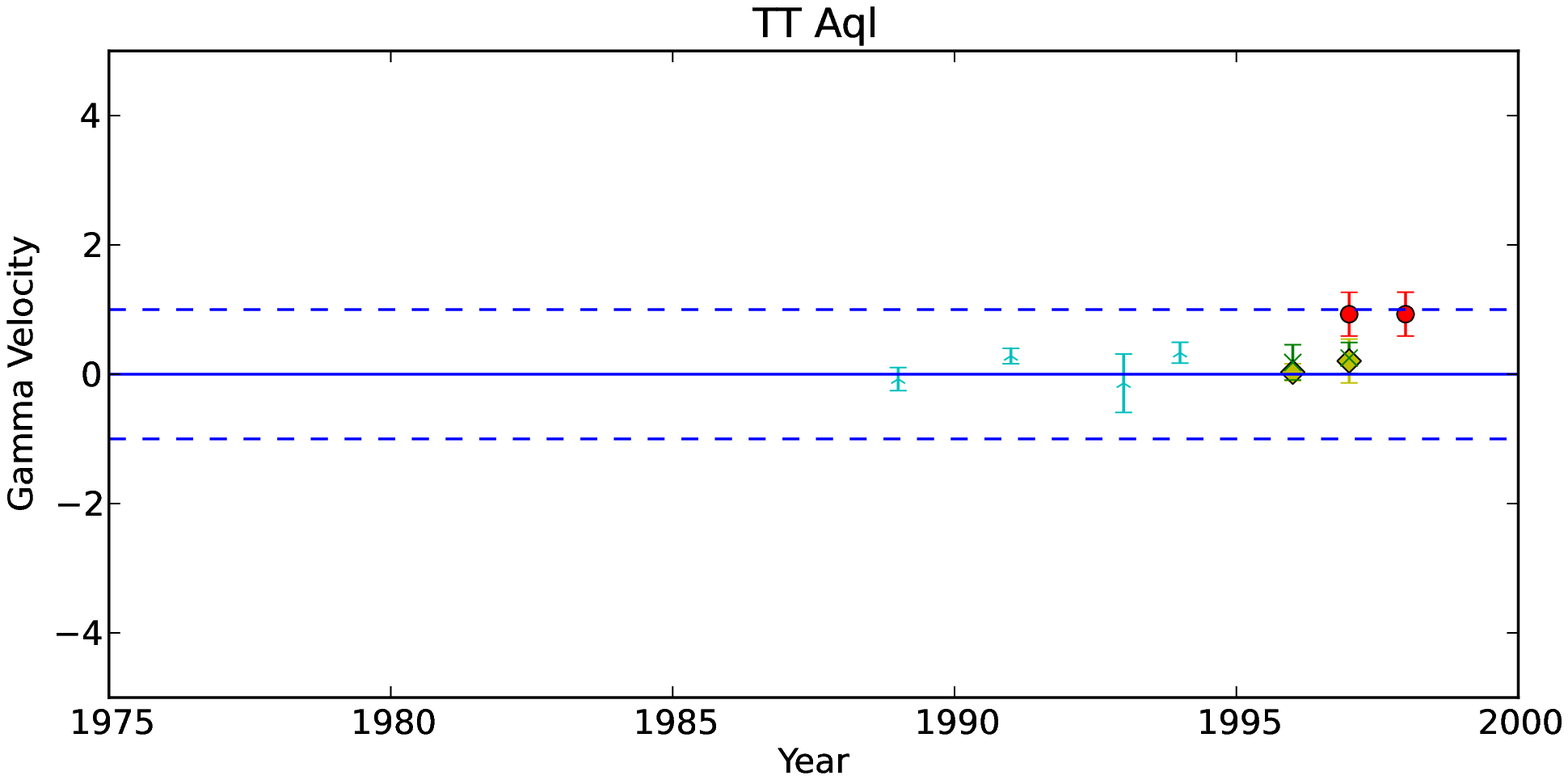}
\includegraphics[totalheight=1.5in]{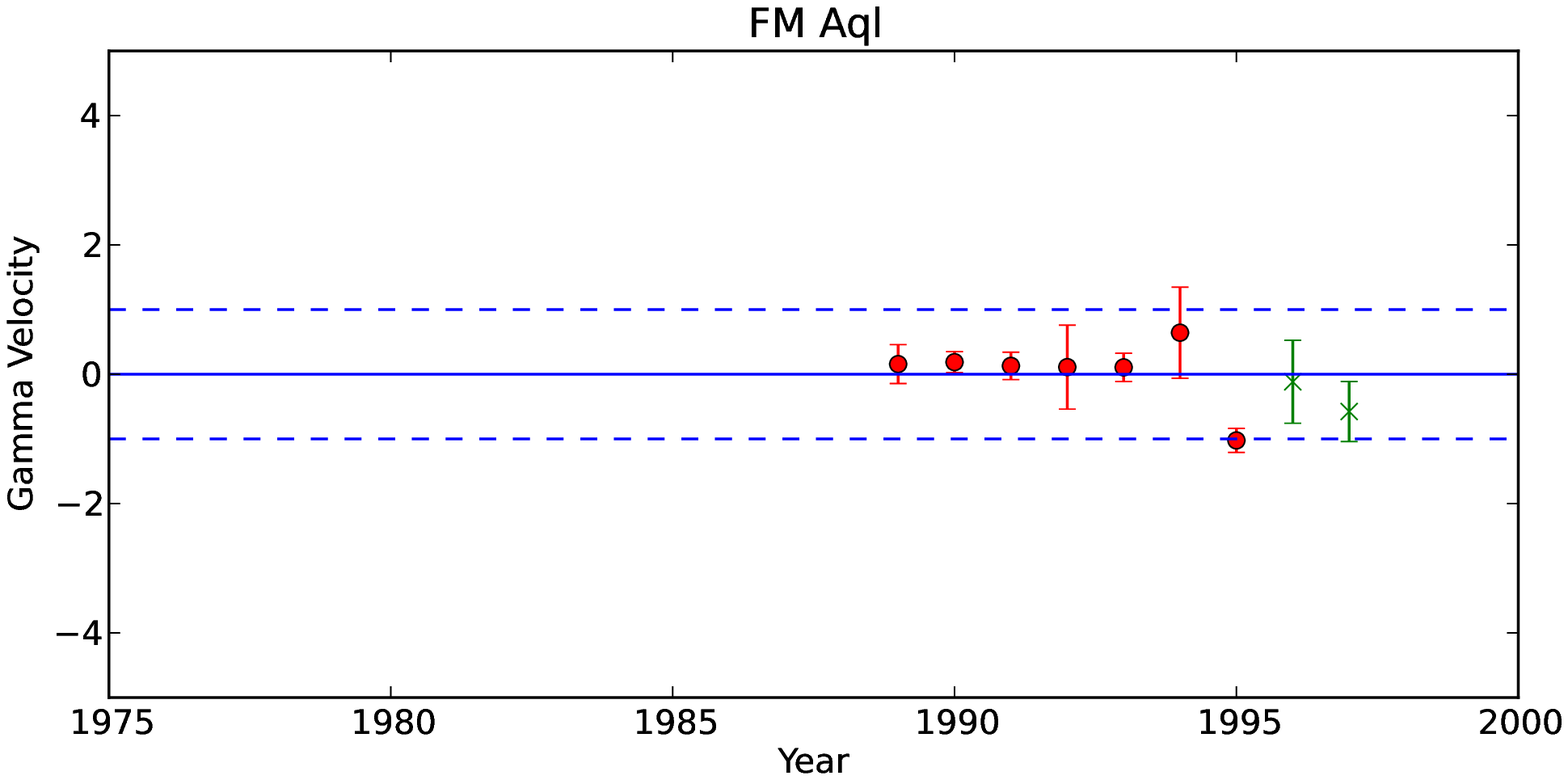} \\
\includegraphics[totalheight=1.5in]{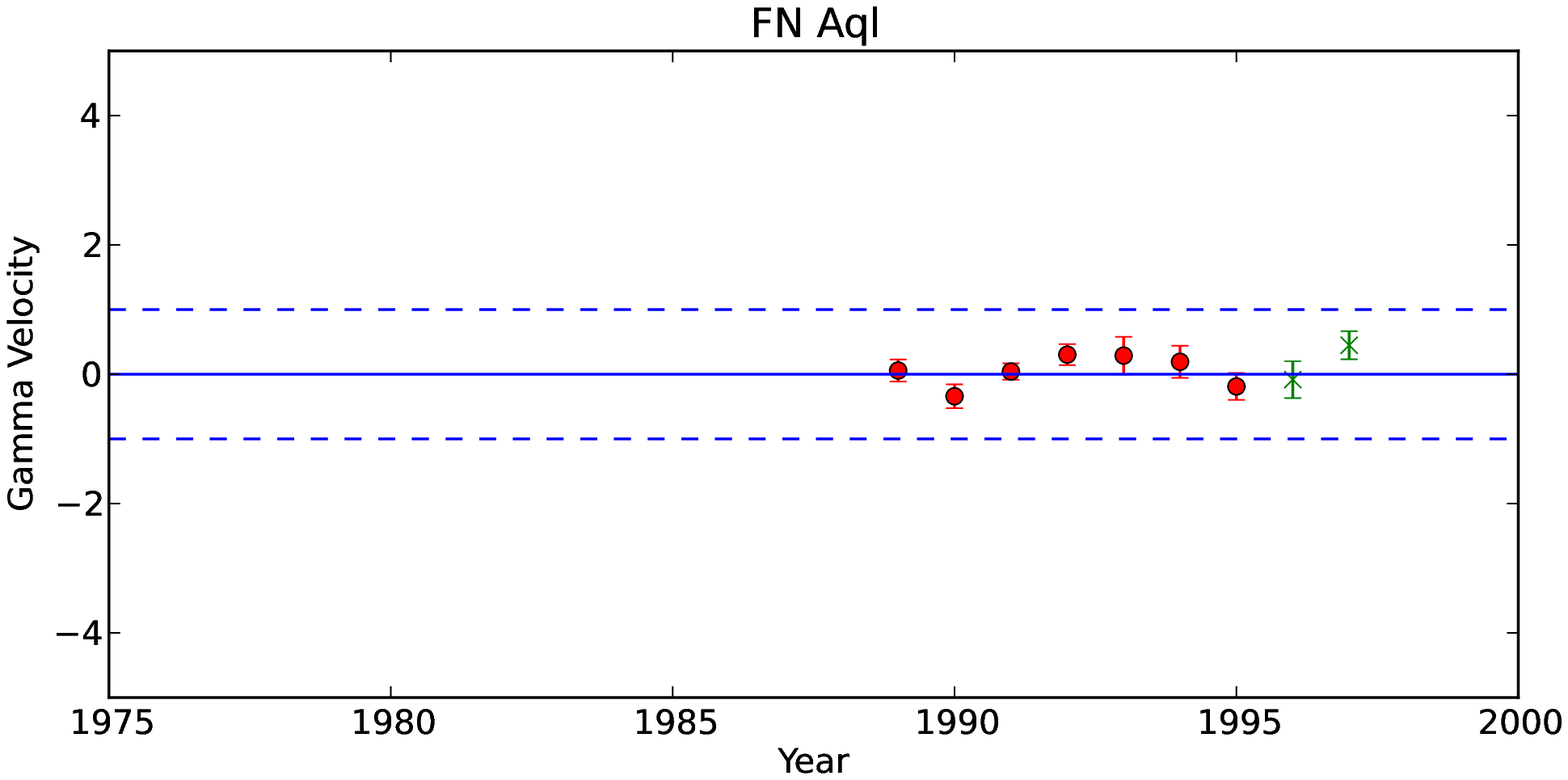}
\includegraphics[totalheight=1.5in]{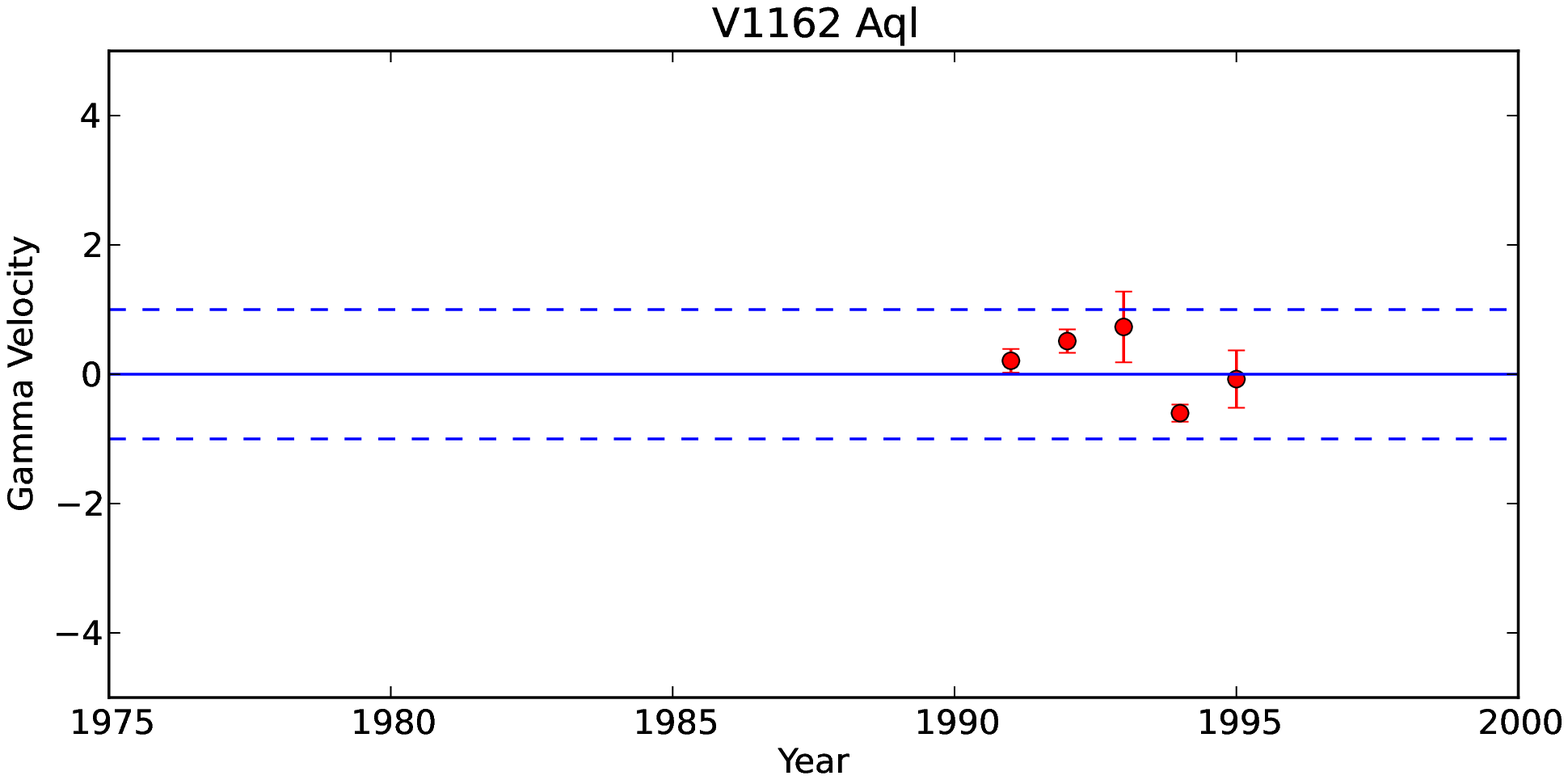} \\
\includegraphics[totalheight=1.5in]{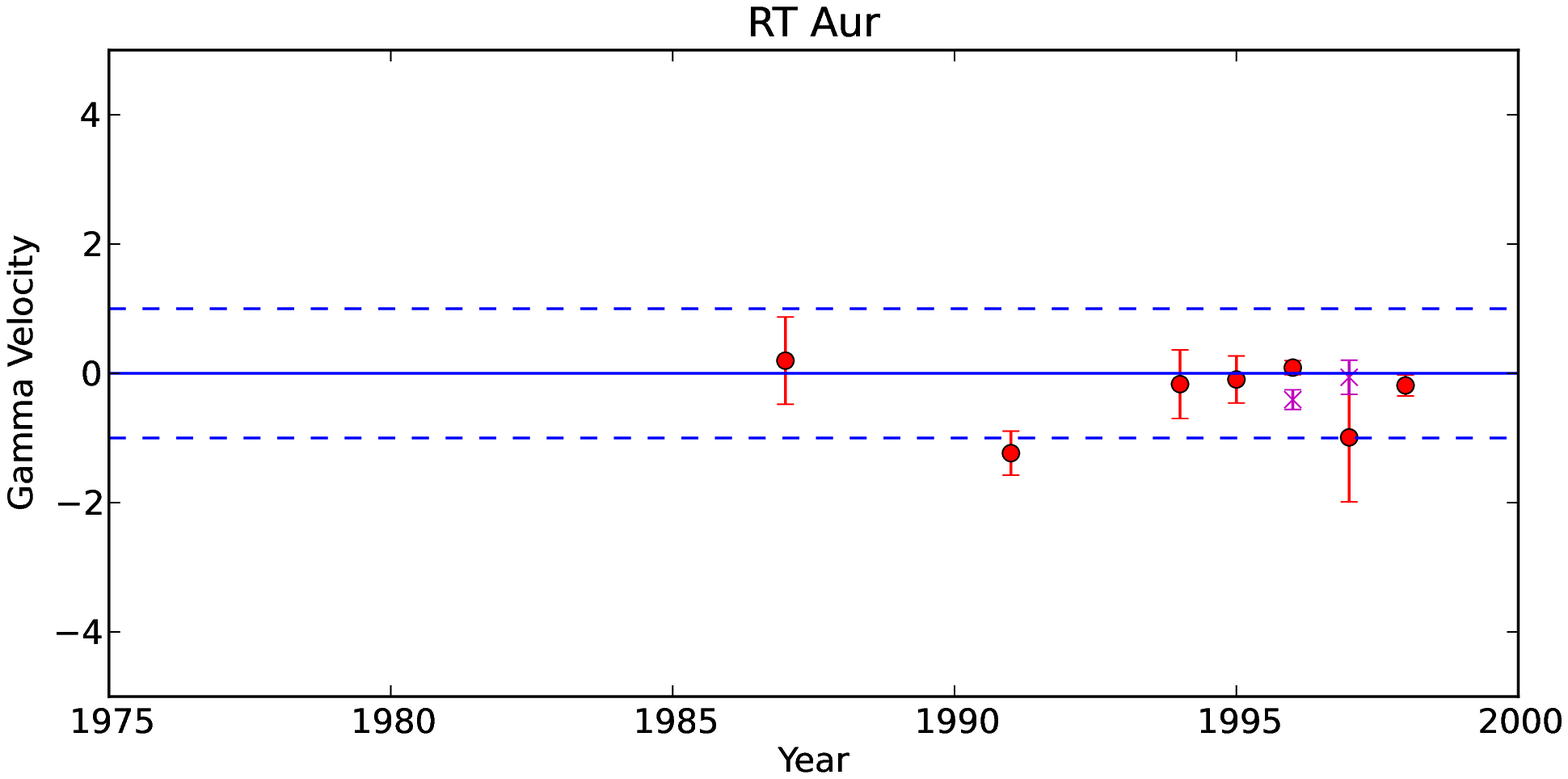}
\includegraphics[totalheight=1.5in]{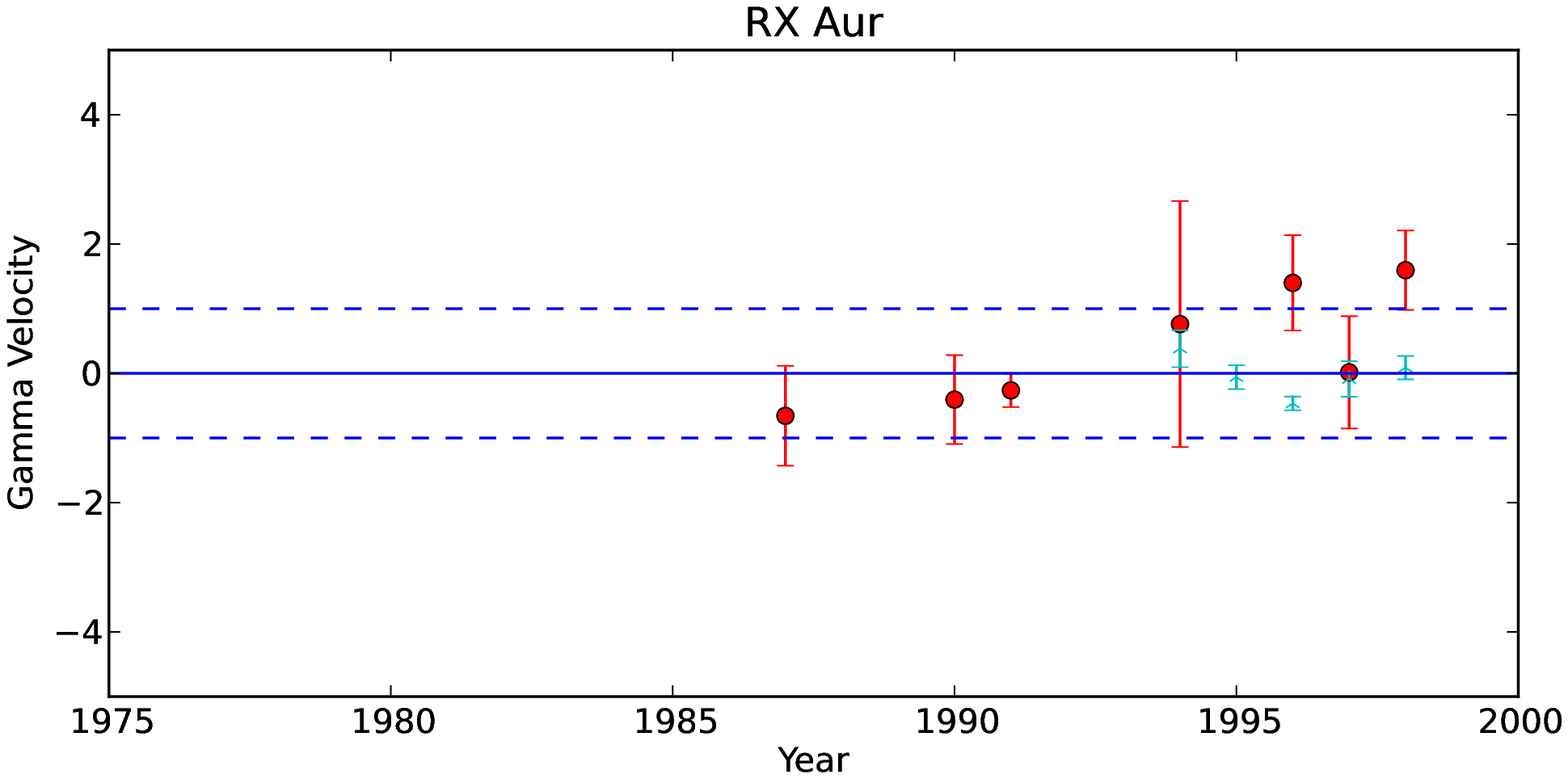} \\

\caption{Annual mean radial velocities corrected for pulsation by year.  Symbols for the
data sources are listed in Table \ref{sources}. All velocities are in 
km s$^{-1}$.  Dashed lines show 
$\pm$ 1 km s$^{-1}$ .\label{vels}}
\end{figure}

The results from this study are presented in Fig \ref{vels}, showing the annual 
mean systemic ($\gamma$)  velocity as a function of year.  Included in the plots is 
the band of $\pm$ 1 km s$^{-1}$.  The symbols for each data source are listed in 
Table \ref{sources}.  Seasons with only one observation have been omitted from the plots.


\begin{figure}

\includegraphics[totalheight=1.5in]{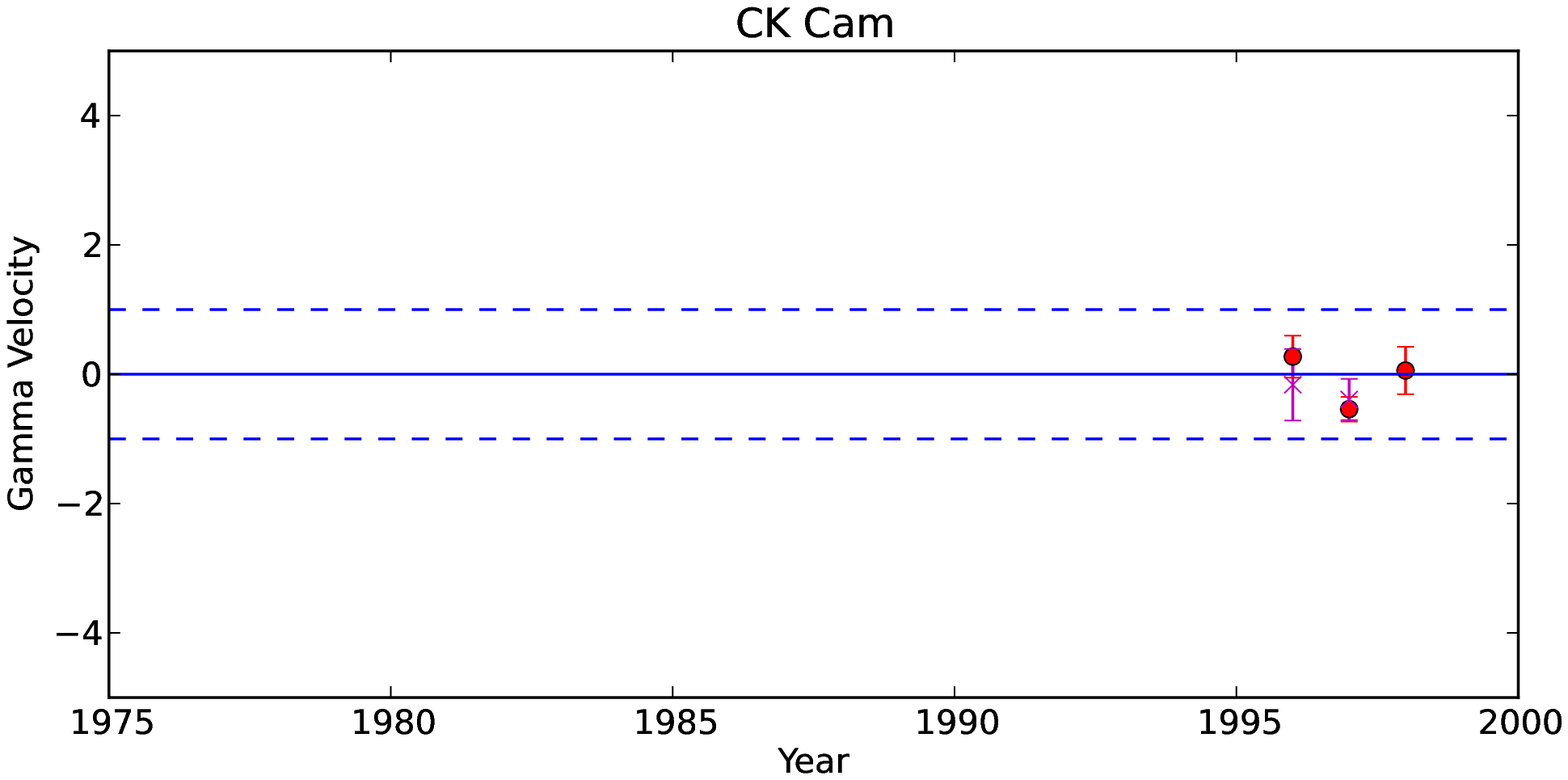} 
\includegraphics[totalheight=1.5in]{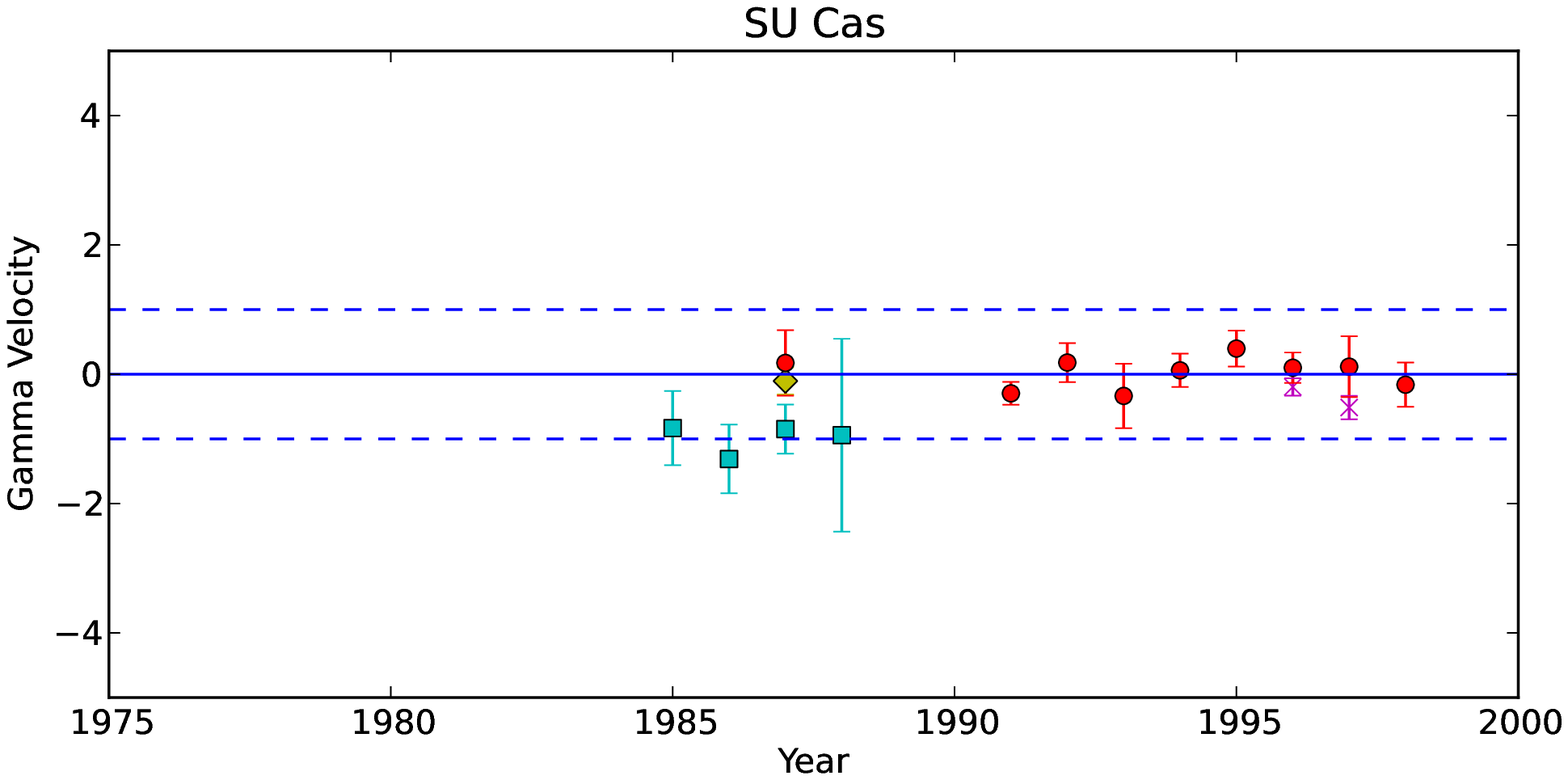} \\
\includegraphics[totalheight=1.5in]{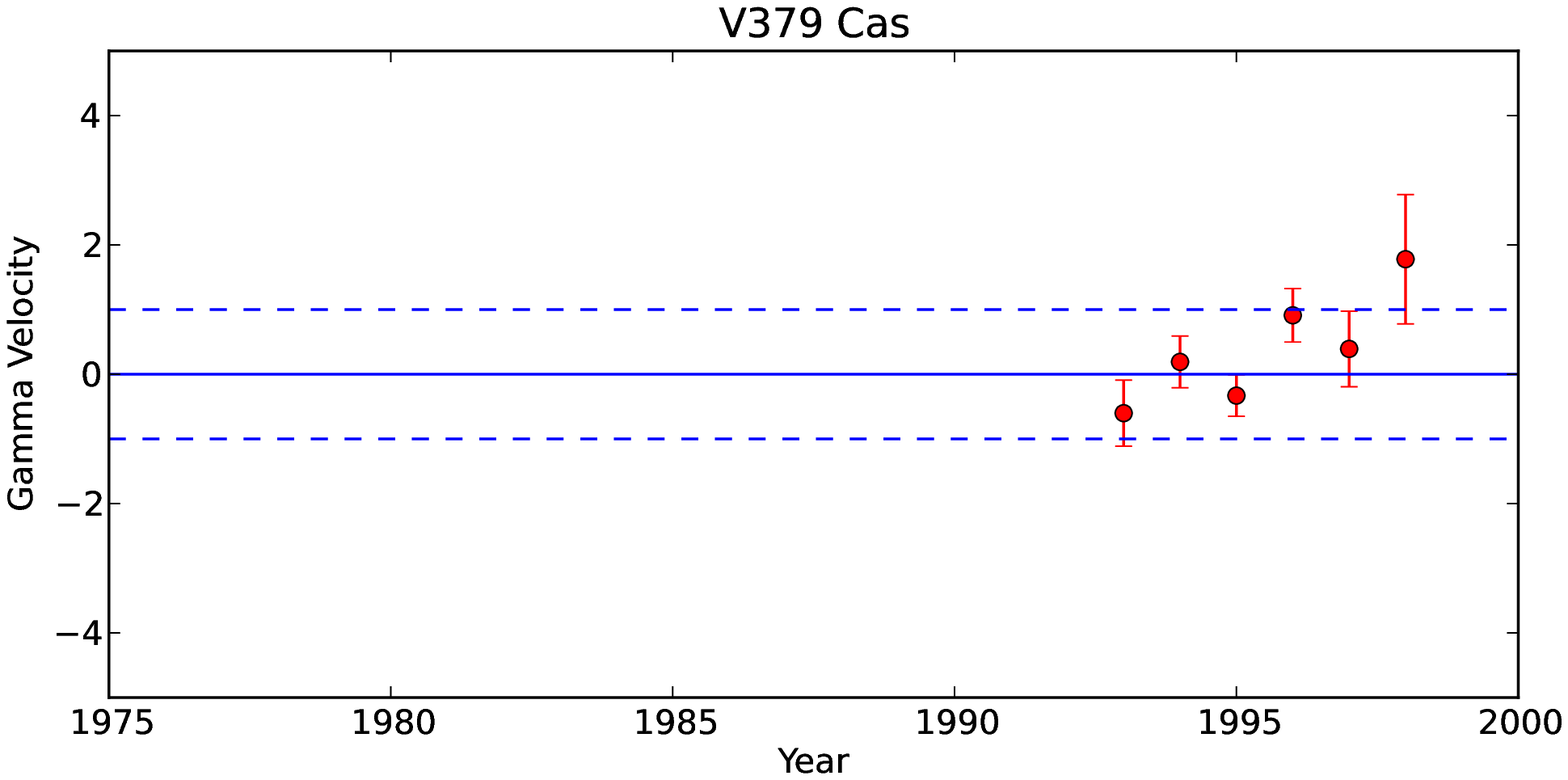} 
\includegraphics[totalheight=1.5in]{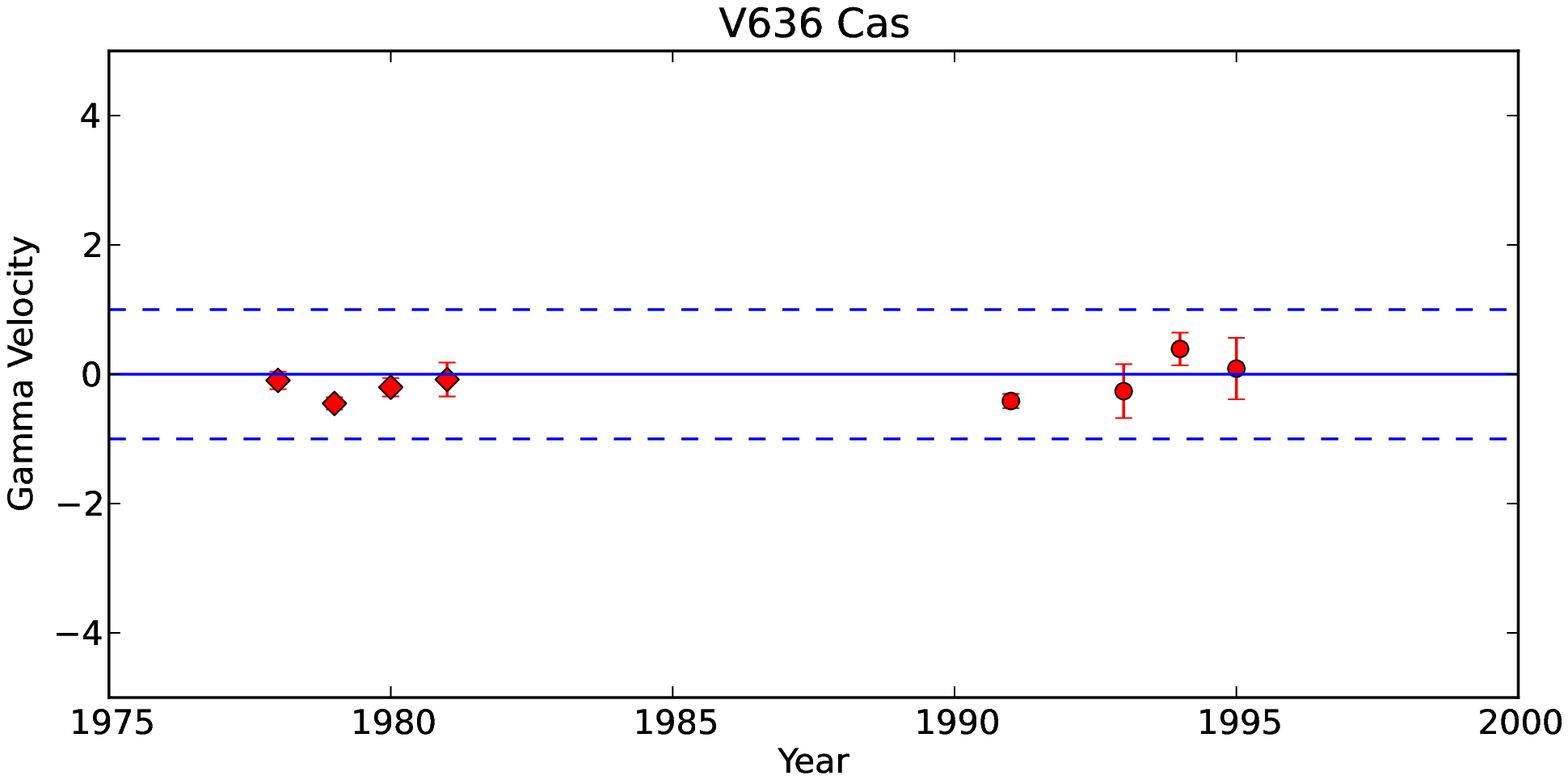} \\
\includegraphics[totalheight=1.5in]{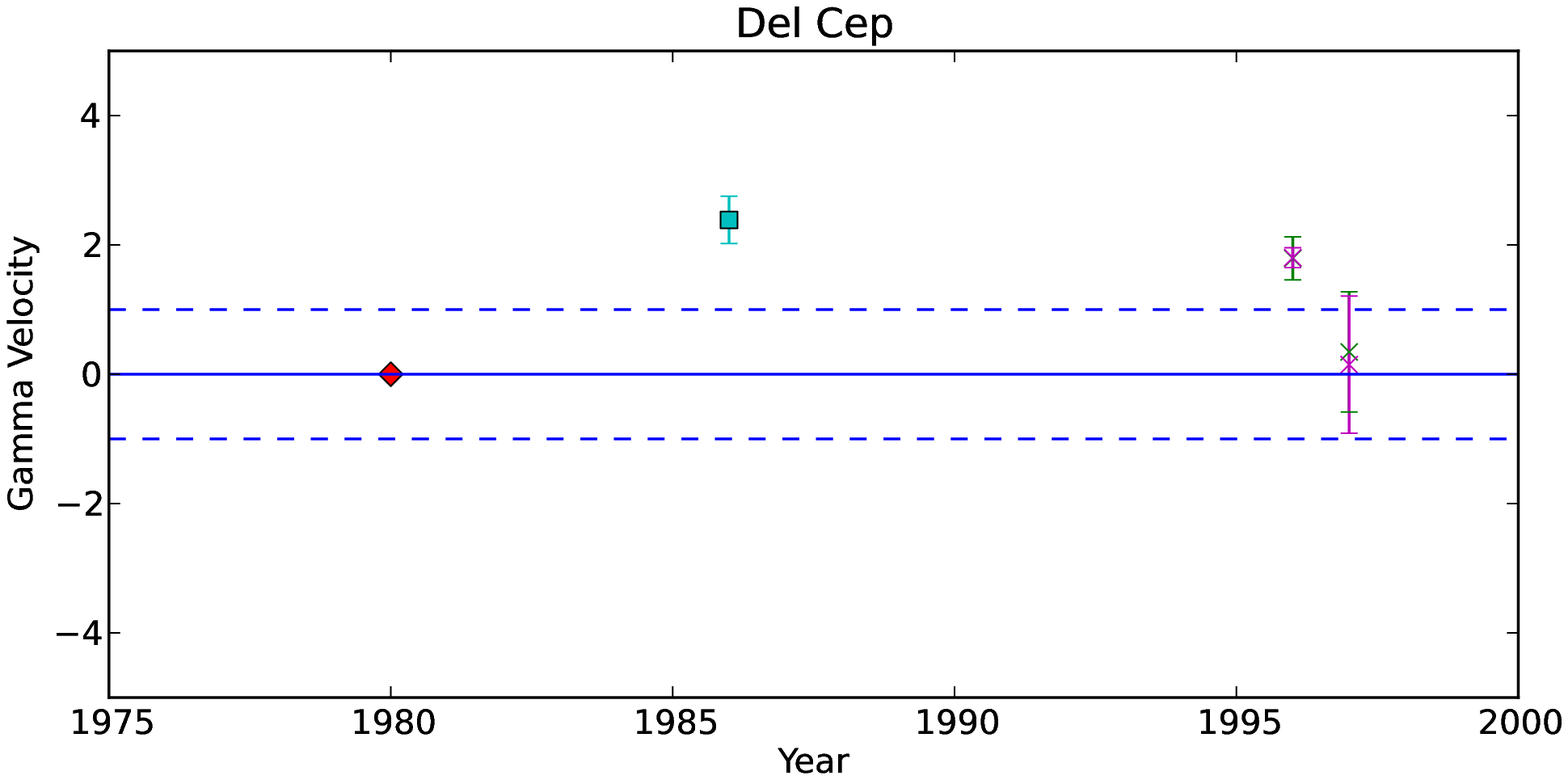} 
\includegraphics[totalheight=1.5in]{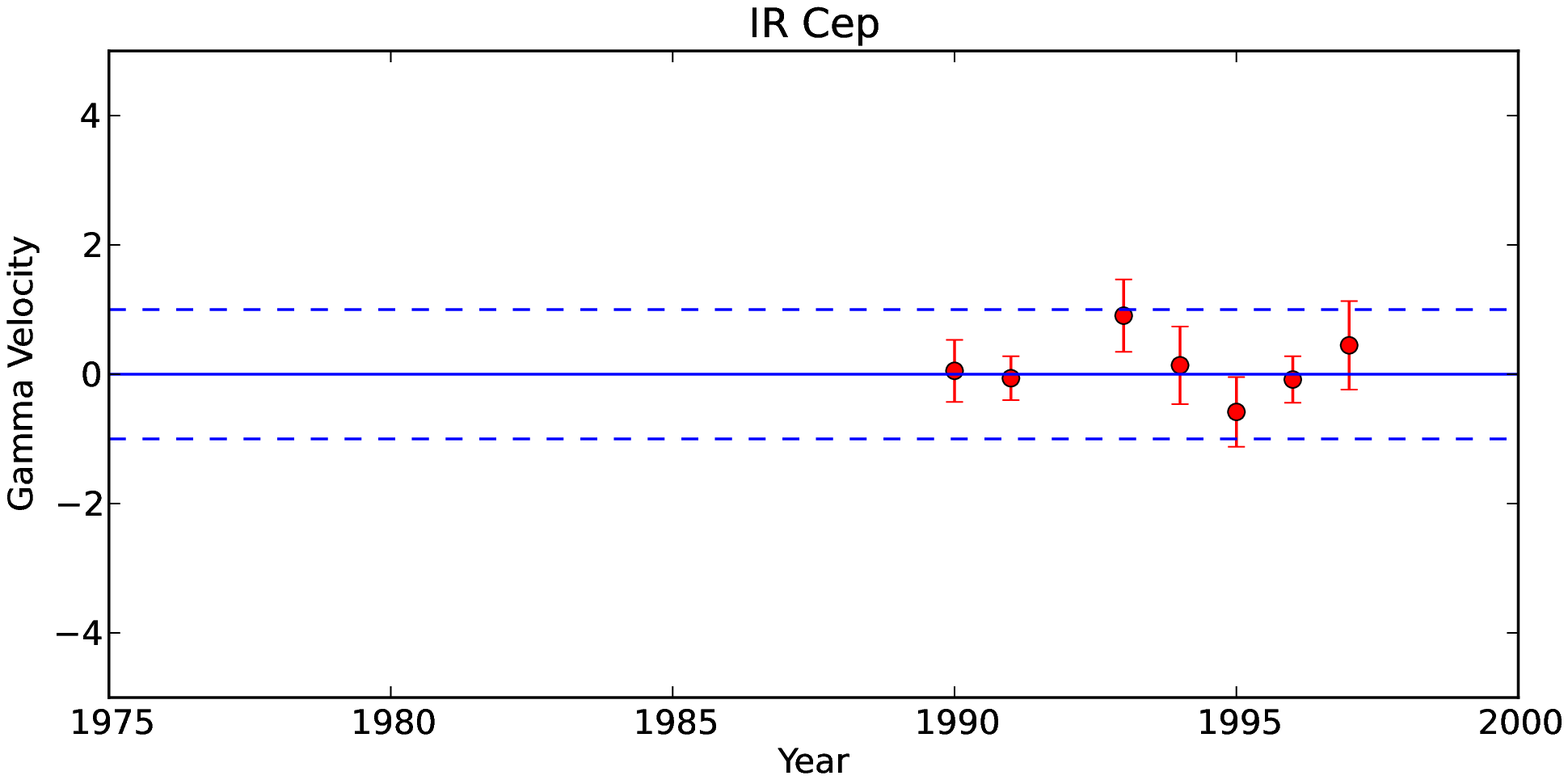} \\
\includegraphics[totalheight=1.5in]{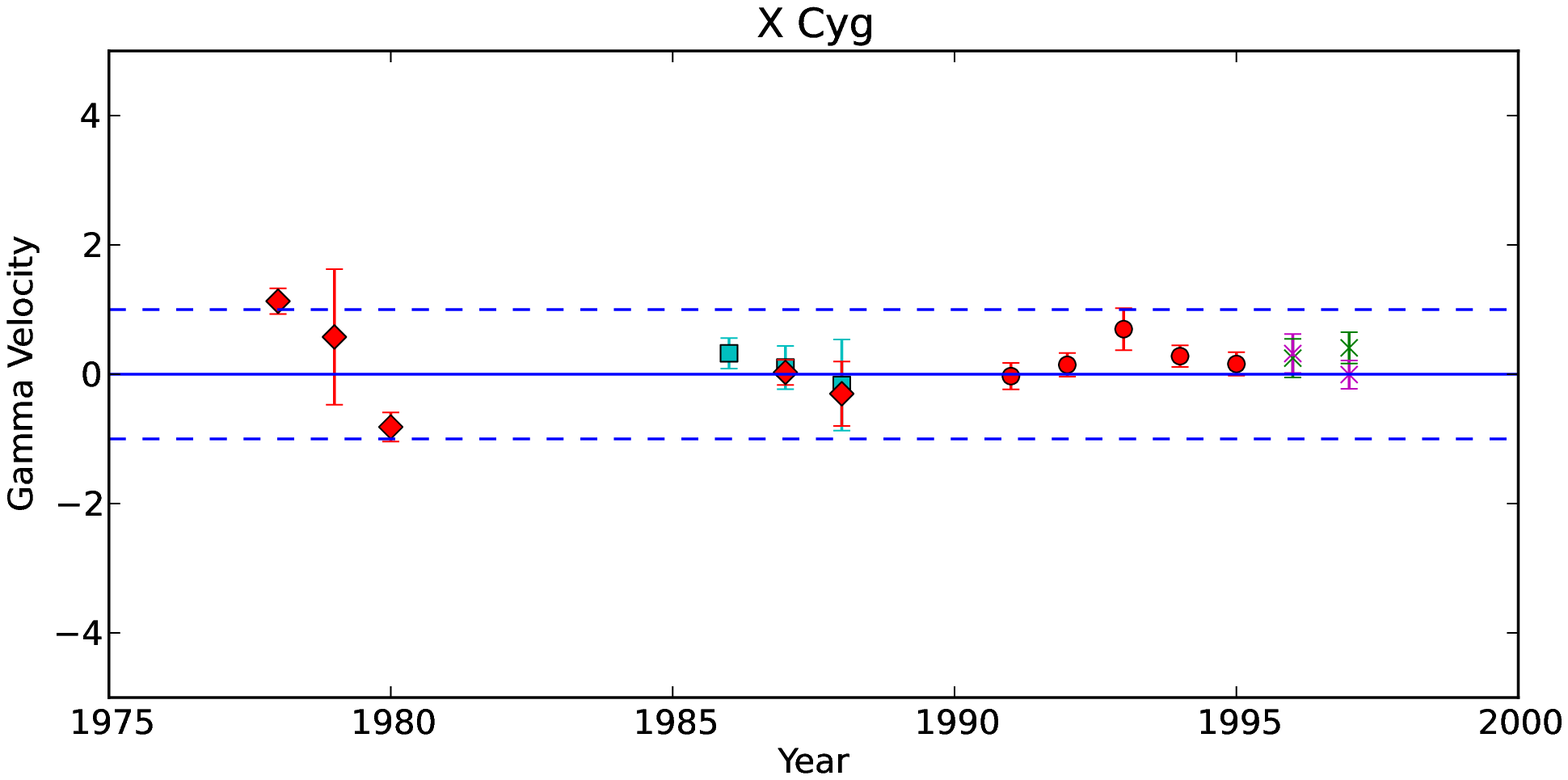} 
\includegraphics[totalheight=1.5in]{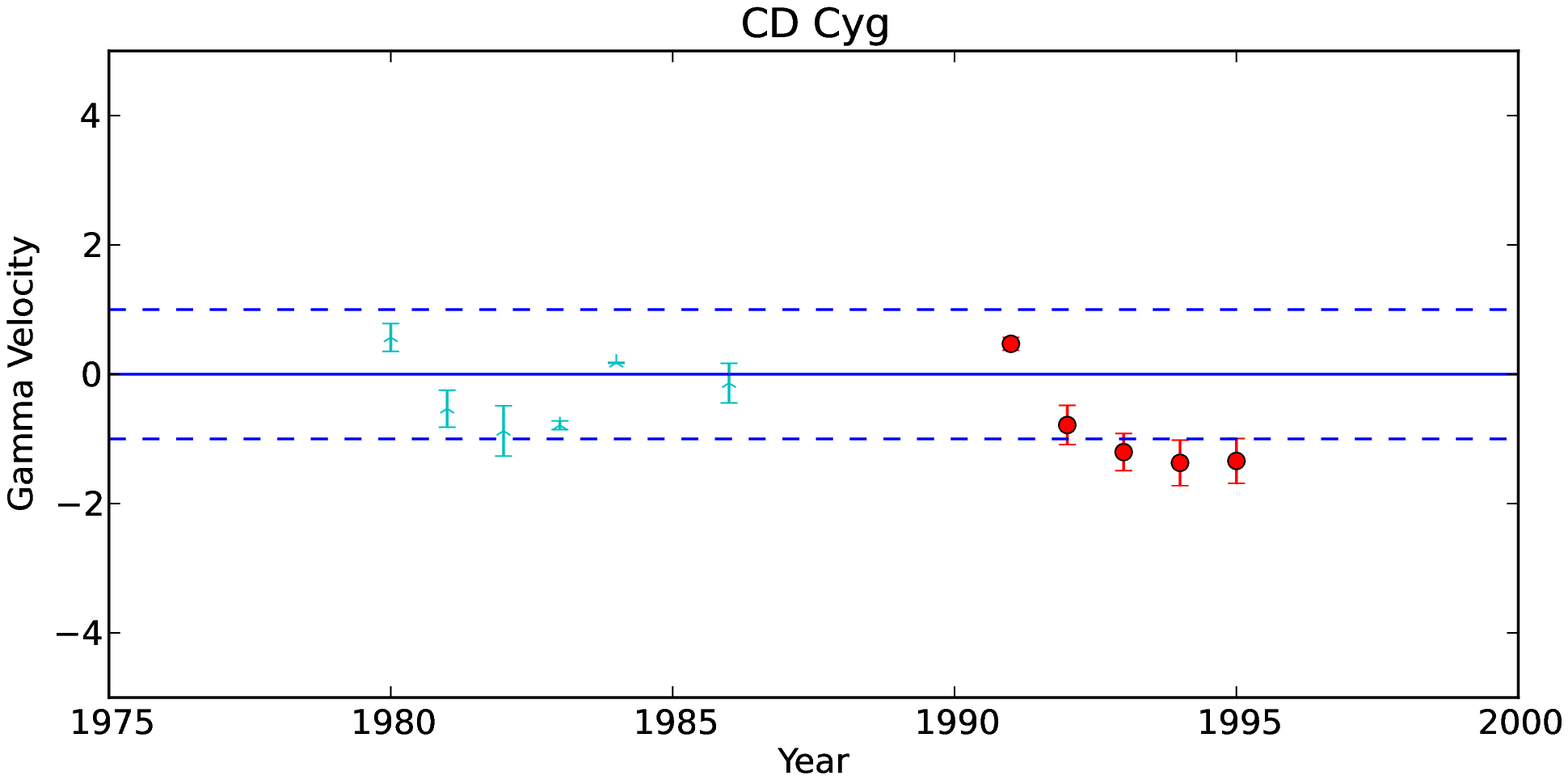} \\

\end{figure}

\begin{figure}

\includegraphics[totalheight=1.5in]{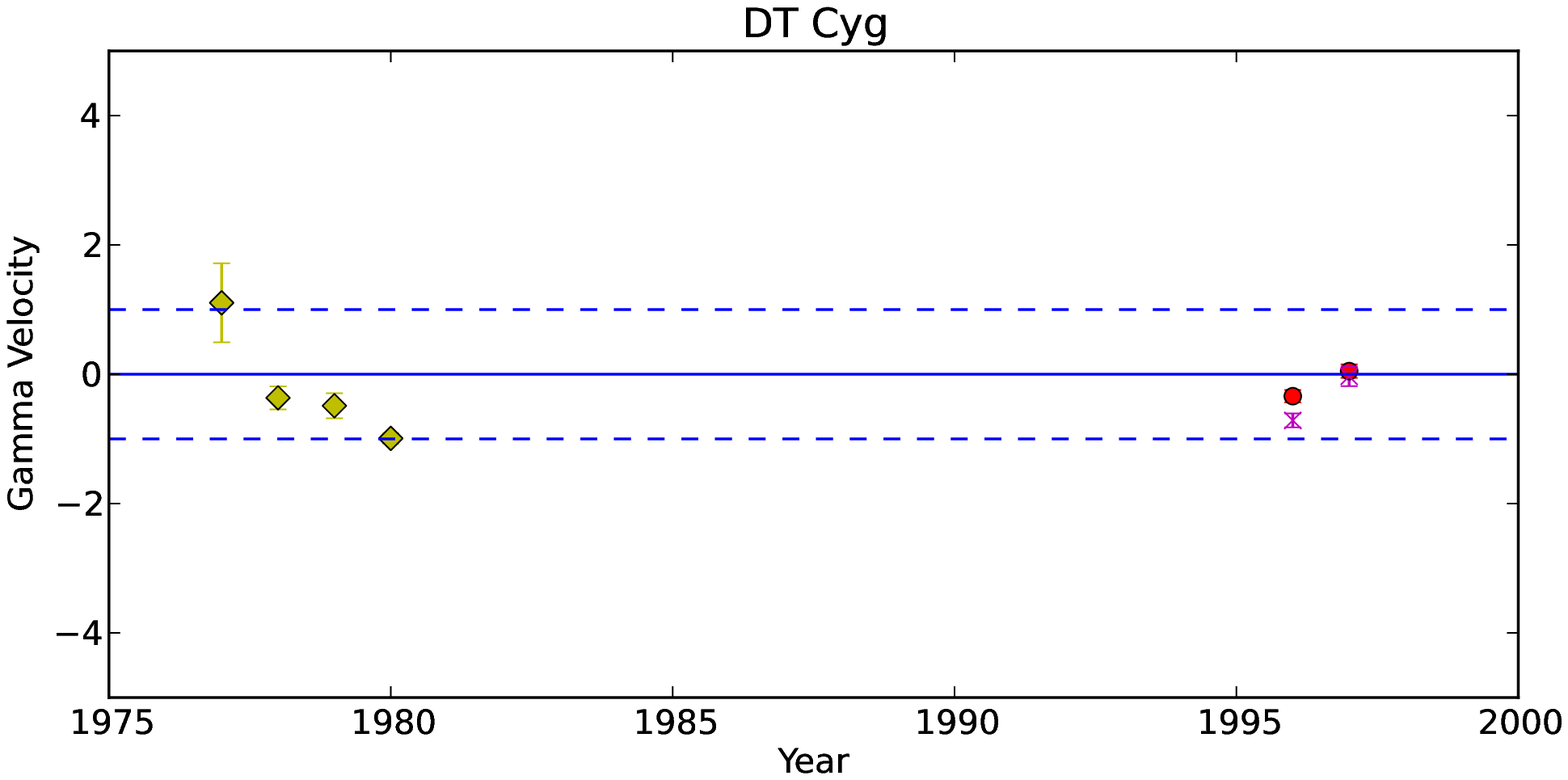} 
\includegraphics[totalheight=1.5in]{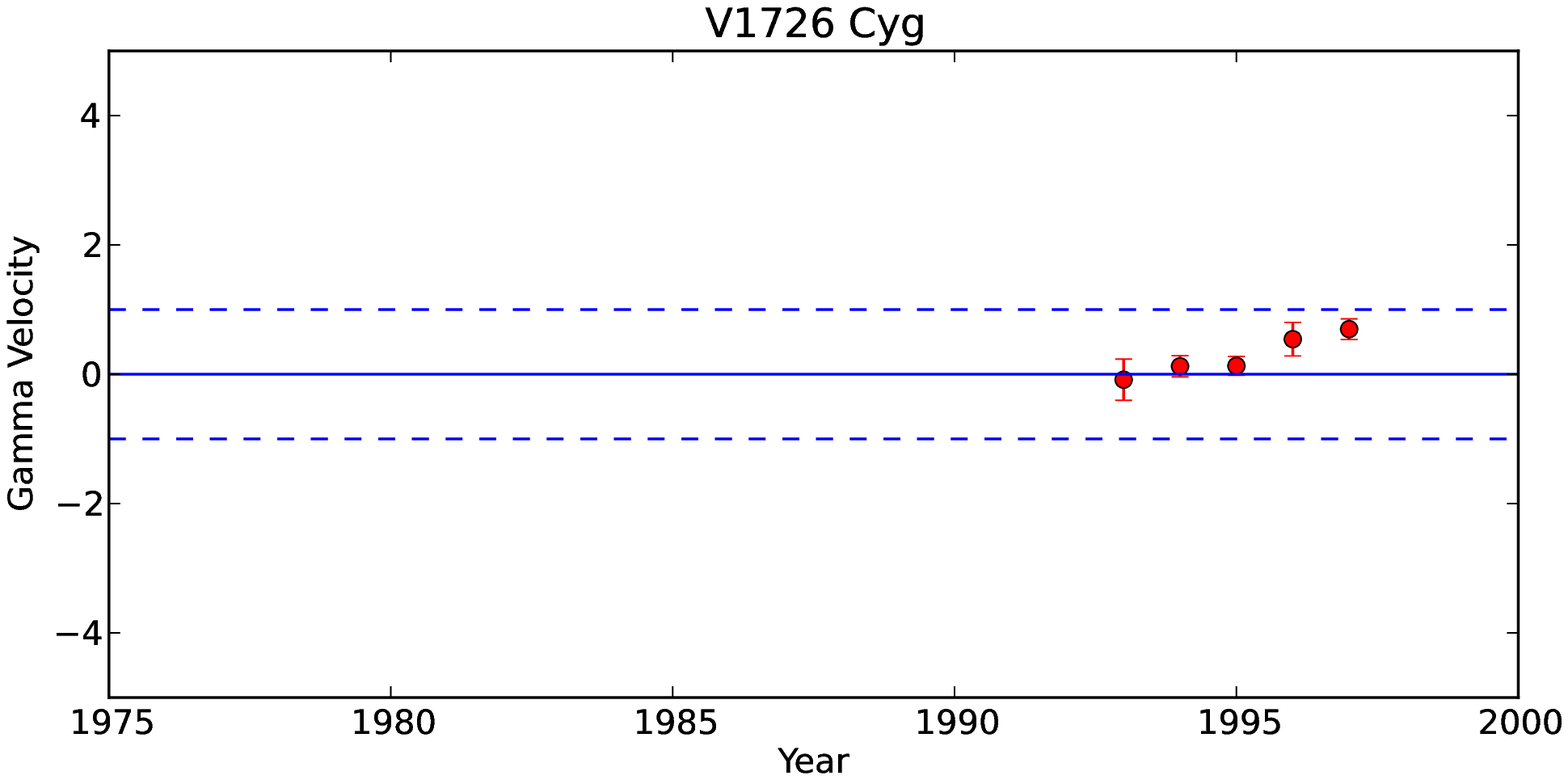} \\
\includegraphics[totalheight=1.5in]{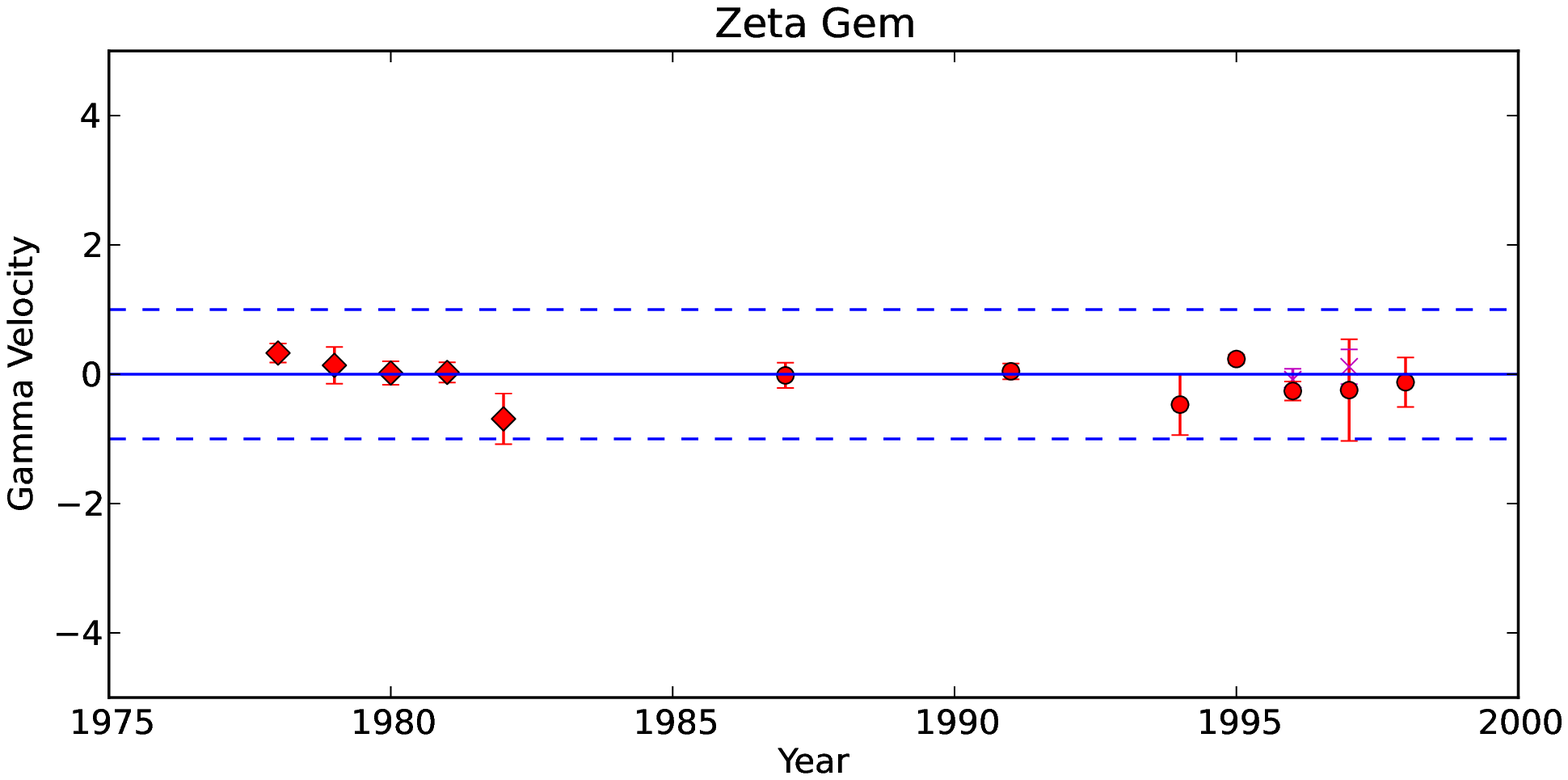} 
\includegraphics[totalheight=1.5in]{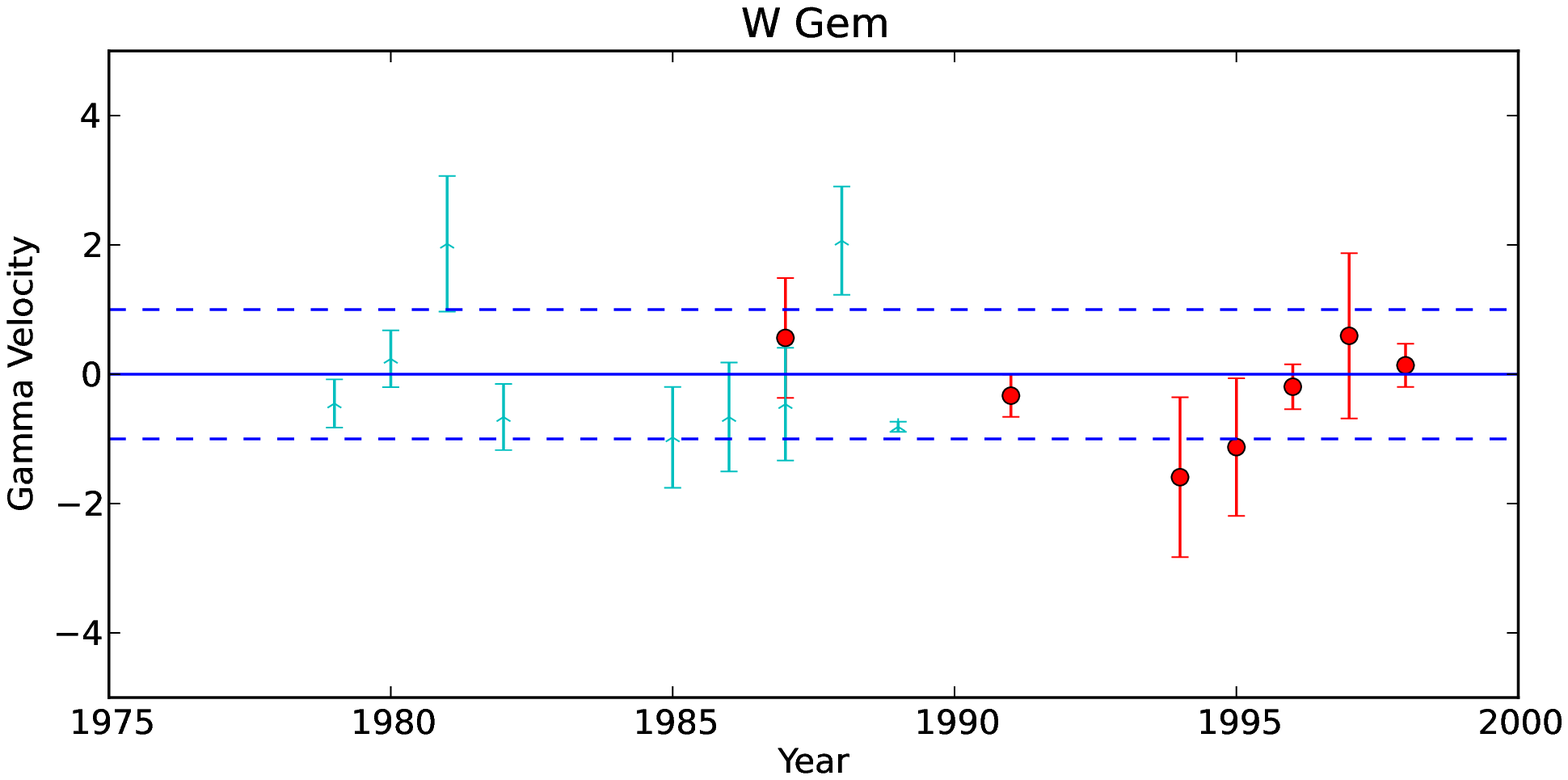} \\
\includegraphics[totalheight=1.5in]{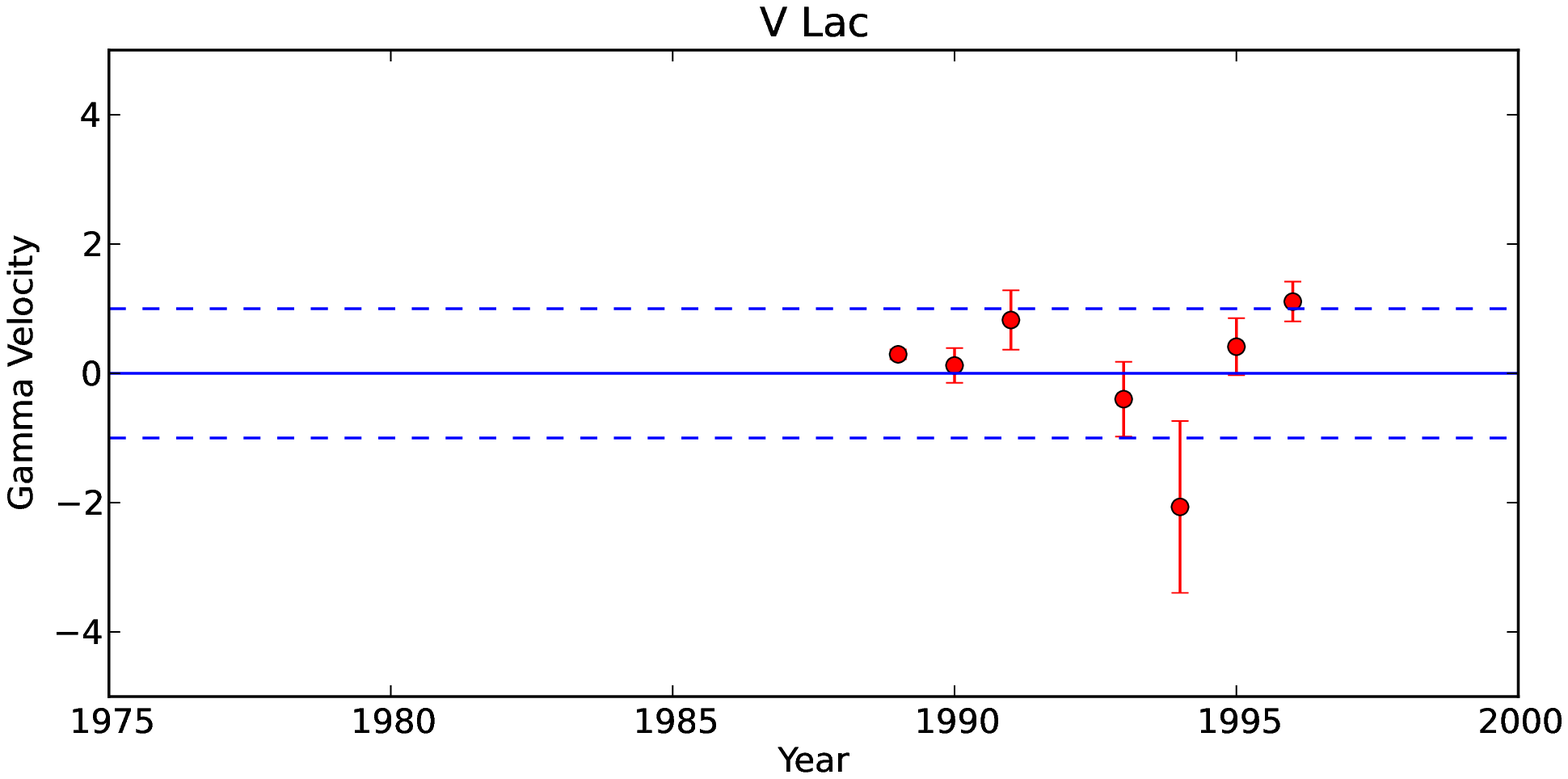} 
\includegraphics[totalheight=1.5in]{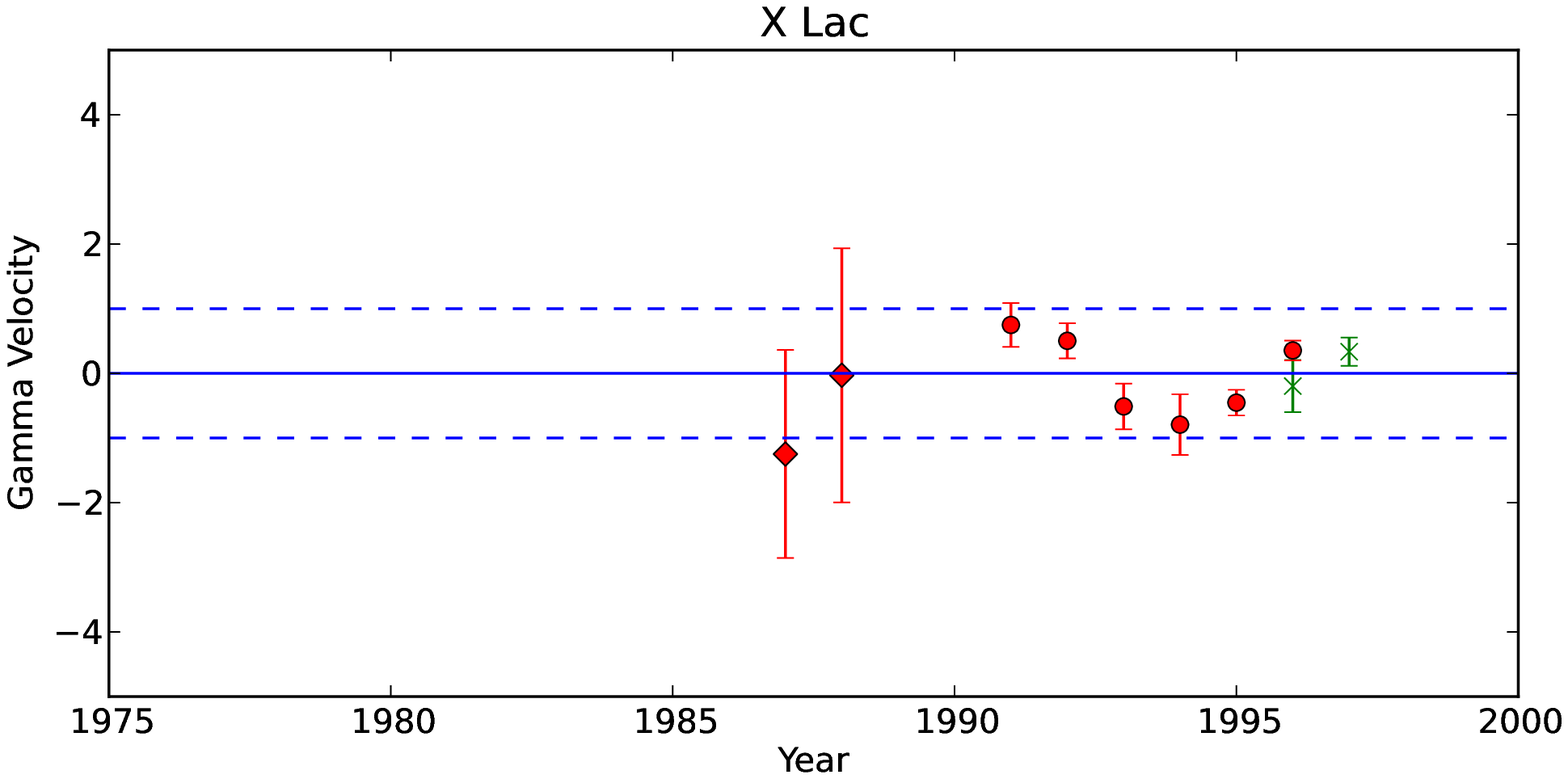} \\
\includegraphics[totalheight=1.5in]{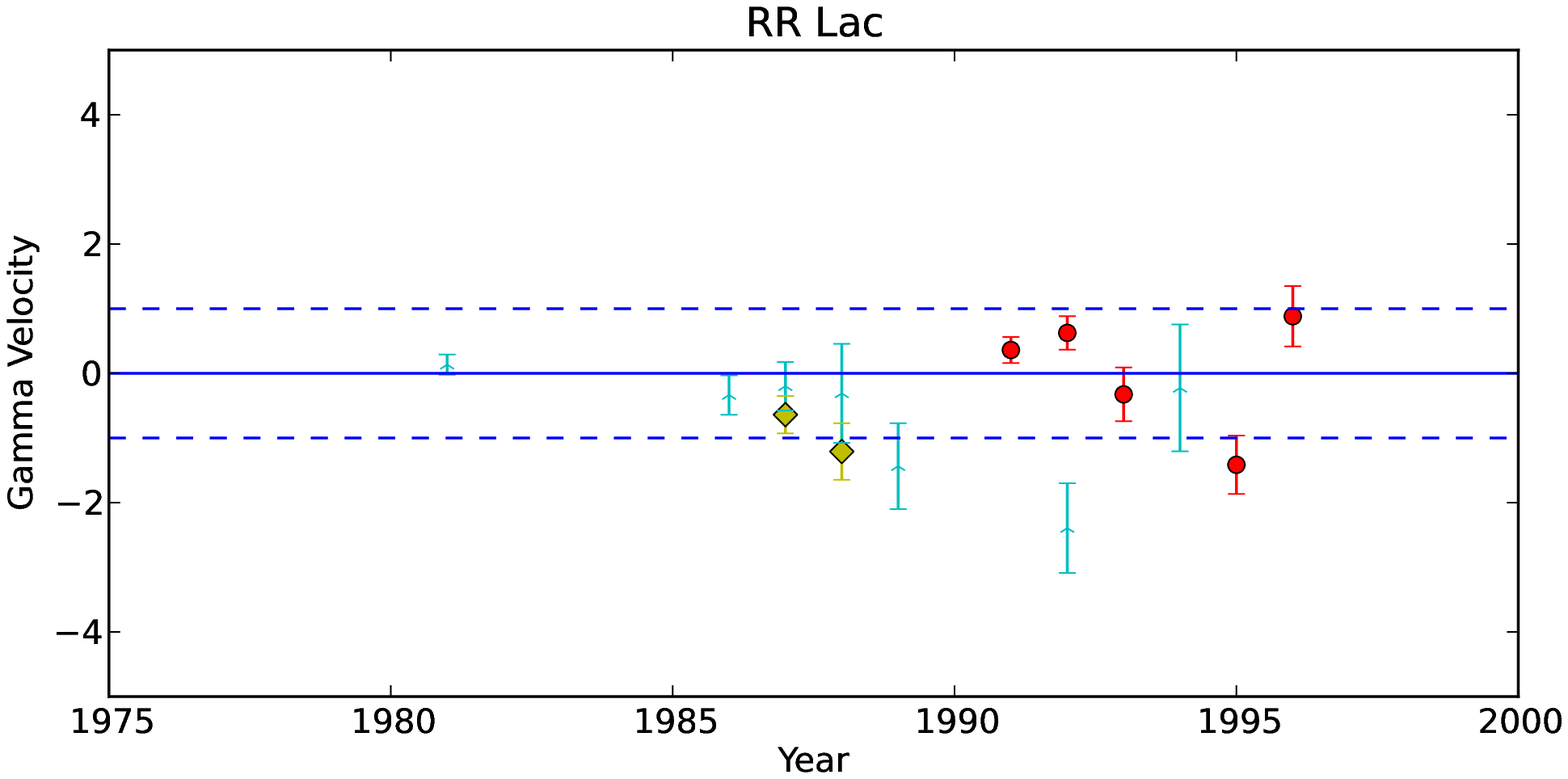} 
\includegraphics[totalheight=1.5in]{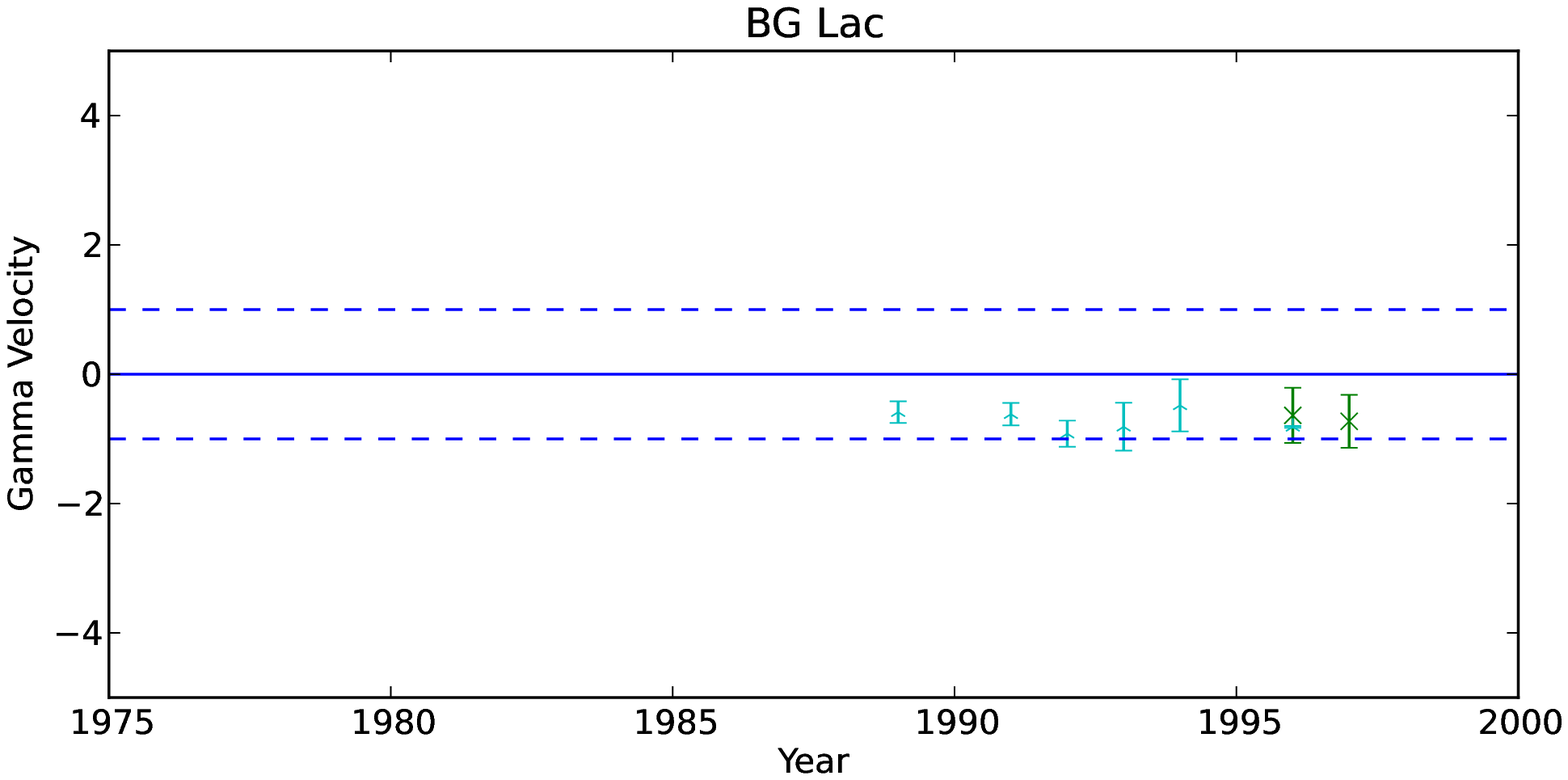} \\

\end{figure}

\begin{figure}

\includegraphics[totalheight=1.5in]{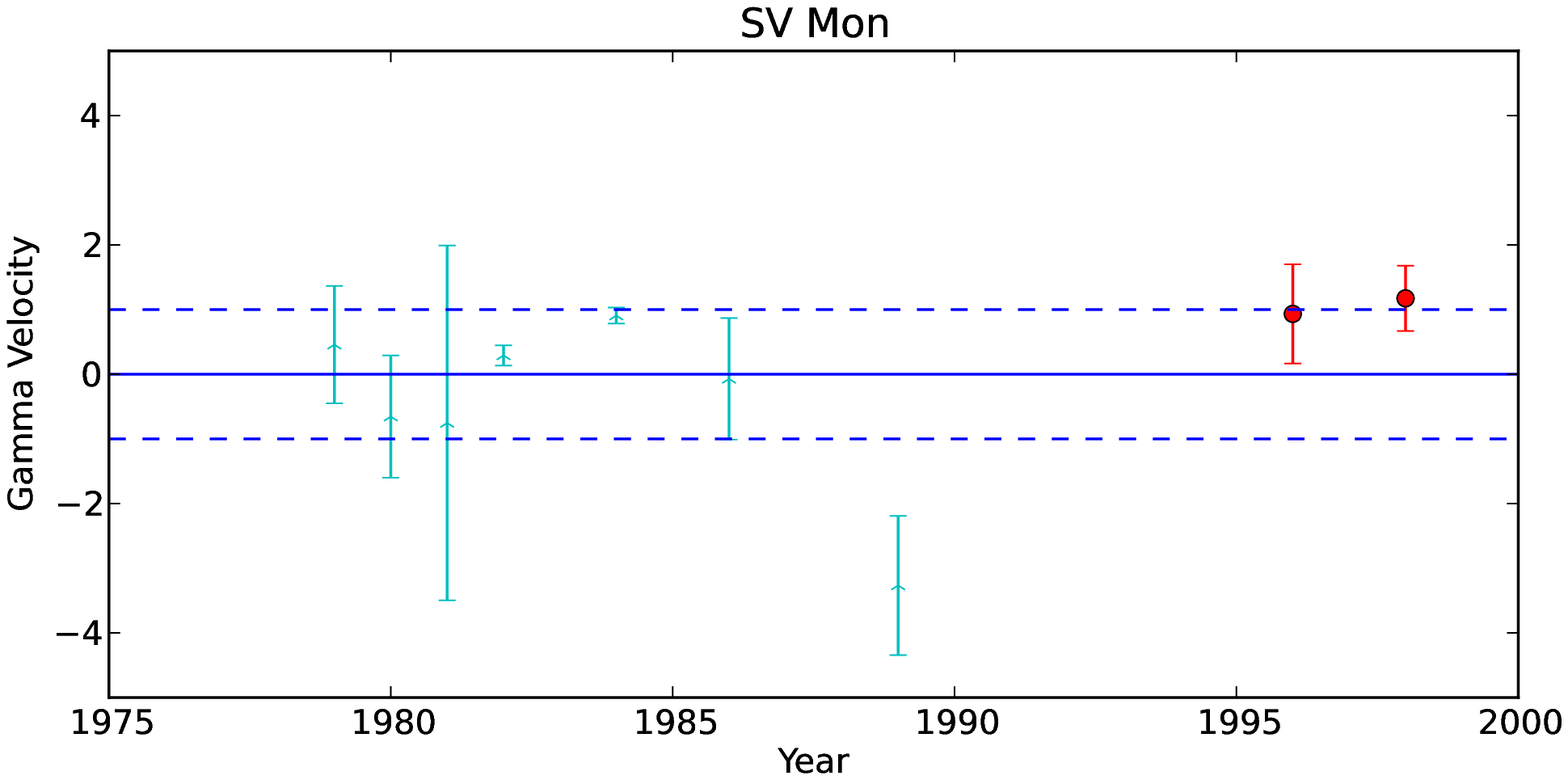} 
\includegraphics[totalheight=1.5in]{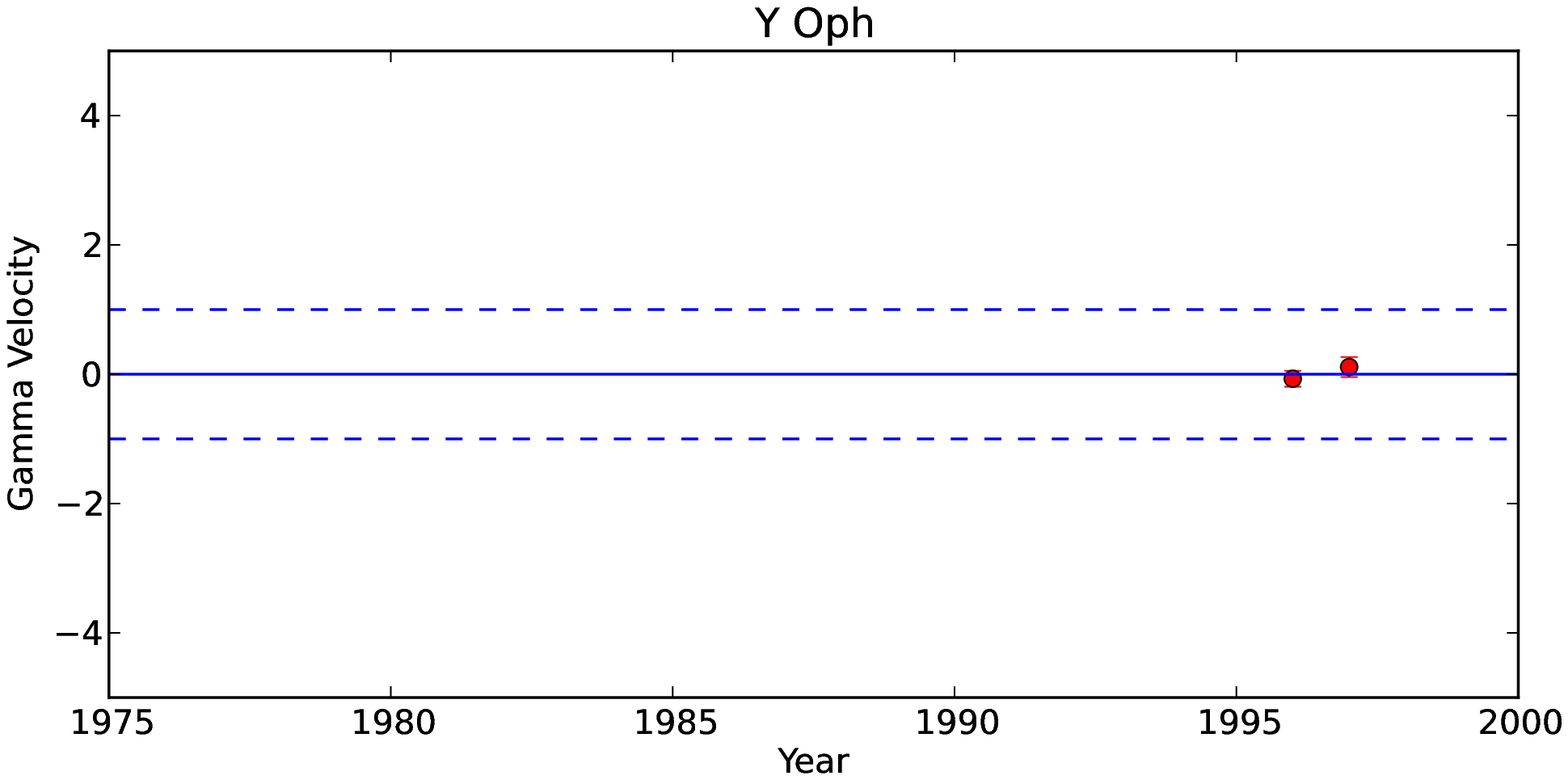} \\
\includegraphics[totalheight=1.5in]{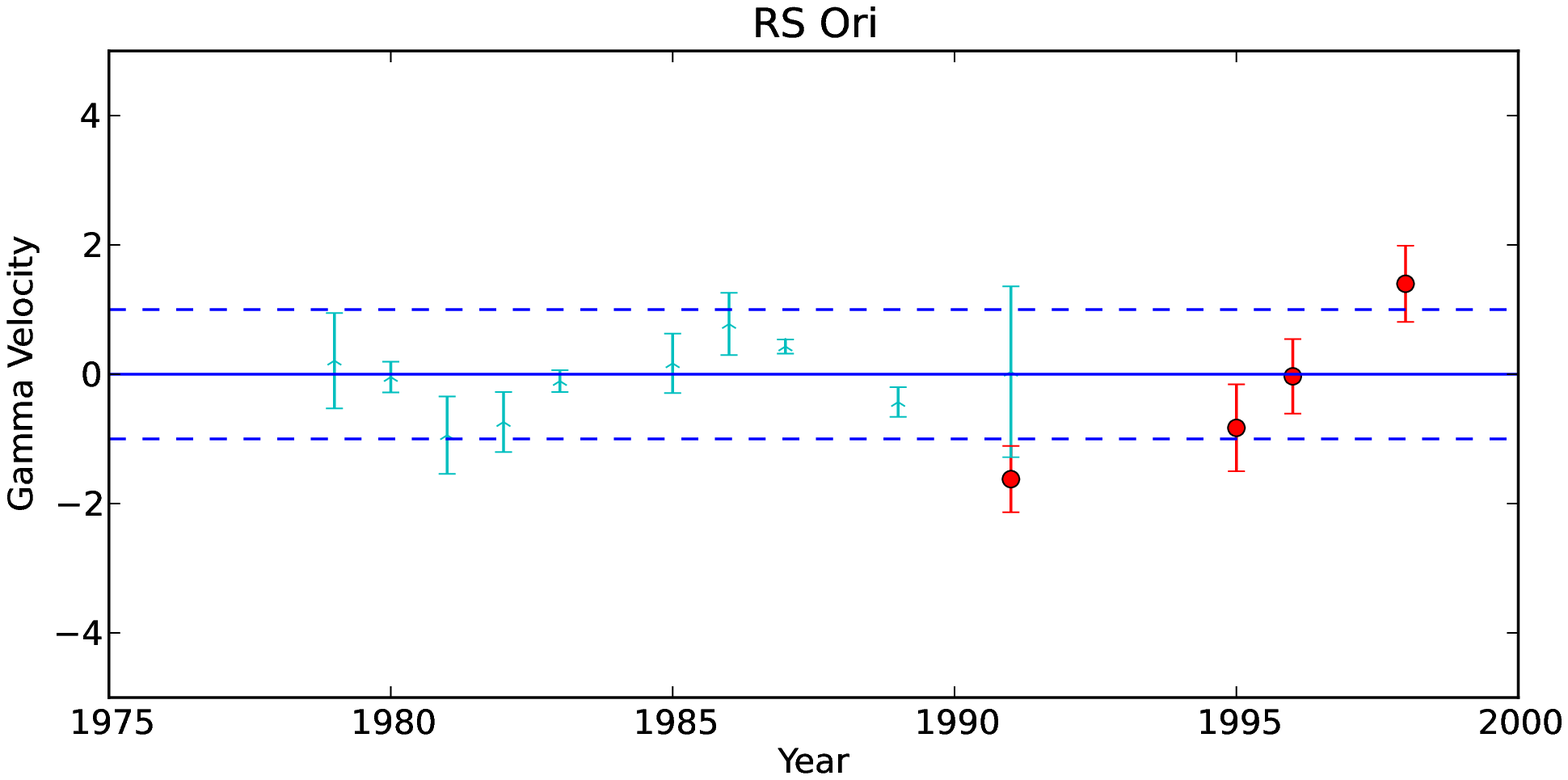} 
\includegraphics[totalheight=1.5in]{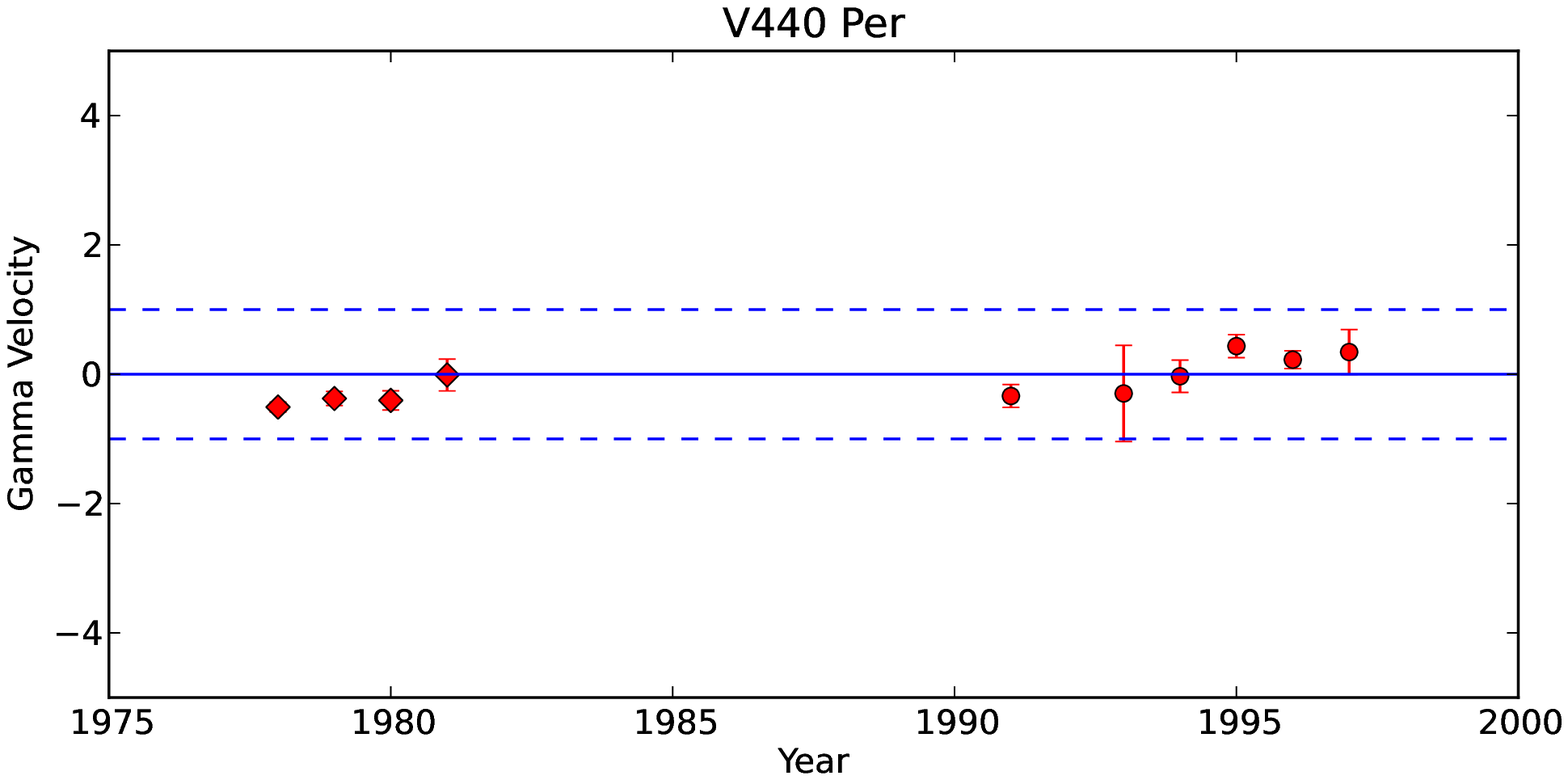} \\
\includegraphics[totalheight=1.5in]{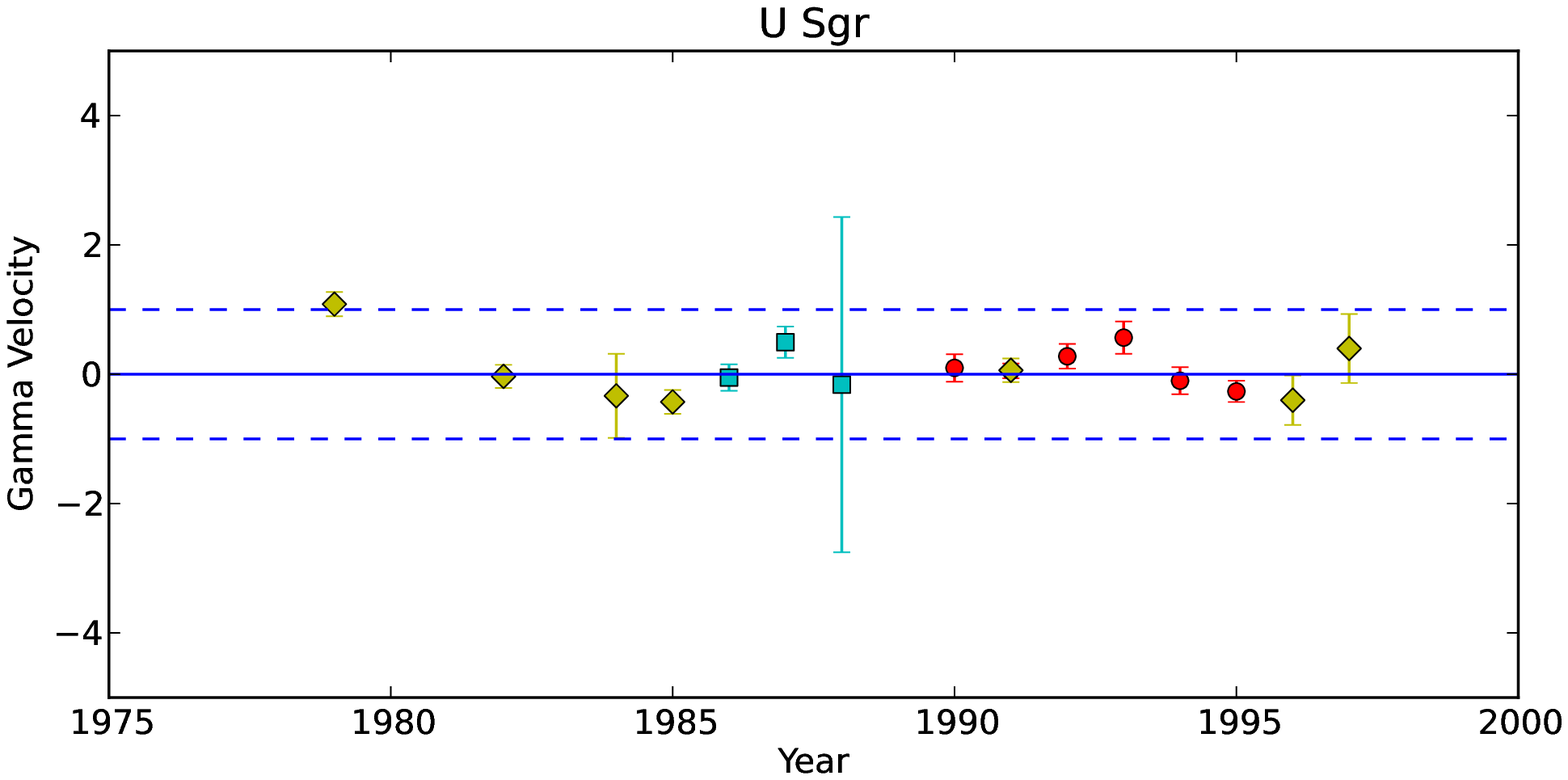} 
\includegraphics[totalheight=1.5in]{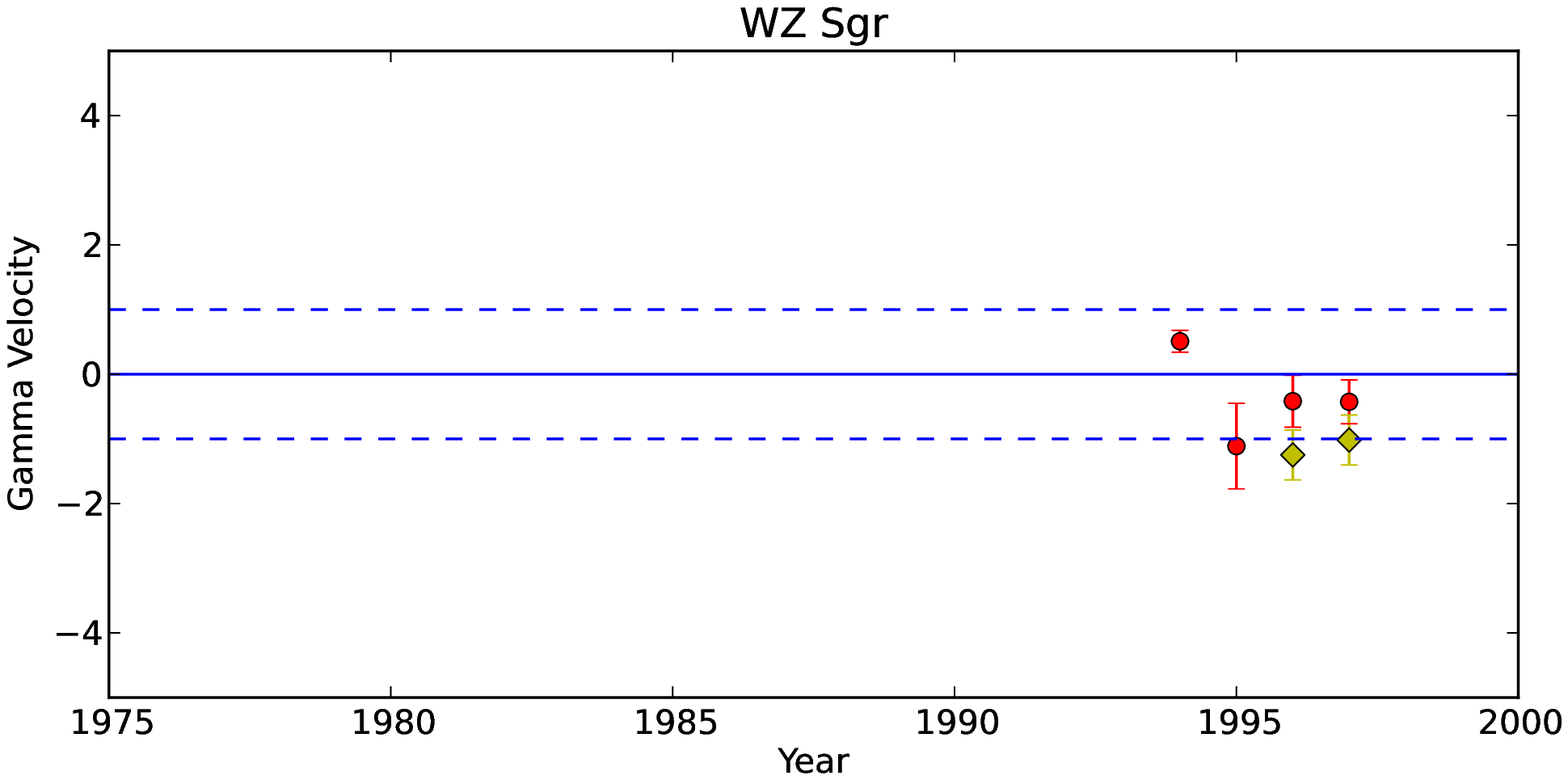} \\
\includegraphics[totalheight=1.5in]{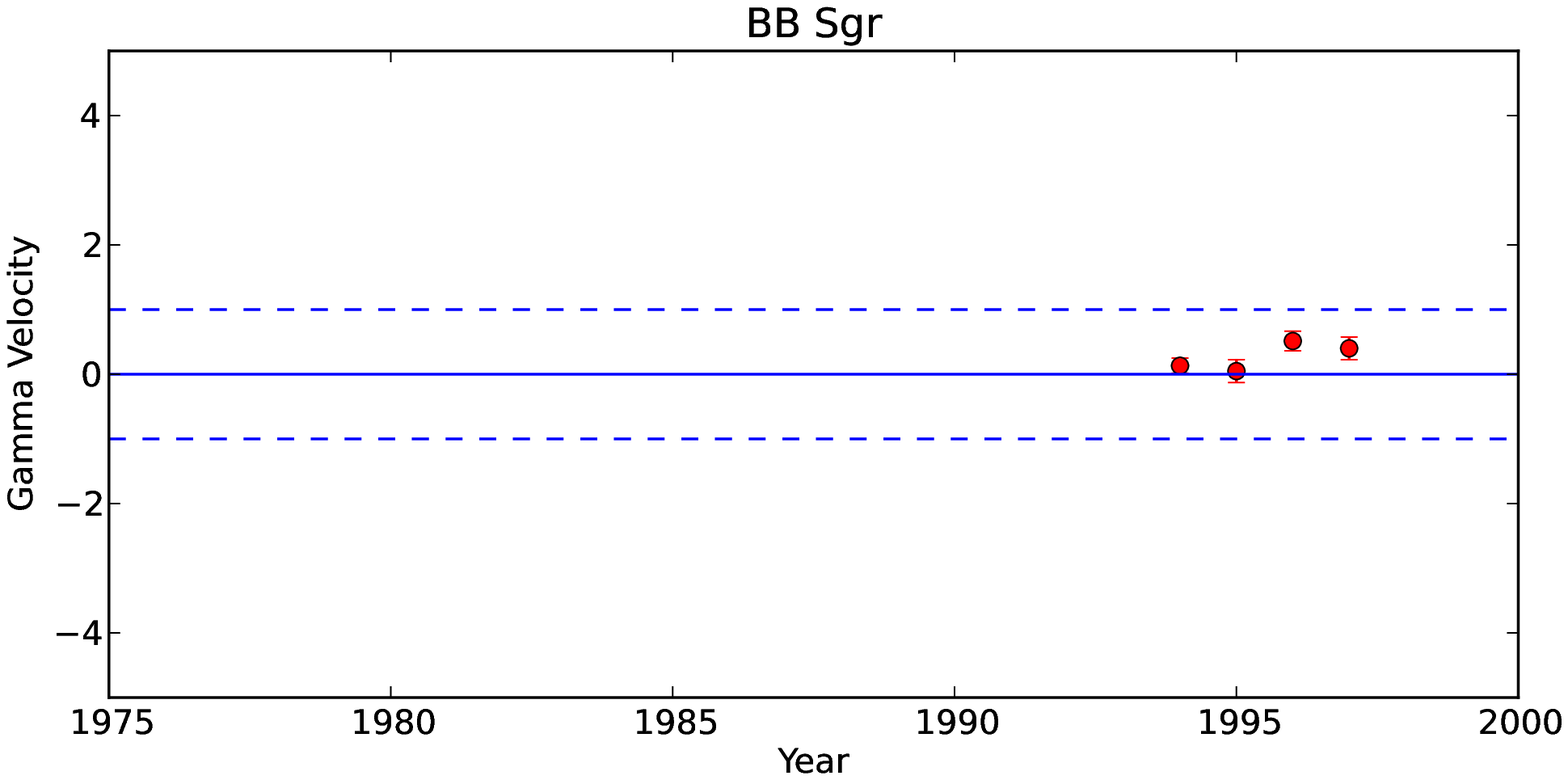} 
\includegraphics[totalheight=1.5in]{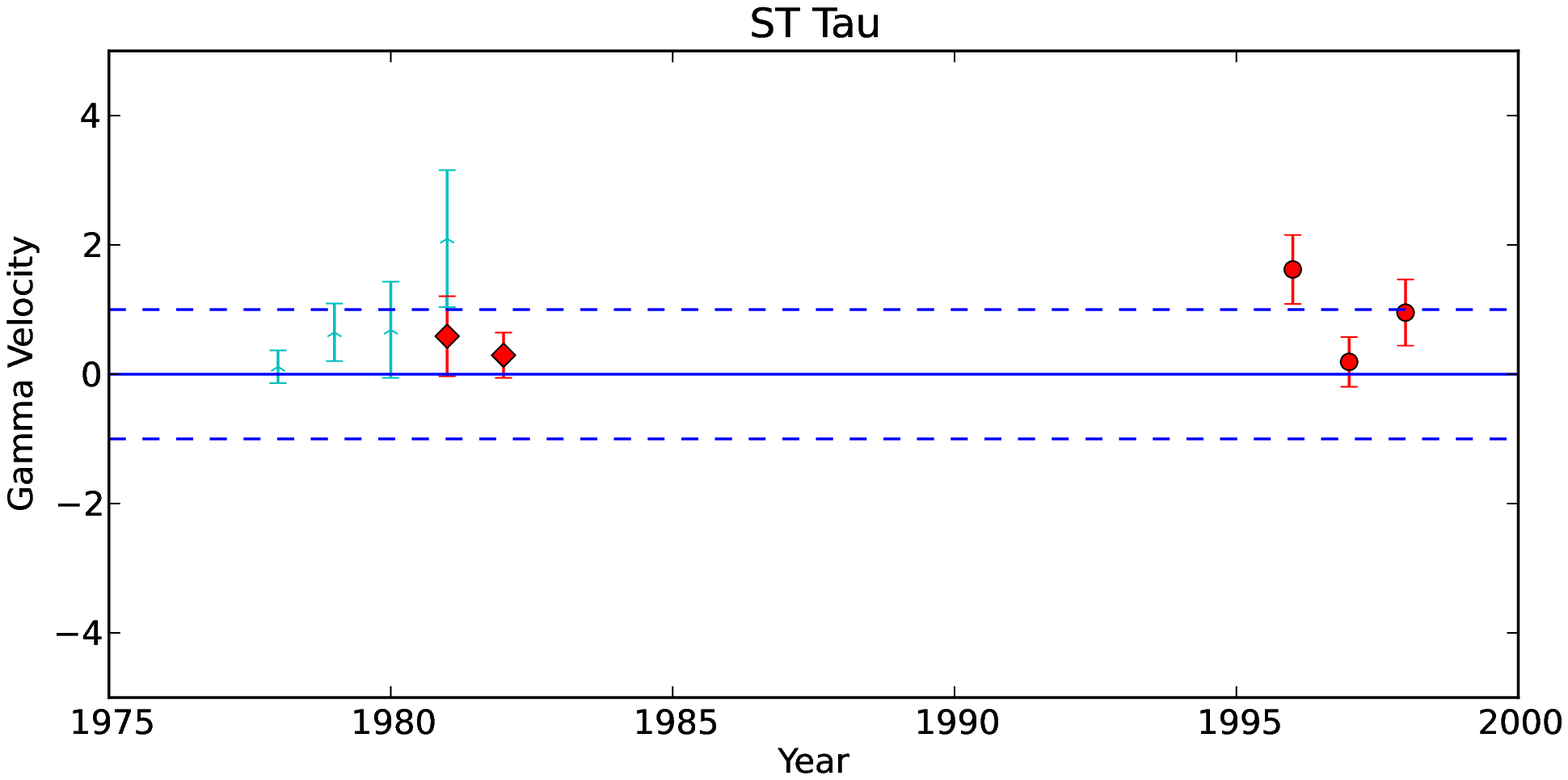} \\

\end{figure}

\begin{figure}

\includegraphics[totalheight=1.5in]{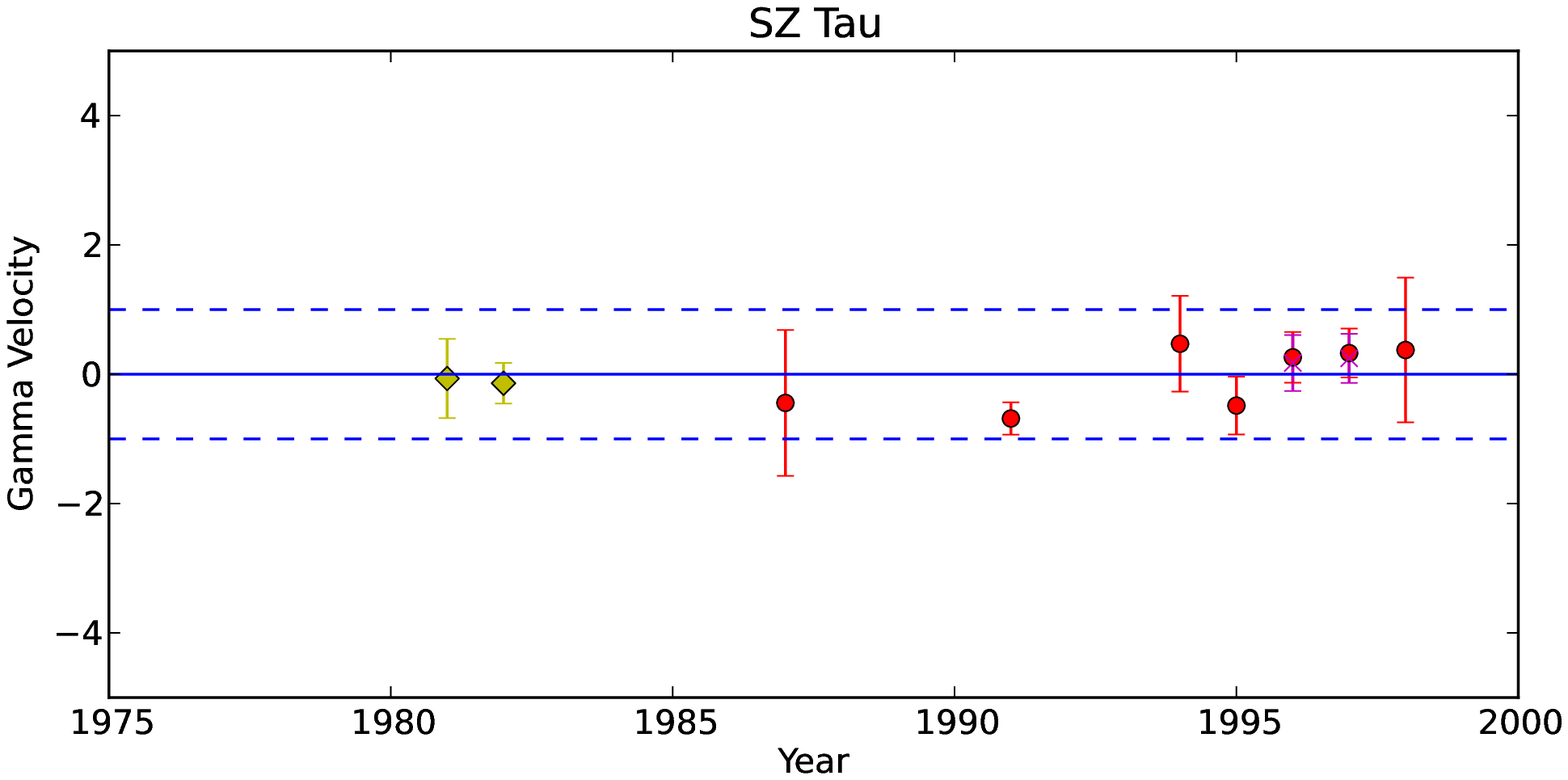} 
\includegraphics[totalheight=1.5in]{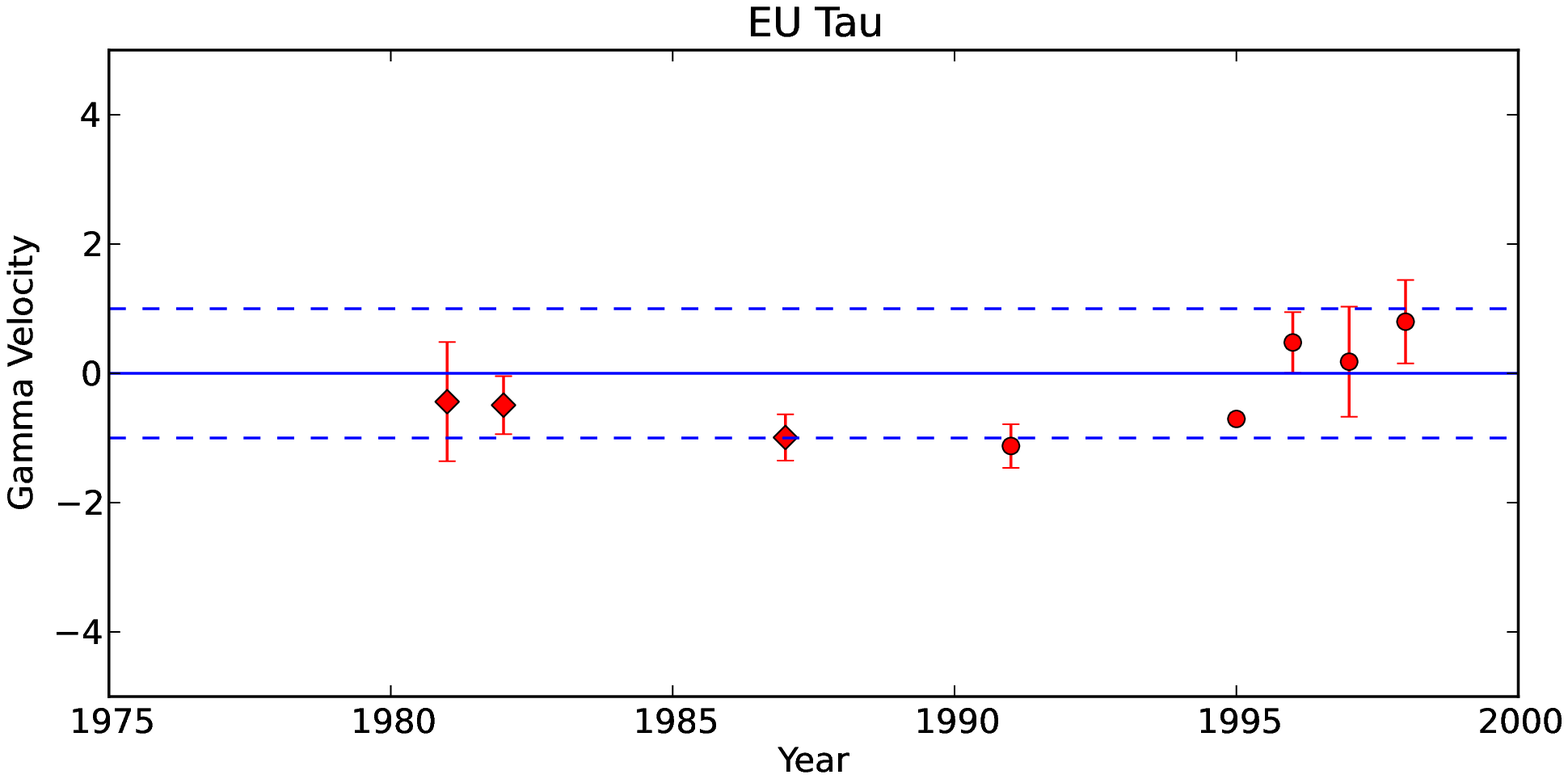} \\
\includegraphics[totalheight=1.5in]{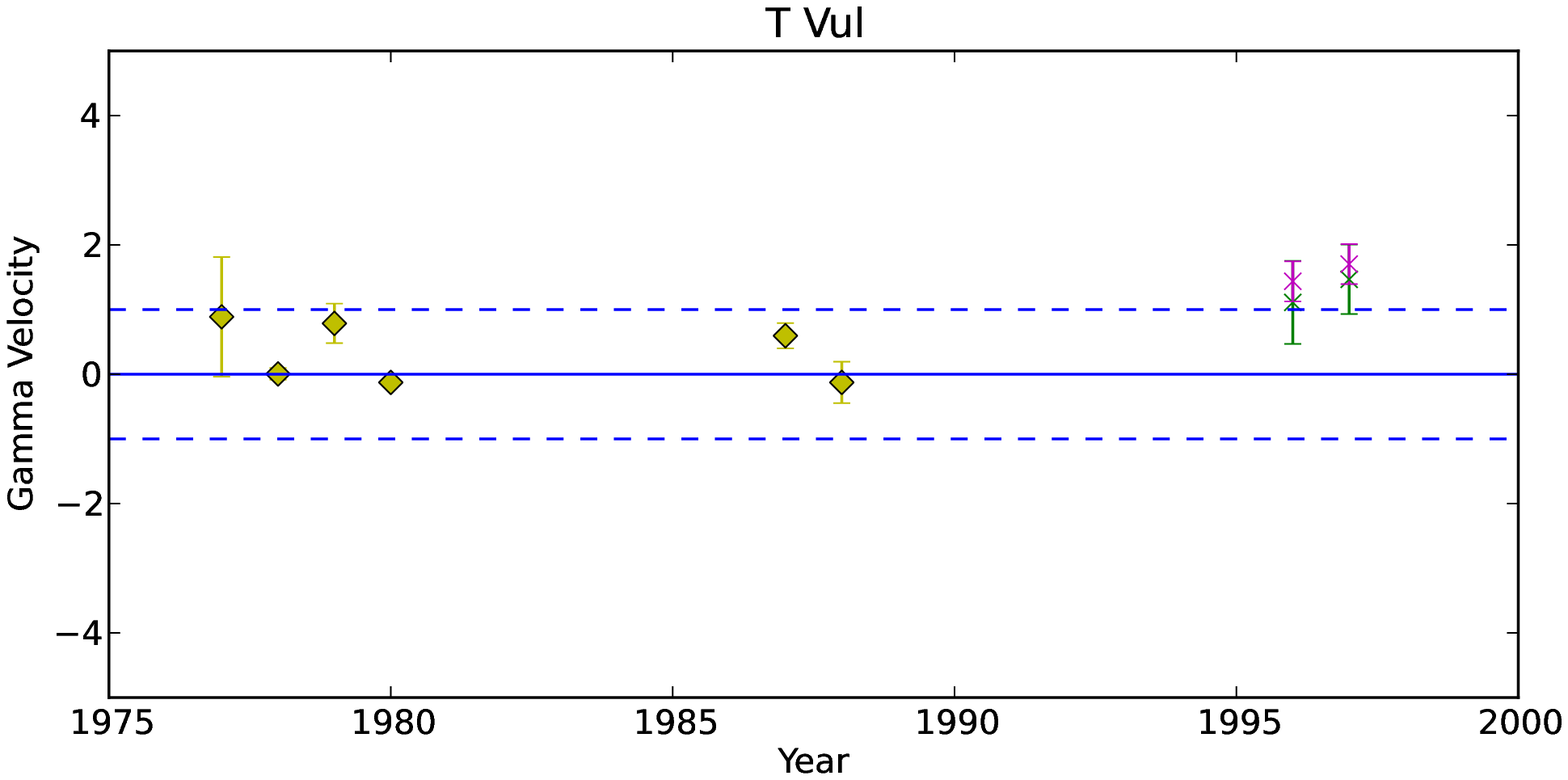} 
\includegraphics[totalheight=1.5in]{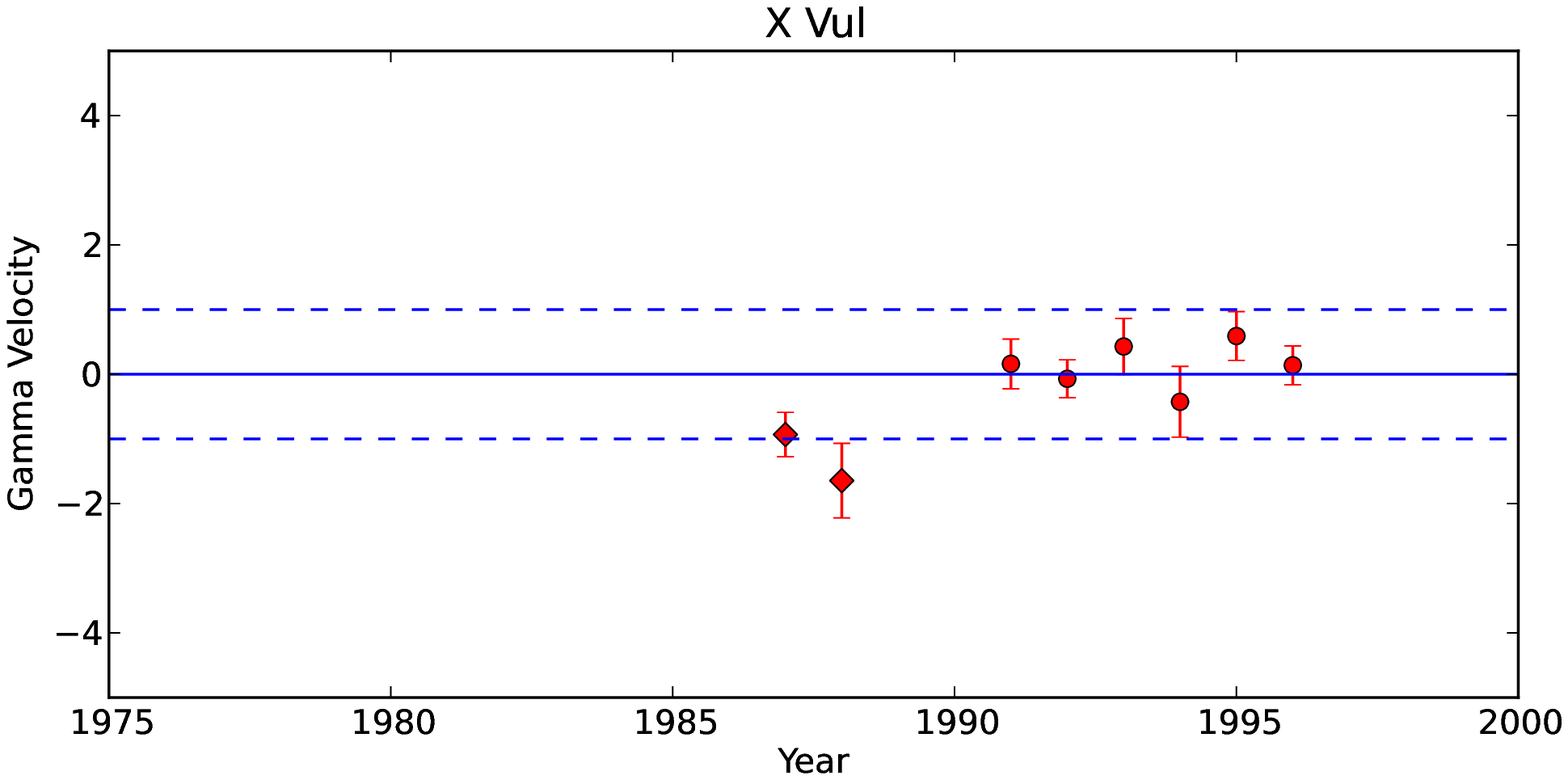} \\

\end{figure}

\hfill\vfill
\eject

\subsection{Window Function}

 Our data sample is drawn from the literature using data which was not
 designed  for evenly spaced long term coverage to identify binaries.   
We have performed the following check to see how thoroughly the
dataset covers the frequency (period) space.  This test was suggested
by the discussion of the optimal observing sequence developed for the
Hubble Space Telescope Cepheid Key Project (Freedman et al. 1995).  We
examined the window functions produced by a Fourier transform of the
dates of observation.  The program  FTCLEAN was supplied by M. Templeton,
written by him based on the CLEAN algorithm of Roberts, Lehar, and
Dreher (1987).  The use of this program was first discussed by
Templeton and Karovska (2009).

\begin{figure}
 \includegraphics[width=10cm,angle=90]{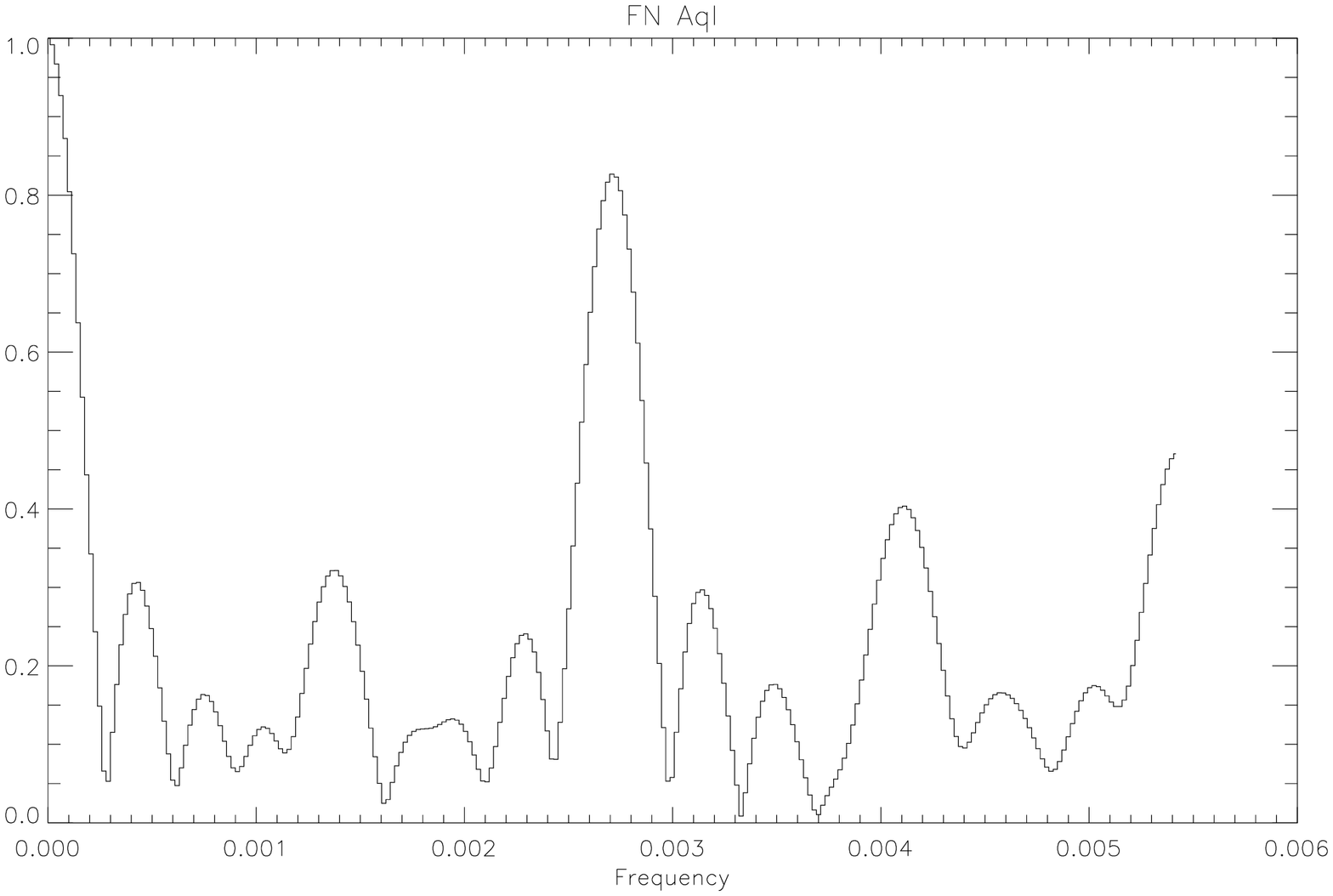}
\caption{The window function for FN Aql. Frequency is in cycles/day.  
The period of interest (1 to 14 years or  
 0.0027 and .0002 cycles/day) is reasonably well covered.
\label{window_FNAql.ps}}
\end{figure}

As an example, Figure~\ref{window_FNAql.ps} illustrates the window function for FN Aql, 
which shows that the frequency space between 1 and 14 years 
(0.0027 and 0.0002 cycles/day)  is reasonably evenly covered.


\subsection{Detection Limits}

In CRaV we search for  Cepheids with orbital
velocity variations with an amplitude larger than 1 km~s$^{-1}$,
that is with a velocity difference between two seasons of at least 2
km~s$^{-1}$. In this section we estimate what combinations of mass
and period would be detected by this criterion.  

We use Monte-Carlo simulations to estimate the fraction of Cepheid
companions that will be detected with our method. For each star in our 
sample we perform the following simulation. We assume a Cepheid mass of 
$5\;M_{\odot}$ and form a grid of eleven values for the mass of the 
secondary and 50 logarithmically spaced values for the semi-major axis
corresponding to periods between 0.5 and 500 years. For each grid point, 
we generate an array of 1000 random lines-of-sight to the system and for 
each line-of-sight we calculate the radial velocity of the Cepheid at 
the actual observation spacing (cadence) for the stars.
For systems with a large semi-major 
axis the orbital period of the system can be much longer than the 
sequence covered by the observations. In this case, the initial position 
of the secondary becomes important, since the radial velocity of the 
Cepheid changes much faster close to periastron than at apastron. Thus, 
we repeat the simulations for 50 evenly spaced initial positions on an 
elliptical orbit. We then calculate the fraction of the total 
simulations for each grid point which predicts a velocity difference between 
the highest and lowest radial velocity $>2$~km~s$^{-1}$. We take this as 
an estimate of the probability of detecting a companion with these system 
parameters based on the radial velocity of the Cepheid.

The Kepler equations for the orbit are solved using the implementation 
of PyAstronomy\footnote{http://www.hs.uni-hamburg.de/DE/Ins/Per/Czesla/PyA/PyA/index.html} 
that is based on the algorithm of Markley (1995). The code 
for our Monte-Carlo simulations is implemented as an IPython notebook
(Perez and Granger 2007), 
 which is available as an electronic file.  In the future, updated 
versions can be found at https://github.com/hamogu/Cepheids.

\begin{figure}
\plotone{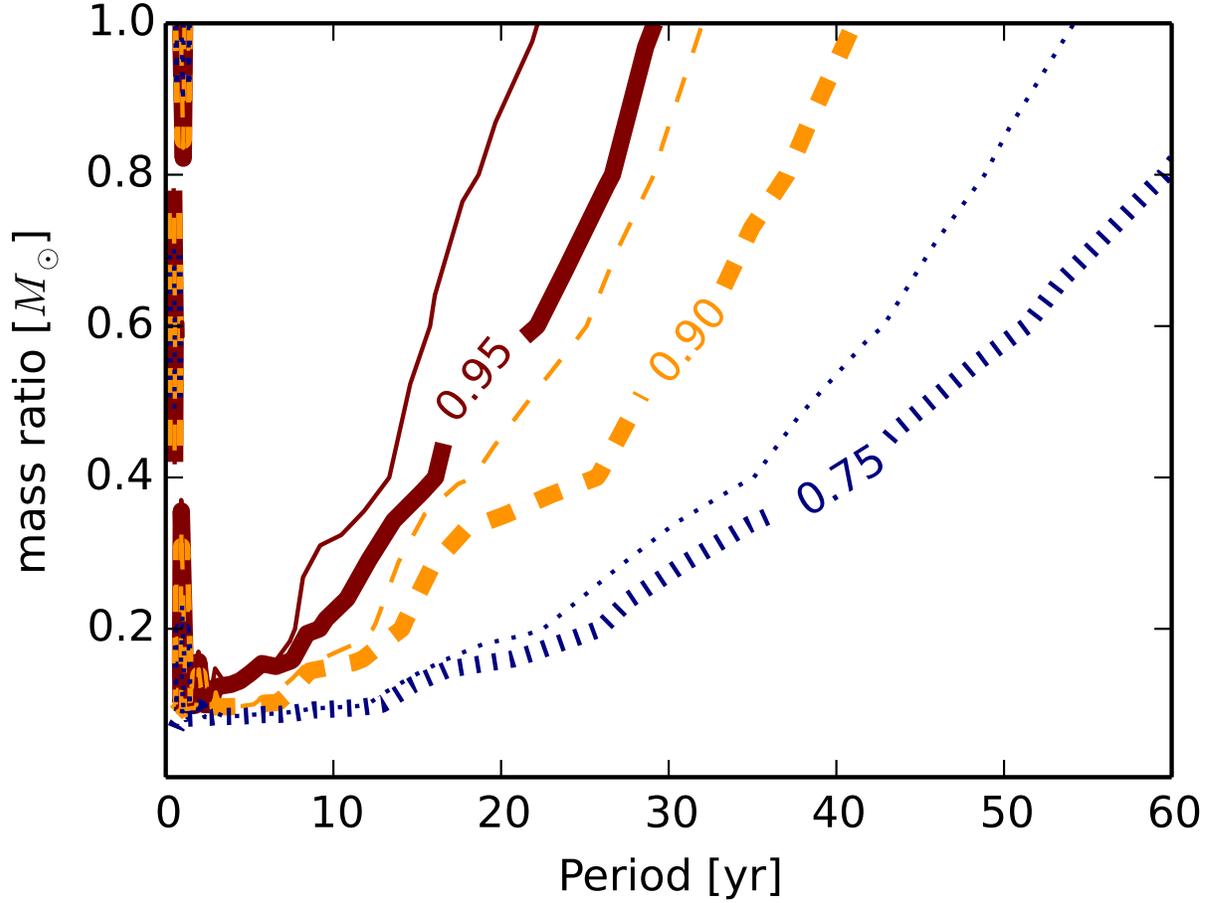}
\caption{ Probability of detecting the 
binary companion of a Cepheid based on the radial velocity of the 
primary. The simulations assume that a radial velocity difference $> 2$~km~s$^{-1}$ 
leads to a significant detection. Further assumptions are discussed in 
the text. The thick lines are contours of the detection probability for 
circular orbits, the thin lines for elliptical orbits with 
$\epsilon=0.5$. The value of the detection probability for the circular 
orbits is labeled in the plot; the labels also apply to the thin lines of 
equal color and line style.
\label{detectionprobability}
}
\end{figure}

We perform this whole set of simulations for each star in our sample 
twice, once with  circular orbits and once with elliptical 
orbits with $\epsilon=0.5$. For the given binary parameters, the probability 
of detecting orbital motion
varies between different stars in our sample, 
because they are observed on different time cadences. We thus average 
the results over the entire sample. 
Figure~\ref{detectionprobability} shows a contour map of the 
averaged detection probabilities. It shows that we are not  sensitive to 
periods below about one year, because the radial velocities used in this 
study are averaged on an annual basis (Fig \ref{vels}).
However, since observations were typically
 made over a period of weeks or months, and the shortest orbital periods 
result in the largest orbital motion, it is unlikely that many systems are 
missed for orbital periods close to 1 year.  
 Except for very low mass 
companions, we expect to find almost all binary systems with periods 
below 10 or 20 years. For higher mass secondaries  
 larger semi-major axes and thus longer periods are detected.
Since the estimates are based on the actual years of observation, the best 
way to increase the detection fraction is to increase the length of the time
 series, which we plan to do in a future paper. 

\begin{figure}
 \includegraphics[width=12cm, angle=0]{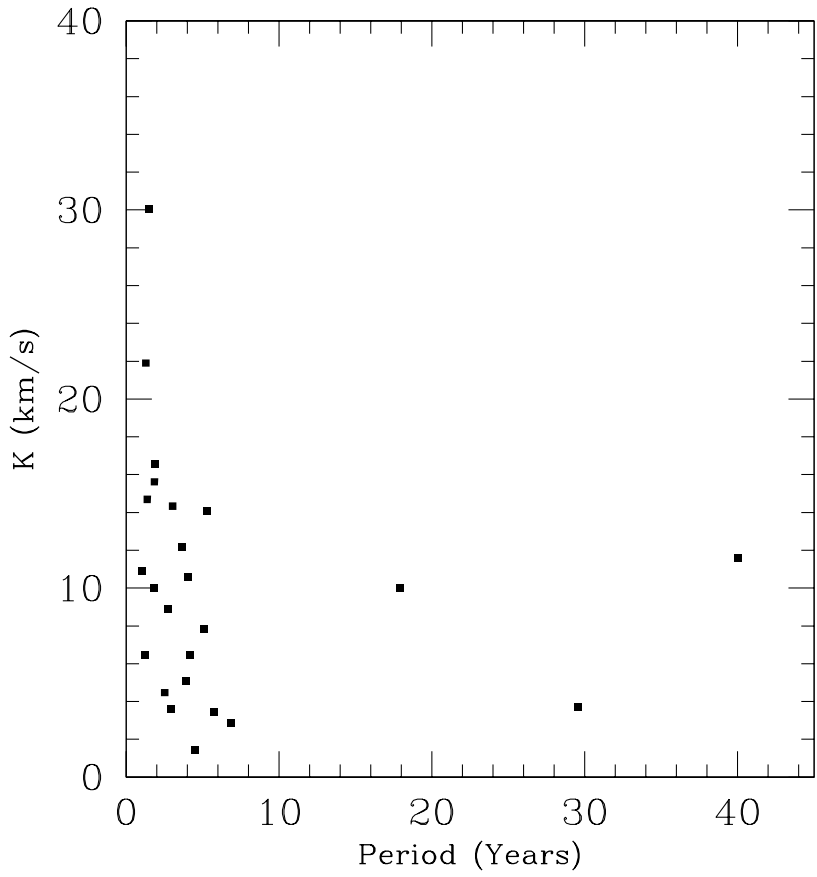}
\caption{The orbital velocity amplitude of  Cepheid orbits as a function of the orbital period
\label{cep.p.k.eps}}
\end{figure}
\begin{figure}
 \includegraphics[width=12cm, angle=0]{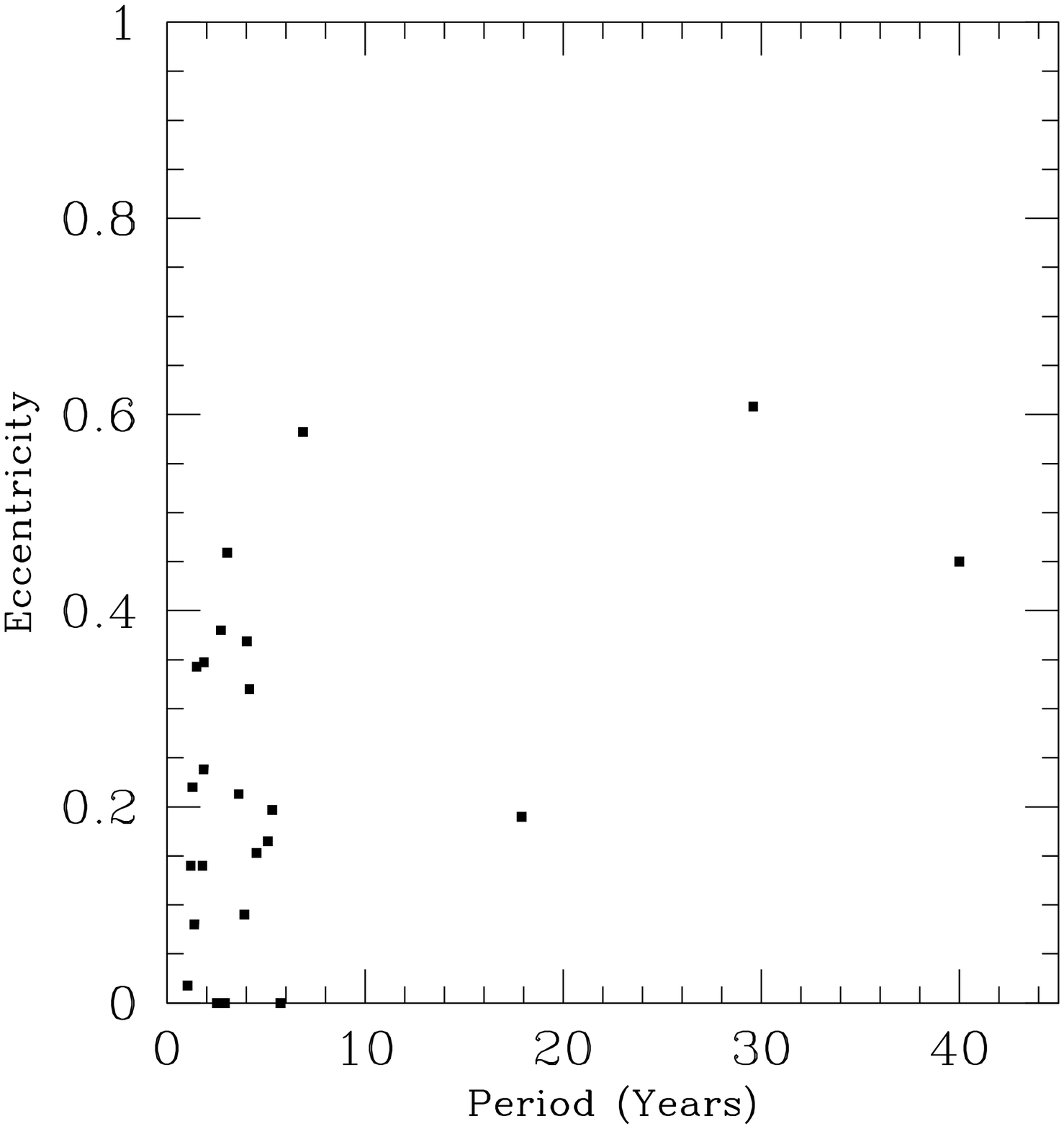}
\caption{The orbital eccentricity of Cepheid orbits as a function of orbital period.
\label{cep.p.ecc.eps} 
}
\end{figure}

To confirm that we have selected reasonable parameters for the
simulation and also  aid in the analysis below, we have compiled a
list of  stars with 
orbits. The list began with the list assembled by 
Szabados\footnote{http://www.konkoly.hu/CEP/orbit.html} and includes orbits 
from Groenewegen (2013).
We have made no attempt to improve orbits, only to select one which gives
reliable parameters. Figs.~\ref{cep.p.k.eps}  and~\ref{cep.p.ecc.eps} 
are from this list of Cepheids with orbits (U Aql, 
FF Aql, V496 Aql, 
RX Cam,  Y Car,  YZ Car,  DL Cas, XX Cen, MU Cep,  AX Cir, SU Cyg, VZ Cyg,
MW Cyg, V1334 Cyg,  Z Lac,  S Mus,  AW Per,  S Sge,  W Sgr, V350 Sgr,  V636 Sco,  
$\alpha$ UMi, FN Vel, and  U Vul).  The eccentricity values in the simulation cover essentially the full range observed in the orbits.


\section{Discussion}

\subsection{Binary Frequency}

We have now derived annual mean velocities from the large 
quantity of accurate velocity measures for Cepheids (Fig \ref{vels}).  
Because of the challenges of maintaining velocity 
zero points over 20 years, as well removing the pulsation velocities sometimes in the face 
of variation in pulsation periods, we use the following criterion to assess the results.  
Annual means that are within 1 $\sigma$ of the $\pm$ 1 km~s$^{-1}$ band in Fig \ref{vels}
are 
considered to show {\bf no} orbital variation.  This judgement was confirmed by the fact that in 
several instances, two sets of observations from different instruments were made during a 
single year.  Typically they are both located within the band, or very occasionally, when 
one set is outside the band, the other is within it, hence indicating no variation.   
(In Fig \ref{vels}, we only include years with more than 1 observation.)  
That limit is realistic, but
 means that we would not identify orbital motion with an amplitude less than 
1 km~s$^{-1}$.  In addition, in the analysis 
for binaries (below) we omitted stars with less than 4 years of observations 
(SZ Aql, Y Oph, and CK Cam).

 The star showing the largest indication of orbital motion in Fig. 2 
is $\delta$ Cep
itself.  This suggests putting it on a watch list.  However, there are a number of previous
high resolution spectral studies of $\delta$ Cep for further comparison.  
Specifically, Shane (1958) discusses three sets of spectra taken with the Mills three 
prism spectrograph at Lick Observatory (10 \AA/mm at 4500 \AA) 
in 1907, 1923, and 1950, all providing good 
phase coverage (Table \ref {delcep}).  All three series have a mean velocity 
within 0.1 km~s$^{-1}$ of the mean of -16.1 km~s$^{-1}$.  
There is a further  10 \AA/mm series from Dominion Astrophysical Observatory  
in 1975-1978 (Wallerstein 1979).  He finds them to have the same systemic velocity 
within the errors.  We have also determined the systemic velocity directly for 
Bersier et al. (1994), Barnes et al. (2005), Storm et al. (2004), and Kiss and Vinko 
(2000) in the same way as described for SV Vul and 
S Vul in Appendix A, all compiled in  Table \ref {delcep}.
Bersier data is the most negative, as shown in Fig \ref{vels},
 but the spread is only about 
2 km~s$^{-1}$, our detection interval.  

After the completion of the discussion above, Anderson, et al. (2015) presented new 
high precision radial velocities (0.015 km sec$^{-1}$) for $\delta$ Cep.  They concluded that  
it is a binary with an amplitude of 1.5 km sec$^{-1}$, a period of 6.0 years, and 
a high eccentricity (0.647).  We had concluded there is little evidence for 
orbital motion beyond our survey goal of $\pm$ 1 km~s$^{-1}$.  The new results, particularly
the low amplitude and large eccentricity, mean it is not surprising that additional
very high quality data was needed for a confident orbital detection.  (We have retained 
non-binary listing in Table  \ref{mult.stat} and the statistics.) More important, this 
serves to emphasize that our binary frequency is only an upper limit because other 
low amplitude systems no doubt remain undetected.


\begin{deluxetable}{llll}
\footnotesize
\tablecaption{$\delta$ Cep Velocities \label{delcep}}
\tablewidth{0pt}
\tablehead{
\colhead{Year} & \colhead{Sys. Vel $\pm$ $\sigma$} &  \colhead{Source} \\ 
 \colhead{(approx.)} & \colhead{km~s$^{-1}$}  & \colhead{} \\  
}
\startdata

                               
  1907    &     -16.1 $\pm$  0.1  &  Shane (1958) \\
  1923    &     -16.2 $\pm$ 0.1  &   Shane (1958) \\
  1950    &     -16.1 $\pm$ 0.1  &   Shane (1958) \\
  1977    &     -16.       &   Wallerstein (1979) \\
  1980 &  -17.4 $\pm$ 0.1  &  Bersier et al. (1994) \\
  1986  & -15.3 $\pm$ 0.2 &  Storm et al. (2004) \\
  1996  & -15.9 $\pm$  0.1 &  Barnes et al. (2005) \\
  1996 &  -15.6 $\pm$  0.1 &  Kiss and Vinko (2000) \\
             
\enddata

\vskip .1truein

\noindent

\end{deluxetable}

We have proceeded as follows to assess the fraction of binaries among
Cepheids.  From the Cepheid database\footnote{http://www.astro.utoronto.ca/DDO/research/cepheids/}
we have generated
a list of Cepheids north of -20.9$^o$, and ordered them by decreasing
luminosity (Table \ref{mult.stat}).  None of the Cepheids 
in the current study
were found to have orbital motion with an amplitude  larger than 1 km
s$^{-1}$, and these stars have been marked x in Table \ref{mult.stat} .  Three
stars with observations spanning less than 4 years (SZ Aql, Y Oph, and
CK Cam) are considered to have insufficient velocities, and left blank
in the table.  On the other hand, stars known to be binaries showing
orbital motion are marked ``bin'' in the table.    References are
given in the Konkoly Cepheid orbit table\footnote{http://www.konkoly.hu/CEP/orbit.html}.

Several designations in Table~\ref{mult.stat} need additional discussion.  V496
Aql and VZ Cyg have orbits from Groenewegen (2013).
X Pup and XX Sgr have recently been discussed by Szabados et al. (2012),
 and they postulate low amplitude orbital motion in both.  For
X Pup, recent data shows little change in velocity.  Among the older
data, we have deferred the velocities from Caldwell et al. (2001) 
to a later paper in order to cross check against other southern
stars.  Thus, we leave the X Pup blank (unknown) in Table  \ref{mult.stat}, however
future analysis and observations may lead to reclassification.  For XX
Sgr, the data during the era of the current project indicate no
orbital motion.  Only older data at lower dispersion (Joy, 1937) have
discordant velocities. Again, it is listed as blank (unknown) in 
Table  \ref{mult.stat}. 
V1344 Aql (V = 7.77 mag) has been examined for orbital motion by
Szabados et al. (2014).  Data from the era of the current project shows
essentially no variation, and even older data shows only very small
radial velocity shifts, so we classify this star as unknown in 
Table  \ref{mult.stat}.


\clearpage

\begin{deluxetable}{llrlr}
\footnotesize
\tablecaption{Multiplicity Status  \qquad\qquad\qquad\qquad \label{}}
\tablewidth{0pt}
\tablehead{
\colhead{Star} & \colhead{$<$V$>$} & \colhead{$P_{\mathrm{pulsation}}$} & \colhead{Status*}  & \colhead{$P_{\mathrm{orbit}}$} \\
\colhead{} & \colhead{mag} & \colhead{d} & \colhead{}  & \colhead{yr} \\
}

\startdata




 Alp UMi   &    1.982   &     3.970  &   bin & 29.6  \\ 
 Eta Aql    &     3.897  &      7.177 &  x  &  \\    
Zeta Gem   &      3.918  &     10.151 &  x  &   \\   
 Del Cep    &     3.954  &      5.366 &  x  &   \\   
  FF Aql   &      5.372 &       4.471 &    bin & 3.9  \\ 
  RT Aur    &     5.446  &      3.728 &  x  &    \\  
   S Sge   &      5.622  &      8.382  &   bin & 1.9  \\ 
   T Vul   &      5.754  &      4.435 &  x   &    \\ 
  DT Cyg    &     5.774  &      2.499 &  x   &    \\ 
V1334 Cyg   &      5.871  &      3.333 &   bin & 5.3   \\
  SU Cas    &     5.970  &      1.949 &  x  &     \\ 
   T Mon    &     6.124  &     27.025  &   bin & $\geq$ 90  \\ 
   Y Oph    &     6.169  &     17.127 &    &     \\ 
V440 Per    &     6.282  &      7.570 &  x  &     \\ 
   X Cyg    &     6.391  &     16.386 &  x  &     \\ 
   U Aql   &      6.446  &      7.024  &   bin & 5.1  \\ 
  SZ Tau   &      6.531  &      3.148 &  x   &    \\ 
   U Sgr   &      6.695  &      6.745 &  x   &    \\ 
  SU Cyg    &     6.859  &      3.845 &    bin & 1.5  \\ 
  BB Sgr    &     6.947  &      6.637 &  x   &    \\ 
   W Gem     &    6.950  &      7.914 &  x   &    \\ 
   U Vul   &      7.128  &      7.991  &   bin & 6.9    \\
  TT Aql    &     7.141  &     13.755 &  x  &     \\ 
V636 Cas    &     7.199   &     8.377 &  x  &     \\ 
  SV Vul    &     7.220  &     44.995 &  x  &     \\ 
  YZ Sgr    &     7.358  &      9.554 &     &     \\ 
V350 Sgr   &      7.483  &      5.154 &    bin & 4.0   \\
  AW Per   &      7.492  &      6.464  &   bin & 40.0   \\
  CK Cam   &      7.58    &     3.295       &   &       \\
  RX Aur   &      7.655  &     11.624 &  x   &     \\
  RX Cam   &      7.682   &     7.912  &   bin & 3.1   \\
V496 Aql    &     7.751  &      6.807  &  bin  & 2.9   \\
V1344 Aql   &      7.767  &      7.478  &    &     \\
  IR Cep    &     7.784   &     2.114 &  x   &    \\ 
V1162 Aql   &      7.798  &      5.376 &  x  &    \\ 
  WZ Sgr    &     8.030   &    21.850 &  x   &    \\ 
  EU Tau    &     8.093   &     2.102 &  x   &     \\
  RY CMa    &     8.110   &     4.678 &      &    \\
  SS Sct    &     8.211  &      3.671  &     &    \\
  ST Tau    &     8.217   &     4.034 &  x  &   \\
  SV Mon     &    8.219  &     15.233 &  x  &    \\
  FM Aql    &     8.270  &      6.114 &  x   &  \\
  FN Aql    &     8.382  &      9.482 &   x &   \\
   X Lac    &     8.407  &      5.445  &  x &   \\
  RS Ori    &     8.412  &      7.567  &  x &   \\
   Z Lac    &     8.415  &     10.886  &    bin & 1.0  \\
   X Pup    &     8.460  &     25.961  &     &     \\
V526 Mon   &      8.597  &      2.675  &  x  &     \\
  SZ Aql   &      8.599 &      17.141  &     &     \\
  RW Cam    &     8.691  &     16.415  &     &     \\ 
  RR Lac    &     8.848  &      6.416 & x    &   \\
   X Vul    &     8.849  &      6.320 & x   &   \\ 
  XX Sgr   &      8.852  &      6.424   &   &     \\ 
  BG Lac    &     8.883  &      5.332  & x  &       \\
   V Lac   &      8.936  &      4.983  &  x &      \\
  CD Cyg   &      8.947  &     17.074  &  x &      \\
  VZ Cyg   &      8.959  &       4.864  &  bin  & 5.7   \\  
   S Vul   &      8.962  &     68.464   & x  &   \\ 
  GQ Ori   &      8.965  &      8.616  &    &     \\ 
  DL Cas   &      8.969  &      8.001  &    bin & 1.9  \\
V379 Cas   &      9.053  &      4.306 &   x   &   \\

\enddata

* bin = binary; x = single

\vskip .1truein

\noindent
\label{mult.stat}
\end{deluxetable}





\begin{table}
\caption{ Binary Fraction  \qquad\qquad\qquad\qquad }
\label{bin.frac}
\begin{tabular}{llrlll}
\hline
\hline
Star \# & $<$V$>$ & \# Single & \# Binary & \% Binary & \% Sample \\
Tab. 6 & mag & Tab. 4 & Tab. 6 &  &     \\
\hline
\multicolumn{6}{c}{ P$_{\mathrm{orbit}} > 1$ yr } \\
\hline

1 -- 20 & $<$ 6.9 & 12  & 7  & 37 $\pm$ 11  & 19/20 = 95\% \\  
21 -- 40 & 6.9 -- 8.2 & 10 & 5 & 33 $\pm$ 12  & 15/20 = 75\% \\
41 -- 61   & 8.2 --9.0 & 13  & 3  & 19 $\pm$  10  & 16/21 = 76\% \\
Total    &  &  35  &  15 & 30 $\pm$ 7  & 50/61 =  82\%  \\
\hline
\multicolumn{6}{c}{ P$_{\mathrm{orbit}} \/ 1-20$ yr } \\
\hline
1 -- 20 & $<$ 6.9 & 12  & 5  & 29 $\pm$ 11  &  \\  
21 -- 40 & 6.9 -- 8.2 & 10 & 4 & 29 $\pm$ 12  &   \\
41 -- 61   & 8.2 --9.0 & 13  & 3  & 19 $\pm$  10  &  \\
Total     &  &  35  &  12 & 26 $\pm$ 6  &     \\
\hline
\end{tabular}
\end{table}

Because the quantity and quality of data often decreases for fainter stars, we have adopted 
the following strategy  to assess the results.  We have ordered the Cepheids from the brightest to 
the faintest (Table \ref{mult.stat}).  We have divided the data into three groups of $\simeq$ 20 stars in this list and 
derived the binary fraction for each group separately.  Table  \ref{bin.frac}
contains the results.  The first 3 
rows are for the groups, with the magnitude range of the stars in Column 2.  Cols 3, 4, and 5 contain 
the number of stars for which velocities are available in our sample, the number of binaries in 
the group and the fraction which are binaries respectively.  
We have estimated the uncertainty in the ratios using equation 1 from Alcock et al. (2003)
The final column shows the percentage of
 the group which was sampled in the top section of the table.  (In the bottom half where a subset
of the data is discussed, it is omitted.)
  The final row gives the results for the whole sample.  We have retain the classification of 
$\delta$ Cep as single (which we originally concluded).  Adding it to the binary 
group (Anderson, et al. 2015) would raise the binary fraction for the brightest 20 stars to 42\%.

As can be seen 
in Table \ref{bin.frac} (top), the percentage of binaries {\it appears} to decrease as  the groups become fainter.  We consider 
this to be only an indication of the difficulty in detecting binary motion in the fainter sample,
however it alerts us to a possible selection effect.  
 A similar decrease in apparent binary fraction for fainter stars 
for all period/separation ranges was found by Szabados (2003). 
The three groups, however, give the same fraction within the 1 $\sigma$ errors.  The two groups 
of the brightest stars give very close results, within 0.2  $\sigma$.  Combining them, we get
a binary fraction of  35 $\pm$ 8\%.  Comparing this with the binary fraction from the 
faintest group provides weak statistical evidence of a selection effect in fainter stars,  
 
We want to consider a further refinement to the statistics in Table  \ref{bin.frac}.  
In the final
column in Table 6 we have added the orbital period from Table 8.  Since the velocity 
studies in this project (Table \ref{bin.frac}) cover the range of 1 to 20 years, 
the bottom of 
Table \ref{bin.frac} shows the statistics when binaries with longer periods are 
omitted.  Again, the results from the two brightest groups are very similar.  Combining 
them, we get a binary fraction of 29 $\pm$ 8\% (20 $\pm$ 6\% per decade of 
orbital period), which becomes our preferred 
fraction.  


Determining 
the frequency of binaries/multiples for intermediate mass stars in the 
period range 1 to 20 years  is the aim 
of this study.  As discussed in the introduction, this radial velocity study is part of 
a group of three studies, including an HST study of resolved companions and an X-ray 
study of low mass companions of B stars.  The sample of spectroscopic binaries containing 
Cepheids  includes systems with separations of approximately 2 to 13 AU.  Thus, the binary 
frequency derived here (29\%) pertains to only part of the whole range of separations.  A more 
comprehensive value for binary fraction will be discussed in the study of resolved 
companions.   In this section, we will compare our result with those of other studies. 


{\bf O Stars:}
A number of recent studies have discussed the massive O stars.
Mason, et al. (2009) found 30\% of O stars to be spectroscopic binaries (excluding their questionable 
spectroscopic binaries) which rises to 75\% when visual companions are included. This, of course 
includes shorter orbital periods than the present study, down to a few days.  
Thus the higher radial velocity accuracy of Cepheids and the resulting binary frequency of 29\% for 
only periods of 1 to 20 year implies a significantly larger fraction of spectroscopic binaries
for the whole period range of the progenitor population of main sequence B stars.
Sana and Evans (2011) find a spectroscopic binary fraction from Milky Way open clusters for O 
stars of 44\% over the whole range of periods.  Since only about $\simeq$15\% of these (their 
Fig. 3)  have periods longer than a year, again, our binary fraction 29\% over 
this  period range implies a larger percentage over the entire period range.  
Alternately, the binary fraction of Sana and Evans (2011) for periods less than a year becomes 37\%.  
This would be added to the binary fraction we find for periods 
longer than a year, resulting in 66\% for the whole period range.    

Kobulnicky et al. (2014) report on a sample of 48 O and early B stars in the Cyg OB 
association including binaries with periods up to 5000 days.  They combine their results 
with those of Sana et al. (2012) and  Garmany, Conti, and Massey (1980).   They stress the large 
incompleteness factor they derive for the longer periods, the period range covered
in the present study.  In fact, the results of our study nicely confirm the incompleteness 
corrections they derive.  For instance, if the 29\% binary fraction we derive for the period 
range   $1-20$ years ($\log P =2.6-3.9$ for $P$ in days) is added to the cumulative fraction as a 
function of period (their Fig. 31), the binarity fraction reaches $\simeq$ 60\%, close to 
their estimated total.  Again, the period range covered by the Cepheid sample and the high
level of completeness (Fig.~\ref{detectionprobability}) underscores the value of the Cepheid velocities.  
Caballero-Nieves et al. (2014) have  performed a survey of OB stars in the Cyg OB association 
with the Hubble Space Telescope Fine Guidance System and find resolved companions as close as 
23 mas.  They find a binary frequency of 22 to 26\% in the period range 20 to 20,000 
years, the period range immediately larger than that of 
the Cepheid sample.  They have the important result that this frequency 
establishes a downturn in the frequency as a function of period, as also found by Evans et al. (2013)
for Cepheid systems with mass ratios $\geq$ 0.4.

{\bf B Stars:}
Several recent studies have been done of stars about the same mass as the Cepheids.
Chini et al. (2012) break their spectroscopic survey down into smaller spectral type bins.  
Their O3 to B1.5 group has a spectroscopic binary frequency of approximately 70\%.  
Their B2 to B6 group corresponds to the mass range of the Cepheid sample (Table 
\ref{bin.frac}), and 
the binary frequency has fallen to approximately 50\%.  For their lower mass B stars 
(B7, B8, B9), the 
frequency is approximately 15\%.  Again, the Cepheid frequency of 29\% for the limited period 
range implies a  higher frequency than the B7, B8, B9 stars for the whole period range.
Abt, Gomez, and Levy discussed B2 to B5 stars. 32 of their sample of 109 (29\%) were 
actually observed to have orbital motion.  They made a correction for incompleteness which 
increased the fraction to 59\%.  The radial velocity sample, however, was only 
sensitive to periods less than a year.  That  fraction  
should be added to the fraction in the Cepheid sample, where only periods longer than a
year are found.   

{\bf A Stars:}
For stars slightly less massive than Cepheids, two recent studies have provided results:
 De Rosa et al. (2014) and Kouwenhoven et al. (2007).
De Rosa et al. (2014) found a multiplicity fraction of $\geq$44\% from a 
combined spectroscopic and imaging 
survey over the full period range for a volume limited sample of A stars.   
Again, the binary fraction from 
the limited period range of the Cepheids implies a larger total binary fraction.   
Kouwenhoven et al. (2007) analyzed the binary frequency for intermediate 
mass stars in Sco OB2.  They used the spectroscopic observations of Levato et al. 
(1987), and found 30\% of the sample of 53 binaries with orbits.  Again, all the 
orbits have periods (much) less than a year.  This fraction thus should be added to 
the Cepheid fraction.  They found a further 43\% to have variable radial velocities. 
At least some of these are low amplitude binaries, however some of them may be variable 
because of pulsation.
Kouwenhoven et al. also discussed the spectroscopic observations of Brown and 
Verschueren (1997).  For their sample of 71 stars, 24\%  are spectroscopic binaries. 
A further 39\% have variable radial velocities, again likely to be divided between 
orbital motion and pulsation.   

{\bf G Stars:}
For solar mass stars, the most recent determination comes from Raghavan et al. (2010),
who  found a multiplicity frequency of 44 \%.  Because these stars are nearby and 
sharp-lined, the corrections for incompleteness are much smaller than for more 
massive stars.  Thus, the current study as well as other studies above confirm
a smaller multiplicity fraction than for more massive stars.


Thus, this survey of Cepheid velocities in the period range of 1 to 20 years 
finds a binary frequency which consistently implies a
 higher frequency (at all period ranges) than found in 
studies of main sequence stars. This is reasonable because
of the high accuracy of the Cepheid velocities compared with other massive stars.   
The periods discussed here are  
longer than typically investigated in the main sequence studies, which is exactly where the 
orbital velocities are smaller, and hence companions of main sequence stars will be missed.  
Presumably a  
large fraction of short period binaries would be discovered in the main sequence studies, but the
velocity accuracy of this study is required to find the longer period systems.
  Thus, we stress that the Cepheid sample fills in the period range between
the short periods usually detected in massive star studies and the long periods of 
resolved companions.

Neilson, et al. (2015) compared the fraction of Cepheids in spectroscopic binaries 
with a binary population synthesis model, and found generally good agreement.  The number
favored here has been slightly revised from the fraction they used (35\%), but the 
general agreement still holds.    

As mentioned above, 
 there are two very important effects which modify the binary fraction in 
 the Cepheid sample: dynamical evolution 
and mergers/binary interaction.  Any system with more than two stars may be unstable to dynamical 
interaction  between the components which
sometimes leads to the ejection of one star (typically the
smallest). This study provides no information about this topic.  The second effect was 
discussed by 
Sana, et al. (2012) who concluded that three quarters of all O stars will either merge
(20-30\%) or have interactions resulting in stripping  the envelope or accreting 
mass because 
of Roche lobe overflow as they evolve off the main sequence.  Binaries in the Cepheid sample 
(Table 8)
 may have suffered no consequences from this.  However, there may be some systems 
in the list of single stars (Table~\ref{mult.stat}) which began life as a short period binary, 
but  have now merged, thus moving from the binary list to the single list.   The obvious 
and simplest result of this is to increase the number of currently single stars and decrease
 the binary fraction.





\subsection{Mass Ratios}

We have stressed that  Cepheids provide a well characterized sample to investigate the 
properties of spectroscopic binaries,  specifically, in the period range of one to 20 years. 
In addition to providing the binary frequency, this group has another valuable property.  Because
the primary (Cepheid) has evolved to become a cool supergiant, hot companions can be observed directly in
the ultraviolet.   Ultraviolet spectra, specifically from the International Ultraviolet 
Explorer (IUE) satellite, provide either an uncontaminated spectrum of the companion or an 
upper limit.  Thus the temperature can be directly observed, and from that a mass inferred.  We 
have derived masses for the companions in this way in Table 8.
The sample used is the 
sample of stars with orbits (including both N and S hemispheres.)    
Masses of the companion (Table 8,
Col 2) are taken from Evans et al. (2013), the IUE survey (Evans 1995), or the Hubble Space 
Telescope measurement  of Polaris (Evans et al. 2013), as indicated in Col. 6.   
 Masses for the Cepheids (Col 3)  are derived as in Evans et al. (2013), 
using the Leavitt
law (Period--Luminosity relation) from Benedict et al. (2008) and  the mass--luminosity 
relation based on  the models of Prada Moroni et al. (2012) with moderate convective 
overshoot. The periods of the overtone pulsators ($\alpha$ UMi and V1334 Cyg) have been 
``fundamentalized'' for the mass calculation.  
Mass ratios, q = M$_2$/M$_1$ (Col 4) are calculated from the Cepheid and companion 
masses. When a companion mass has a range, a mass from the center 
of the range was used.

A few stars require extra comments.  MU Cep, MW Cyg, VZ Cyg, and FN Vel were not observed by 
IUE because they are  fainter than the survey limit, so they lack mass ratio information, 
and are listed at the bottom of Table 8.

The mass ratio data concentrates on orbital periods of 30 years or less because that includes 
the most of the known orbits, and also corresponds 
reasonably well to the period range which the 
present  radial velocity data are sensitive to. 
However, AW Per with a slightly longer orbital period is included. We also made an 
exception and included W Sgr, even though the mass of the hottest companion (Col 2) is 
actually of a resolved companion, not the companion of the spectroscopic orbit 
(Evans et al. 2009).  This means the mass ratio is actually an upper limit to 
the ratio of the spectroscopic system.

Figure 7 shows the distribution of the mass ratios for the Cepheid systems.  
The q data from Table 8
  were divided into bins of 0.2 for q $>$ 0.4.  For smaller q values
(both observed values and upper limits), the data was summed in one q bin  0.0 to 0.4 
(total count of 10). The location of the data for the lowest q value is shifted
slightly for clarity.  Poisson errors are shown.   Although the sample is fairly small, it is clear that there 
is no preference for equal mass companions in the Cepheid sample.  We stress that this 
sample contains systems with periods longer than a year, so this says nothing about 
studies of shorter period systems which have been found in some studies to 
show a preference for equal mass companions.
Creating a single bin for stars with q $<$ 0.4 may disguise a rise in frequency 
for small q values.  If even half of the 6 stars which have only an upper limit belong
to the 0.2 to 0.4 bin, the number in that bin rises to 7, larger than the
0.4 to 0.6 bin.  Even with a simpler division into two bins, 0--0.5 and 0.5 -- 1.0 because 
of small number statistics, the result still favors small q systems.  The two bins 
have 14 and 6 systems, respectively.  
In addition, from the detection analysis (Fig. 4), any undiscovered
binaries are likely to be in the  0.0 to 0.2 bin.  Nevertheless,     
the  mass ratio distribution is relatively flat, and similar to that found by Sana and Evans
(2011) for more massive O stars over the whole range of periods  they studied. 

As we get more and more information about binary/multiple systems, we can 
investigate the finer detail of the
properties.  We have used the data of Raghavan et al.
(2010) to create a comparison with solar mass stars.  Specifically, we have used 
their Figure 17 (the spectroscopic binaries shown as +) to create a distribution of mass 
ratios for the period range 1 year to 30 years.  This is shown in Fig. 7 for comparison, 
scaled by the size of the sample.  The  mass ratio distribution for the 
solar mass stars is relatively flat for 
q $<$0.8. similar to that for the Cepheids.  For larger q, the solar mass stars have 
a much larger fraction of equal mass pairs than the Cepheids.     

The purpose of CRaV, and other studies of binary properties is to identify 
characteristics which relate to star formation, and how properties vary depending
on the stellar mass, and the separation of the systems.  In the comparison in Fig. 7 
between 5$M_\odot$ Cepheids and solar mass stars, it may be that the apparent differences 
are the result of asking the wrong question.  For instance, the companions are also drawn 
from very different populations (typically 3 $M_\odot$ vs 0.5 $M_\odot$).  From the 
Fig. 17 in Raghavan et al. (2010) it is clear that the distribution of q changes 
from short period systems (large q) to longer period ones (spread in q to smaller 
values).  In this sense, the distribution of q for solar mass stars would be more 
similar for the group of longer period stars.  

A brief comparison with the previous distribution of mass ratios for Cepheids (Evans 
1995) is in order.  The current distribution of q is 
flatter for two reasons.  First, accumulating
evidence favors somewhat smaller Cepheid masses.  Second, a realization of the large
number of triple systems makes the derivation of lower limits more questionable.  

One further point should be mentioned in discussing the binary fraction of Cepheids.  
In addition to the alteration in the binary systems when a post-main sequence short
period binary undergoes Roche lobe overflow, there may be mass loss during the Cepheid
phase itself.  See, for example the summary in Neilson et al. (2012).  
While mass loss at this phase is not as dramatic as in massive main sequence stars, 
it could affect both the mass 
ratios and the orbits.  
In addition, comparison with main sequence progenitors will be altered.  Resolution
of these questions awaits further information on mass-loss at the Cepheid phase.

\section{Summary}

We have examined velocity data for Cepheids to determine the binary/multiple fraction.
The velocity data have a typical accuracy of 1 km~s$^{-1}$ per observation, 
and were obtained over a period of 
20 years.  After correcting the observed velocities for the pulsation velocity, annual means were 
formed.  We have assessed our detection limits and found a high probability of recognizing binary 
systems with mass ratios as small as 0.1 and periods as long as  20 years. Further data extending the time series 
will increase the the period/separation range. 
A binary fraction was formed by combining this sample with Cepheids known to be members
of binary/multiple systems. The binary fraction appears to decease at fainter magnitudes. However, 
we take this to be an indication of undiscovered binaries at fainter magnitudes.  
Therefore, we conclude that the most 
reliable binary fraction of 29 $\pm$ 8\% (20 $\pm$ 6\% per decade of orbital period)
 comes from the  sample of the 40 
brightest stars, which has the most
complete information. The range of separations that the Cepheids sample probes is approximately 
2 to 20 AU, which is poorly studied for other massive stars.  
Comparing the binary fraction in this limited period/separation range with the other 
recent determinations of binary frequency confirms that the high accuracy velocities of the Cepheids 
results in a higher binary frequency.  The distribution of
the  mass ratio q is also examined from spectral 
information about the companions in the ultraviolet, and found to be flat as a function of q.    

Future  work is planned to extend the time span in the Northern hemisphere, and also to 
make a similar analysis of Cepheids in the Southern hemisphere.

\clearpage

\begin{deluxetable}{lcrrrlr}
\footnotesize
\label{mas.rat}
\tablecaption{Mass Ratios  \qquad\qquad\qquad\qquad \label{}}
\tablewidth{0pt}
\tablehead{
\colhead{Star} & \colhead{M$_2$} & \colhead{M$_1$} & \colhead{q} &  
\colhead{$P_{\textrm{pulsation}}$}  & Ref &\colhead{$P_{\textrm{orbit}}$}    \\
\colhead{} & \colhead {$M_\odot$} & \colhead {$M_\odot$} & \colhead{} &  
\colhead{$^d$} &  & \colhead{yr} \\
}

\startdata


SU Cyg & 	 3.2  &	     4.7  &	  0.68  & 3.845	& 1  & 1.5 \\ 
$\alpha$ UMi &	 1.26 &	 4.7   &	     0.27  &	     3.970 & 3 & 29.6 \\
FF Aql  &	 1.6-1.4  &	 4.9   &	    0.31  &   4.471 & 2 & 3.9 \\
Y Car  &	 2.5  &	    4.9   &	     0.51   &	  4.640  & 2 & 2.7 \\
V1334 Cyg &	 4.0   &	  4.9 &	  0.82  & 4.722	& 1 & 5.3 \\ 
V350 Sgr &	 2.5  &	    5.1  &	   0.49  & 5.154	& 1 & 4.0 \\ 
AX Cir & 	 5.0  &	     5.2 &	   0.96   & 5.273 	& 1 & 17.9 \\
AW Per  &	4.0   &	    5.4 &	   0.74    & 6.464	  &  1 & 40.0    \\   
V636 Sco &	 2.4   &	   5.6  &	    0.43  &  6.797	& 1 & 3.6 \\
V496 Aql  & $<$ 1.9  & 5.6    & $<$0.34 & 6.807 & 4  & 2.9 \\
U Aql &	 2.3  &    5.7  &	  0.40   & 7.024	 &	1 & 5.1 \\
W Sgr  &     $<$ 2.2   &	    5.8  &	 $<$  0.38  &  7.595	& 1 & 4.5 \\
RX Cam  &	 2.2  &	    5.8  &	  0.38   &  7.912 & 1 & 3.1 \\
U Vul &	  $<$2.1 &	   5.8  &	     $<$0.36    &	  7.991  & 4 & 6.9 \\
DL Cas  &	 2.5   &	   5.8  &    0.43 &    8.001  & 2 & 1.9 \\
S Sge  &	1.7-1.5  &	 5.9    &	   0.27  &	     8.382 & 2 & 1.9 \\
S Mus  &	5.3   &	    6.2  &	  0.85  &  9.660	& 1 & 1.4 \\
Z Lac  &	$<$1.9   &	  6.4  &	      $<$0.30  &    10.886 & 2 & 1.0 \\
XX Cen  &	 $<$2.1 &	   6.5   &	    $<$0.32   &	   10.953 & 2 & 2.5 \\
YZ Car & 	$<$2.9-2.2  &	 7.7  &	    $<$0.32  &	   18.166  & 2 & 1.8 \\

& & & & & \\
MU Cep   &	&	& & 3.768 & & 4.2 \\	
VZ Cyg 	 & & &	& 4.864 & & 5.7	 \\                   
FN Vel  & & &	& 5.324	& &  1.3 \\
MW Cyg   & & & & 5.955  & & 1.2	 \\

\enddata

\vskip .1truein

1. Evans et al. 2013

2. Evans 1995 + M$_{Cep}$

3. Evans et al. 2008 

4. Evans 1992 + M$_{Cep}$

\noindent

\end{deluxetable}

\begin{figure}
 \includegraphics[width=\columnwidth]{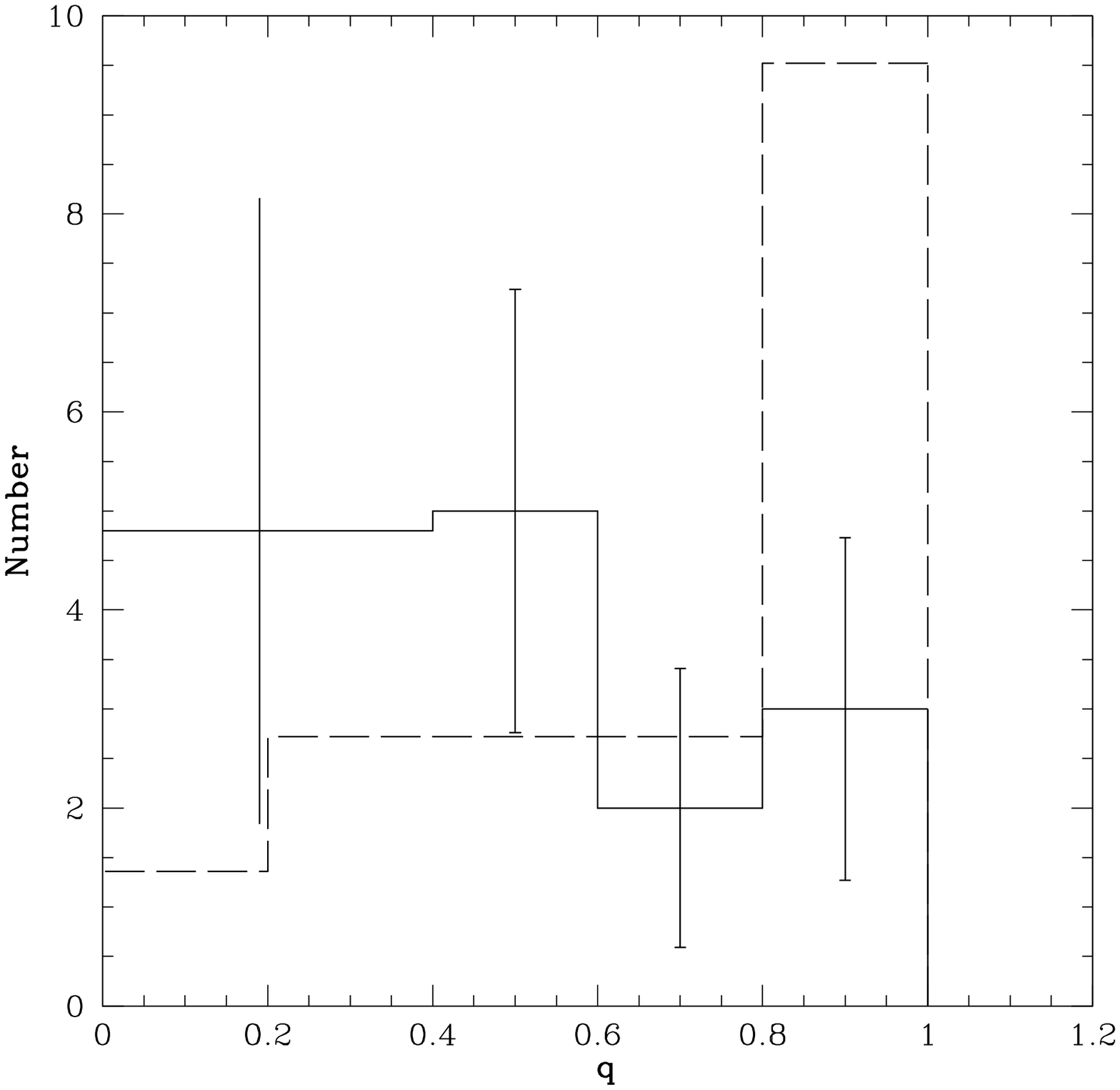}
\caption{The distribution of mass ratios.  Solid line: Cepheids; dashed line: solar mass 
stars (Raghavan et al. 2010), scaled as described in the text. 
\label{}}
\end{figure}


\section{Appendix A: Period Variations}

The stars which have period variations more complex than a constant period or a
constantly changing period are discussed here, with figures showing the 
``O--C'' (Observed minus Computed) diagrams.  The representation in the
text and figures is the same 
for all stars: C is the JD of maximum light, E is the epoch.

\subsection {EU Tau}

\begin{figure}
 \includegraphics[width=\columnwidth]{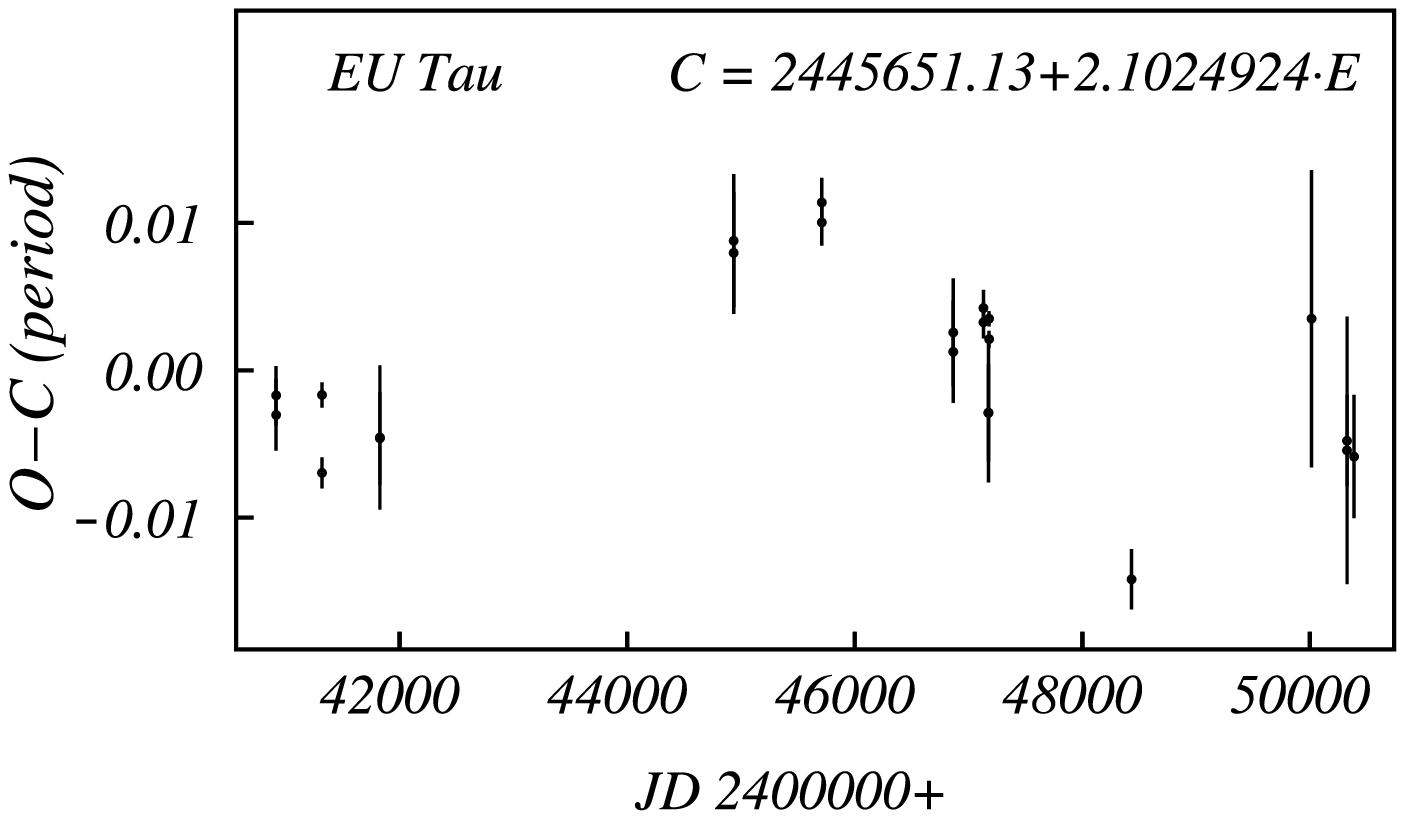}
\caption{The period variation (Observed minus Computed O-C) diagram for EU Tau.
 Time is in days; O--C is in phase. In the label,  C is the computed JD of maximum 
light; E is the epoch. See Appendix A for discussion.  \label{eutau.oc}}
\end{figure}

As with many overtone pulsators, EU Tau has erratic period fluctuations
(Berdnikov et al. 1997; Szabados 1983; Evans et al. 2015).  
This is shown in the O-C diagram     of recent photometry 
(Fig.~\ref{eutau.oc}) for:

C= 2445651.13 + 2.1024924 $\times$ E

\noindent In the case of EU Tau, the best fit
for the time period discussed in this study (which begins just before
2,445,000),   is what we use to
phase the velocity curves together: 

C = 2449679.541 + 2.10256181 $\times$ E.















\subsection{ SV Mon}    

\begin{figure}
 \includegraphics[width=\columnwidth]{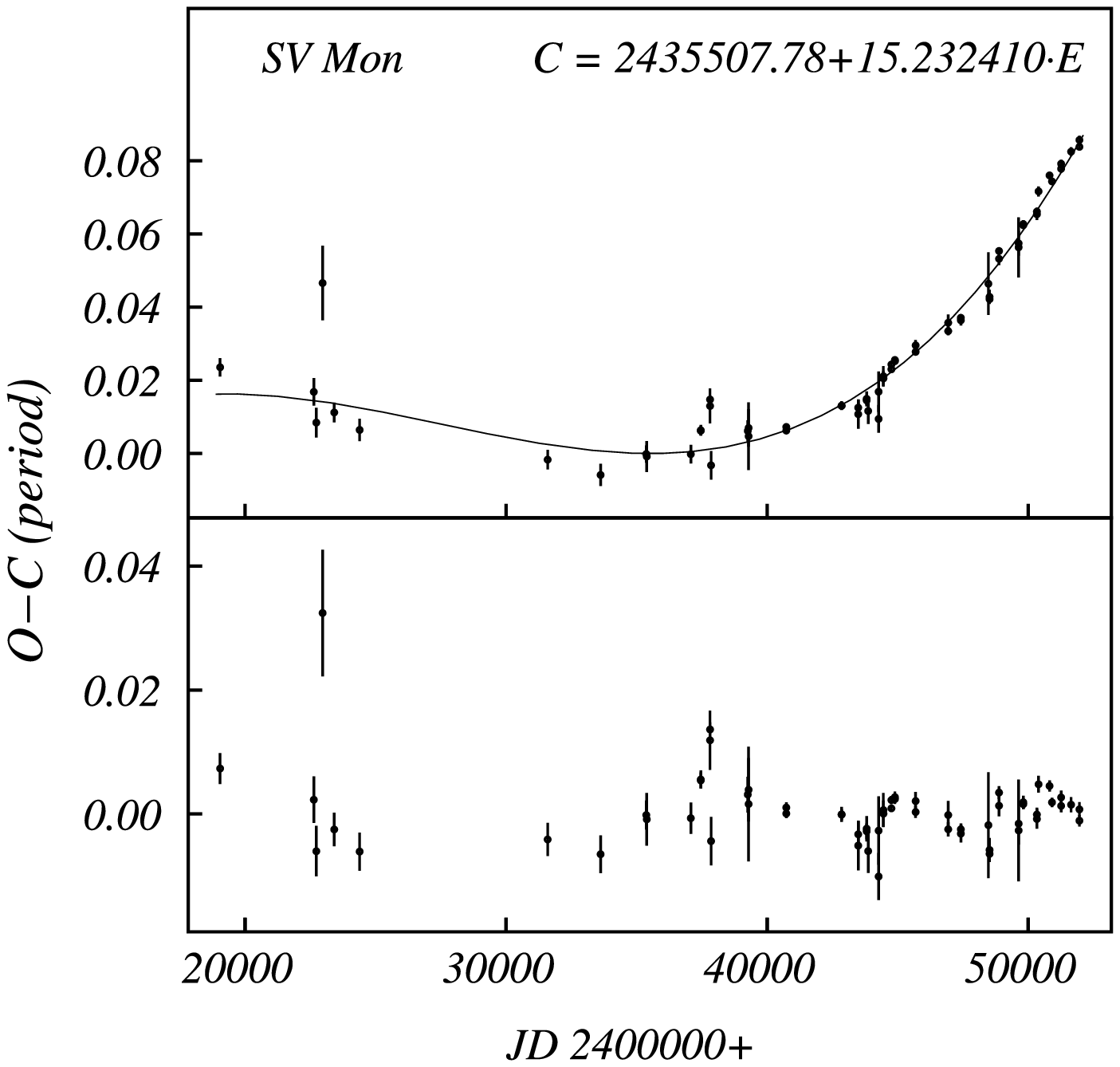}
\caption{The O-C diagram for SV Mon in the same units as Fig~\ref{eutau.oc}.
Top: Data and the cubic fit. Bottom: Residuals from the cubic fit. \label{svmon.oc}}
\end{figure}

SV Mon has a complicated period fitted by Berdnikov with a cubic
equation (Fig. \ref{svmon.oc}) for the long-term variation:  

 C = 2435507.781 +15.23241047$\times$   E
       + 0.660623600 $\times$  10$^{-6}$ $\times$   E$^2$
       + 0.415700058 $\times$ 10$^{-9}$ $\times$  E$^3$ 



\noindent However, for our analysis we used a single period and the Fourier
coefficients fitted to the radial velocities

  C = 2443794.33800+ 15.23278000 $\times$ E

\noindent Again this is adequate for the time period covered by the velocities 
which start just before JD 2,445,000.  
This star has among the largest errors on the annual means, possibly
because of some additional jitter in the period.

\subsection{ST Tau} 


The period used is from Szabados (1977):

                      C = 2441761.96300+  4.03429900 E

\begin{figure}
 \includegraphics[width=\columnwidth]{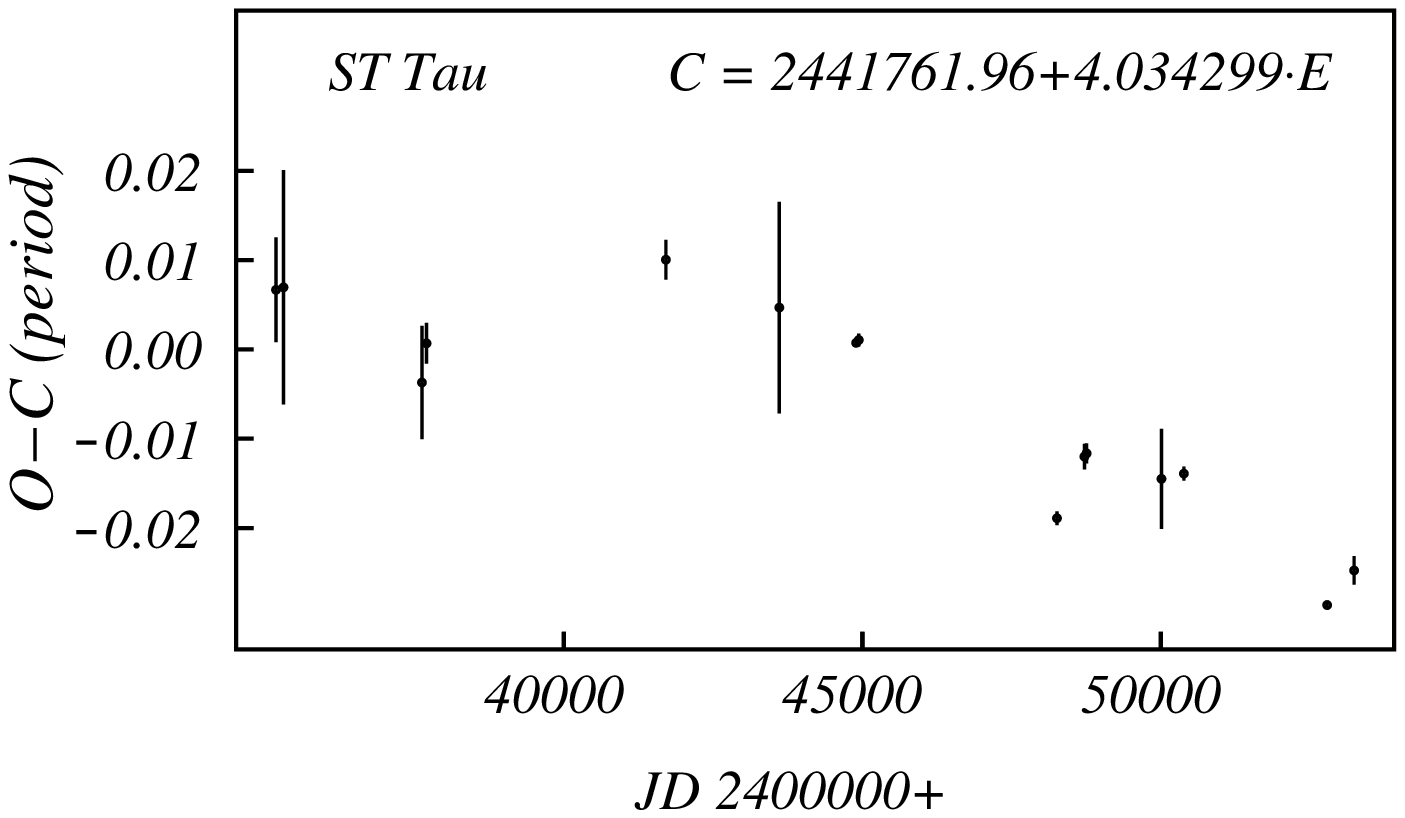}
\caption{The O-C diagram for ST Tau in the same units as  Fig~\ref{eutau.oc}. 
\label{sttau.oc}}
\end{figure}

\noindent This provides a good Fourier series for the velocity curves. 
The O--C diagram (Fig. \ref{sttau.oc}) suggests a possible small period change.  
Some of the scatter in the resulting plot of the annual mean
velocities (Fig \ref{vels}) is probably due to variations in the period 
(Fig \ref{sttau.oc} ).




%






\subsection{X Lac}

\begin{figure}
 \includegraphics[width=\columnwidth]{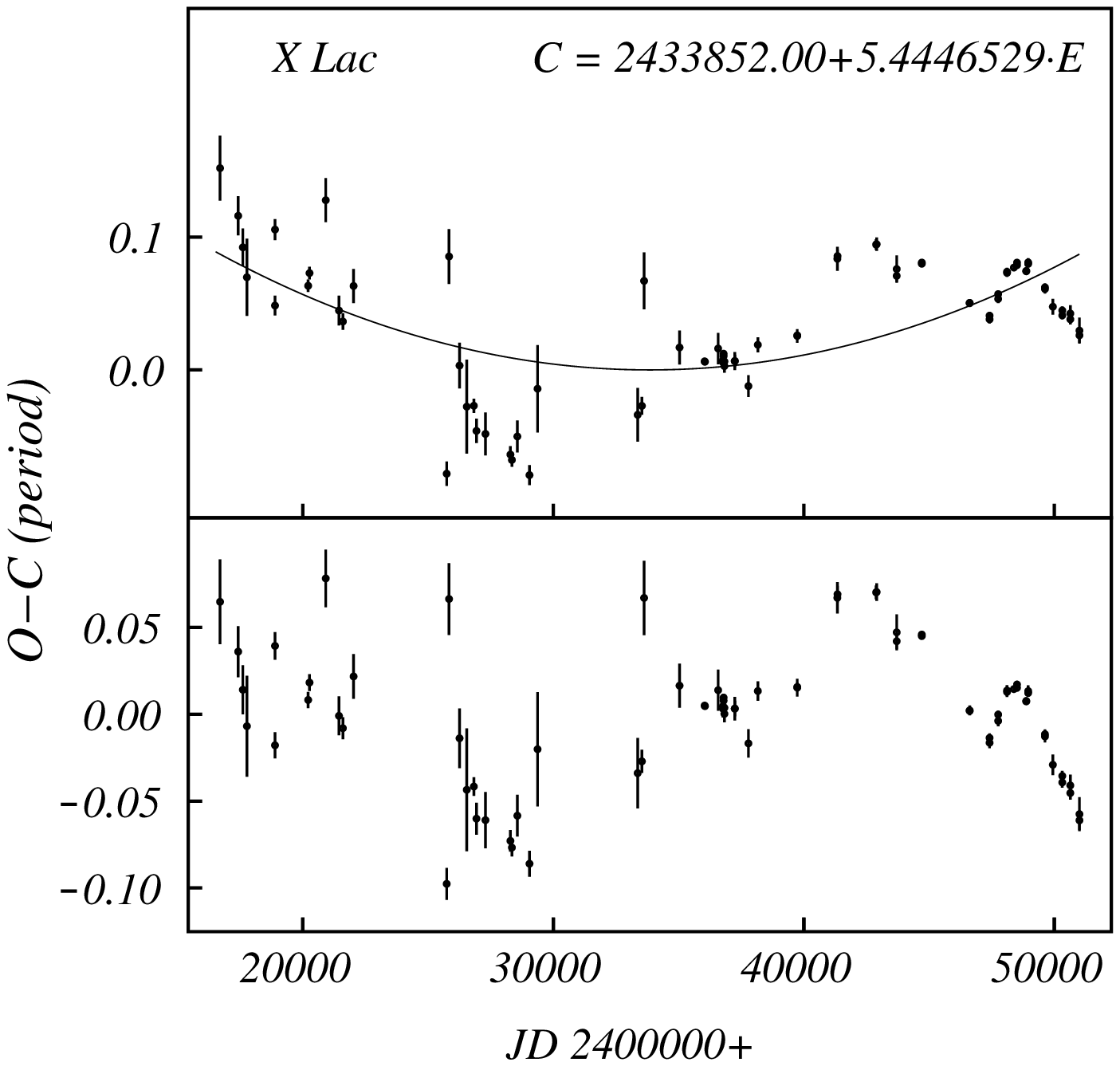}
\caption{The O-C diagram for X Lac in the same units as Fig~\ref{eutau.oc}. 
Top: data plus a parabolic fit; bottom: the residuals from the
  parabola.  \label{xlac.oc}}
\end{figure}

The O-C period change diagram for X Lac from photometry  is shown in  
Fig. \ref{xlac.oc}. It is
evident that the period fluctuations are larger and more erratic than
even the parabolic fit can describe.  Since the time range covered
by velocities is well mapped by photometry, we have approximated the
period changes with several linear sections. With this, we have phased
together the velocity curves, resulting in Fig. \ref{vels}.   

One interesting result of this investigation has been consideration of
X Lac.  It has a period of 5.4 $^d$ and a moderate amplitude (velocity
amplitude c(0) in the equation  Sect. 3.1 is 
25 km~s$^{-1}$).  These two characteristics are do not immediately suggest
overtone pulsation, as found in the small amplitude, short period `s'
Cepheids.  However, the period varies in an erratic way  (Fig.~\ref{xlac.oc}).
  Overtone pulsators have unusually large period changes,
larger than expected from evolution through the instability strip
(Szabados 1983; Berdnikov et al. 1997; Evans et al. 2015).  As longer
series of photometric data have
been accumulated, it appears in many cases that data that was
previously  fit with a parabola, representing a
changing period, is  actually a series increasing and decreasing
periods.  In the case of X Lac the substantial changes in period
(Fig. \ref{xlac.oc}) raise the interesting
question of whether it may in fact be pulsating in the first
overtone. The most recent discussion of Fourier diagnostics is given
by Storm et al. (2011).  In their plot of Fourier amplitude A$_1$ as
a function of period X Lac falls somewhat below the fundamental mode
sequence.  Similarly, the amplitude ratio R$_{21}$ could be the
continuation of the overtone sequence as a function of period.  The
$\phi_{21}$ --period diagram is more complicated, particularly in the
sample discussed by Kienzle et al. (1999).  The fundamental and
overtone sequences appear to cross near P = 5.44$^d$, so the value for
X Lac provides little information about the pulsation mode.  Storm, et
al. note that both X Lac and V496 Aql have A$_1$ and also R$_{21}$
smaller than the fundamental mode sequence.  While they find the
Fourier parameters inconclusive, the variable period for X Lac is
another overtone characteristic, and we tentatively
classify it as an overtone pulsator.

What makes this particularly interesting is that there are very few known or
suspected overtone pulsators with periods this long.  This is partly
because some of the standard Fourier diagnostics are ambiguous in this period
range. It is therefore valuable to identify possible long period
overtone pulsators in order to have a larger sample to define the
the overtone locus.



\subsection{S Vul}
\begin{figure}
 \includegraphics[totalheight=1.5in]{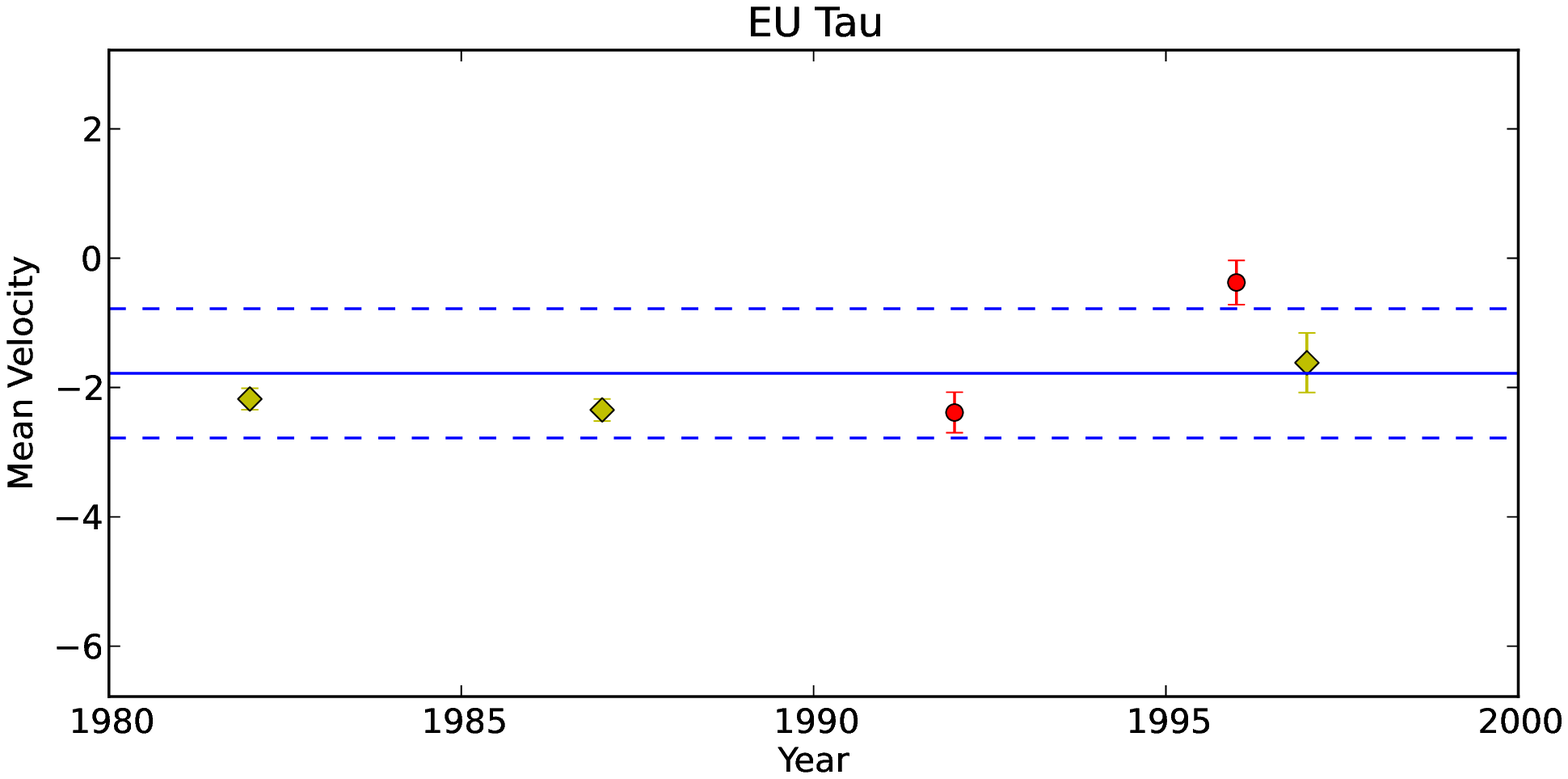}
\caption{The radial velocities corrected for pulsation (by year) 
for EU Tau for the fits to the curve without
  assuming a period (see text). Units and symbols are the same as  Fig~\ref{vels}.
 The results are essentially the same as Fig~\ref{vels}.
    \label{eutau.lb}}
\end{figure}

For stars with a variable period, we sometimes resort to the following
approach. Instead of taking the phasing of the velocity curves from
the period (including a parabolic term as needed) from the photometry,
we solve for both the velocity mean and the phase of maximum light 
from the velocity curve alone.
This means we can only use seasons/years with a well defined curve
containing many  data points.  To confirm this approach,
 the results for  EU Tau
 are shown in Fig \ref{eutau.lb}.  The results are essentially
the same as in Fig \ref{vels} (with a small mean offset) but with fewer 
seasons than the 
standard approach using the photometric period.  

\begin{figure}
\includegraphics[totalheight=1.5in]{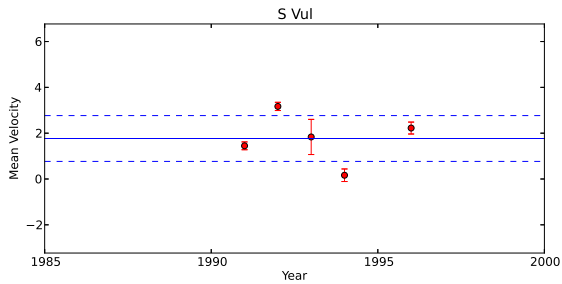} 
\caption{Radial velocities for S Vul corrected for pulsation by year. In this case, data for each 
year has been fitted individually to annual velocity curves.   
Units and symbols are the same as  Fig~\ref{vels}.
\label{svul}}
\end{figure}

S Vul has a period of  68$^d$ with significant period  
variations (Berdnikov 1994).  We have  used 
 individual fits to annual velocity curves.  
Occasionally the means cover more than 1 season, since for this
approach, a larger number of data points are needed to do an accurate
fitting.  
Table \ref{svul.lb} has the values from these fits which are in shown Fig. \ref{svul}

\begin{deluxetable}{llrrrr}
\footnotesize
\tablecaption{ Cepheid Annual Mean Velocities: S Vul  \qquad\qquad\qquad\qquad \label{svul.lb}}
\tablewidth{0pt}
\tablehead{
\colhead{JD } & \colhead{Phase} & \colhead{Mean} & \colhead{N} &
\colhead{Ref }
 & \colhead{JD Range } \\
\colhead{ } & \colhead{} & \colhead{km~s$^{-1}$} & \colhead{} &
\colhead{ }
 & \colhead{ } \\
}

\startdata

  48406.65   &     0.21    &     1.45    &            15  &  Gorynya  &    47309.46 -   48565.06 \\
               &   $\pm$       0.002   &    $\pm$   0.17    &                  &  \\
  48886.61    &       0.22     &    3.17   &             22  &  Gorynya   &   48795.45 -   48889.27 \\
             &       $\pm$      0.002   &     $\pm$  0.18    &                  &  \\
  49228.12    &       0.21    &     1.84    &        19  &  Gorynya   &   49161.53 -   49256.32 \\
              &       $\pm$     0.02      &   $\pm$   0.77     &               &   &       \\
  49569.13   &        0.19   &   0.16        &        10  &  Gorynya &  49499.52 -   49609.28 \\
                &       $\pm$        0.004         &  $\pm$   0.27    &      &      &           \\
  50117.43    &       0.20   &   2.22     &        21  &  Gorynya &  49926.48 -   50309.42 \\
               &    $\pm$    0.003   &   $\pm$    0.26       &    &            &    \\


\enddata

\end{deluxetable}

\subsection{SV Vul}  
\begin{figure}
\includegraphics[totalheight=1.5in]{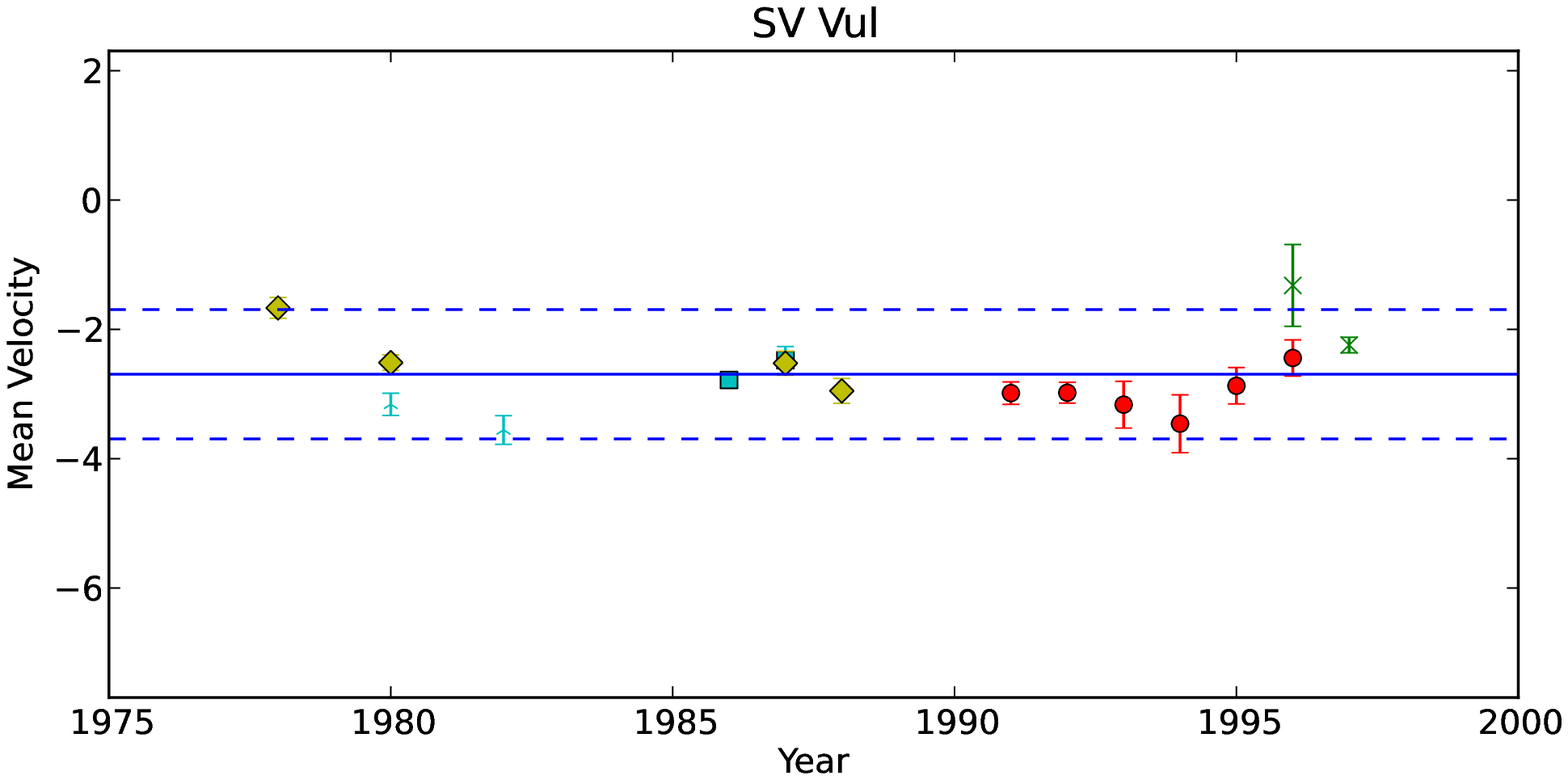} 
\caption{Radial velocities for SV Vul corrected for pulsation by year. In this case, data for each 
year from  individual fits.   
Units and symbols are the same as  Fig~\ref{vels}.
\label{svvul}}
\end{figure}

It also has a long period (45$^d$), and, as with S
Vul, we have derived mean velocities by fitting data from individual
seasons.  Details are given in Table \ref{svvul.lb} and Fig   \ref{svvul}

\noindent


\vskip .1truein

\begin{deluxetable}{llrrrr}
\footnotesize
\tablecaption{ Cepheid Annual Mean Velocities: SV Vul  \qquad\qquad\qquad\qquad \label{svvul.lb}}
\tablewidth{0pt}
\tablehead{
\colhead{JD } & \colhead{Phase} & \colhead{Mean} & \colhead{N} &
\colhead{Ref }
 & \colhead{JD Range } \\
\colhead{ } & \colhead{} & \colhead{km~s$^{-1}$  } & \colhead{} &
\colhead{ }
 & \colhead{ } \\
}

\startdata

  43719.09   &     0.05    &    -1.67        &       25 &   Bersier  &  43681.59 -   43735.44 \\
                &  $\pm$     0.001   &    $\pm$   0.16    &          &       &  \\
  44393.96   &     0.04    &    -3.16         &      17  &   Imbert  &  44168.30 -   44569.28 \\
                &  $\pm$     0.001   &   $\pm$    0.17    &           &      &  \\
  44438.71   &     0.03   &     -2.51       &        30  &  Bersier  &  44184.29 -   44483.40 \\
                &   $\pm$    0.001   &    $\pm$   0.12    &           &      &  \\
  44979.86   &     0.05   &        -3.56     &          20  &   Imbert  &  44902.29 -   45139.58 \\
                &   $\pm$    0.002    &    $\pm$  0.22    &            &     &  \\
  46692.83   &     0.11    &    -2.78       &        47  &    Storm  &  46567.91 -   46787.53 \\
                &  $\pm$     0.001   &   $\pm$    0.08    &           &      &  \\
  47097.69   &     0.10   &     -2.48        &       28  &    Storm  &  46842.96 -   47407.58 \\
                &   $\pm$    0.001   &   $\pm$    0.21    &           &      &  \\
  47142.75   &     0.11   &     -2.52        &       17 &   Bersier  &  47107.26 -   47149.29 \\
                &   $\pm$    0.002   &    $\pm$   0.19    &          &       &  \\
  47457.08  &     0.09   &     -2.95        &       12  &  Bersier  &  47424.32 -   47495.26 \\
                &    $\pm$   0.001  &    $\pm$    0.19    &          &       &  \\
  48537.21   &     0.09   &     -2.99       &        21 &   Gorynya  &  48426.52 -   48600.10 \\
                &   $\pm$    0.002   &   $\pm$    0.17    &            &     &  \\
  48851.04   &     0.06    &    -2.98        &       27 &     Gorynya  &  48745.52 -   48921.19 \\
                &   $\pm$    0.002   &    $\pm$   0.16    &               &  \\
  49210.23   &     0.04    &    -3.16        &       22  &  Gorynya  &  49161.54 -   49256.32 \\
                &   $\pm$    0.004   &     $\pm$  0.36     &          &      &  \\
  49569.31   &     0.02    &    -3.46        &       14  &  Gorynya &   49499.54 -   49609.27 \\
                &   $\pm$    0.002    &    $\pm$  0.45    &            &     &  \\
  49973.38   &    -0.01    &    -2.87         &       12  &    Gorynya  &  49932.40 -   50000.24 \\
                &    $\pm$   0.002   &    $\pm$   0.28     &              &  \\
  50288.13   &    -0.01   &     -2.44        &       15 &   Gorynya  &  50236.40 -   50314.35 \\
                &   $\pm$    0.001  &     $\pm$   0.28     &           &    &  \\
  50288.667   &    -0.004   &     -1.32        &       14  &   Barnes &   49947.85 -   50353.62 \\
                &   $\pm$    0.01   &     $\pm$     0.63  &           &      &    \\
  50647.17   &    -0.04   &     -2.24          &     20  &   Barnes  &  50609.87 -   50709.71 \\
                &    $\pm$   0.001  &     $\pm$   0.12     &          &      &  \\

\enddata

\end{deluxetable}



\acknowledgments

We are grateful for comments from A. Tokivinin and B. Mason which resulted in an 
improved manuscript.
It is a pleasure to thank Beth Sundheim and Bharath Kumaraswamy
for work at the beginning of this 
project. 
We also thank Matthew Templeton for providing the program FTCLEAN used to
generate the window functions, and for advice on its use. 
Comments from an anonymous referee improved the presentation of the paper.
 Support for this work was also provided  from the Chandra X-ray Center NASA 
Contract NAS8-03060 and HST grant GO-12215.01-A.
N.A.Gorynya is grateful to Russian Foundation for Basic Reserach 
grant No. 14-02-00472 for the support of radial velocity measurements.
A.S.Rastorguev is grateful to Russian Scientific Foundation 
grant No. 14-22-0041 for the support of data processing. 
PM acknowledges support from the Polish NCN grant no.
    DEC-2012/05/B/ST9/03932.
Vizier and  SIMBAD were used in the preparation of this study.

\clearpage

\clearpage






\section{Electronic Tables}









In this section we provide the content of long tables which can be
retrieved electronically. 



\begin{deluxetable}{llclr}
\footnotesize
\tablecaption{Annual Corrections   \qquad\qquad\qquad\qquad \label{ann.corr-e}}
\tablewidth{0pt}
\tablehead{
\colhead{Source} & \colhead{Year} & \colhead{Mean} & \colhead{$\sigma$}
& \colhead{N } \\
\colhead{} & \colhead{} & \colhead{km~s$^{-1}$} & \colhead{km~s$^{-1}$}
& \colhead{} \\
}

\startdata


 & & & & \\
   gg &    1986 &   0.16 &   0.19 &    1 \\ 
   gg &    1987 &   0.71* &   0.17 &    6 \\ 
   gg &    1989 &   0.25 &   0.07 &    3 \\ 
   gg &    1990 &   0.41* &   0.10 &    6 \\ 
   gg &    1991 &   0.35* &   0.04 &   20 \\ 
   gg &    1992 &   0.25 &   0.07 &   10 \\ 
   gg &    1993 &  -0.39* &   0.09 &   16 \\ 
   gg &    1994 &  -0.34* &   0.05 &   22 \\ 
   gg &    1995 &  -0.42* &   0.04 &   23 \\ 
   gg &    1996 &   0.08 &   0.04 &   22 \\ 
   gg &    1997 &   0.17 &   0.06 &   17 \\ 
   gg &    1998 &   0.18 &   0.10 &   12 \\ 
 & & & & \\
   b0 &    1978 &  -0.16 &   0.06 &    4 \\ 
   b0 &    1979 &  -0.38* &   0.07 &    4 \\ 
   b0 &    1980 &   0.30 &   0.08 &    4 \\ 
   b0 &    1981 &   0.02 &   0.12 &    4 \\ 
   b0 &    1982 &  -0.69 &   0.39 &    1 \\ 
   b0 &    1987 &  -0.22 &   0.17 &    3 \\ 
   b0 &    1988 &  -0.85 &   0.37 &    3 \\ 
 & & & & \\
   b9 &    1977 &   1.04 &   0.51 &    2 \\ 
   b9 &    1978 &  -0.07 &   0.08 &    2 \\ 
   b9 &    1979 &   0.02 &   0.12 &    3 \\ 
   b9 &    1980 &  -0.15 &   0.03 &    3 \\ 
   b9 &    1981 &  -0.07 &   0.61 &    1 \\ 
   b9 &    1982 &  -0.06 &   0.16 &    2 \\ 
   b9 &    1983 &   0.74 &   0.48 &    1 \\ 
   b9 &    1984 &  -0.33 &   0.65 &    1 \\ 
   b9 &    1985 &  -0.43 &   0.18 &    1 \\ 
   b9 &    1987 &   0.09 &   0.13 &    3 \\ 
   b9 &    1988 &  -0.51 &   0.26 &    2 \\ 
   b9 &    1989 &   0.37 &   0.36 &    1 \\ 
   b9 &    1991 &   0.06 &   0.18 &    1 \\ 
 & & & & \\
   im &    1978 &   0.12 &   0.24 &    1 \\ 
   im &    1979 &   0.06 &   0.26 &    4 \\ 
   im &    1980 &   0.28 &   0.15 &    5 \\ 
   im &    1981 &   0.39* &   0.13 &    6 \\ 
   im &    1982 &   0.86* &   0.13 &    4 \\ 
   im &    1983 &  -0.70 &   0.06 &    2 \\ 
   im &    1984 &   0.18 &   0.01 &    2 \\ 
   im &    1985 &  -0.13 &   0.40 &    2 \\ 
   im &    1986 &  -0.10 &   0.19 &    5 \\ 
   im &    1987 &   0.37 &   0.10 &    3 \\ 
   im &    1988 &   0.77 &   0.56 &    2 \\ 
   im &    1989 &   0.06 &   0.06 &    6 \\ 
   im &    1991 &  -0.00 &   0.10 &    3 \\ 
   im &    1992 &  -1.04 &   0.19 &    2 \\ 
   im &    1993 &  -0.54 &   0.29 &    2 \\ 
   im &    1994 &   0.25 &   0.13 &    4 \\ 
   im &    1995 &  -0.06 &   0.18 &    1 \\ 
   im &    1996 &  -0.81 &   0.01 &    2 \\ 
   im &    1997 &  -0.09 &   0.28 &    1 \\ 
   im &    1998 &   0.09 &   0.18 &    1 \\ 
 & & & & \\
   s4 &    1985 &  -0.83 &   0.57 &    1 \\ 
   s4 &    1986 &   0.30 &   0.12 &    5 \\ 
   s4 &    1987 &   0.10 &   0.17 &    3 \\ 
   s4 &    1988 &  -0.30 &   0.62 &    3 \\ 
 & & & & \\
   kk &    1996 &   0.95* &   0.06 &   10 \\ 
   kk &    1997 &   0.70* &   0.08 &   10 \\ 
 & & & & \\
   ba &    1996 &   0.83* &   0.11 &   10 \\ 
   ba &    1997 &   0.22 &   0.10 &   10 \\

\enddata

\vskip .1truein

*:  corrections incorporated in  Table~\ref{cep.ann-e}

\noindent

\end{deluxetable}

\begin{deluxetable}{lllrrr}
\footnotesize
\tablecaption{Cepheid Annual Mean Velocities  \qquad\qquad\qquad\qquad \label{cep.ann-e}}
\tablewidth{0pt}
\tablehead{
\colhead{Star} & \colhead{Source} & \colhead{Year} & \colhead{Mean} &
\colhead{$\sigma$}
 & \colhead{N } \\
\colhead{} & \colhead{} & \colhead{} & \colhead{km~s$^{-1}$} &
\colhead{km~s$^{-1}$ }
 & \colhead{ } \\
}

\startdata



 & & & & & \\

  Eta  Aql &    b0 &    1983 &   0.74 &   0.48 &    5 \\ 
  Eta  Aql &    gg &    1986 &   0.16 &   0.19 &   25 \\ 
  Eta  Aql &    s4 &    1986 &   0.12 &   0.28 &   25 \\ 
  Eta  Aql &    b0 &    1989 &   0.37 &   0.36 &   33 \\ 
  Eta  Aql &    ba &    1995 &   2.71 &   ---  &    1 \\ 
  Eta  Aql &    kk &    1996 &   0.58 &   0.33 &    8 \\ 
  Eta  Aql &    ba &    1996 &   1.14 &   0.51 &   13 \\ 
  Eta  Aql &    ba &    1997 &  -0.05 &   0.51 &   16 \\ 
  Eta  Aql &    kk &    1997 &   0.71 &   0.30 &    6 \\ 
 & & & & & \\
   SZ  Aql &    b0 &    1996 &  -0.38 &   0.24 &   21 \\ 
   SZ  Aql &    ba &    1996 &  -0.16 &   0.23 &   13 \\ 
   SZ  Aql &    ba &    1997 &  -0.15 &   0.27 &   19 \\ 
   SZ  Aql &    b0 &    1997 &  -0.29 &   0.57 &   10 \\ 
 & & & & & \\
   TT  Aql &    im &    1989 &  -0.07 &   0.18 &   14 \\ 
   TT  Aql &    im &    1991 &   0.28 &   0.12 &    5 \\ 
   TT  Aql &    im &    1993 &  -0.14 &   0.45 &    5 \\ 
   TT  Aql &    im &    1994 &   0.33 &   0.16 &    3 \\ 
   TT  Aql &    ba &    1995 &   0.67 &  --- &    1 \\ 
   TT  Aql &    ba &    1996 &   1.01 &   0.27 &   12 \\ 
   TT  Aql &    b0 &    1996 &   0.03 &   0.13 &   32 \\ 
   TT  Aql &    b0 &    1997 &   0.20 &   0.34 &   10 \\ 
   TT  Aql &    ba &    1997 &   0.25 &   0.24 &   19 \\ 
   TT  Aql &    gg &    1997 &   0.93 &   0.34 &   15 \\ 
   TT  Aql &    gg &    1998 &   0.93 &   0.34 &   15 \\ 
 & & & & & \\
   FM  Aql &    gg &    1989 &   0.16 &   0.30 &    9 \\ 
   FM  Aql &    gg &    1990 &   0.60 &   0.16 &   33 \\ 
   FM  Aql &    gg &    1991 &   0.48 &   0.21 &   23 \\ 
   FM  Aql &    gg &    1992 &   0.11 &   0.65 &    4 \\ 
   FM  Aql &    gg &    1993 &  -0.28 &   0.22 &   13 \\ 
   FM  Aql &    gg &    1994 &   0.30 &   0.70 &   10 \\ 
   FM  Aql &    ba &    1995 &  -1.35 &  --- &    1 \\ 
   FM  Aql &    gg &    1995 &  -1.44 &   0.19 &   16 \\ 
   FM  Aql &    ba &    1996 &   0.71 &   0.64 &   13 \\ 
   FM  Aql &    ba &    1997 &  -0.58 &   0.47 &   16 \\ 
 & & & & & \\
   FN  Aql &    gg &    1989 &   0.06 &   0.17 &   15 \\ 
   FN  Aql &    gg &    1990 &   0.07 &   0.19 &   15 \\ 
   FN  Aql &    gg &    1991 &   0.39 &   0.13 &   19 \\ 
   FN  Aql &    gg &    1992 &   0.30 &   0.16 &    5 \\ 
   FN  Aql &    gg &    1993 &  -0.10 &   0.29 &   14 \\ 
   FN  Aql &    gg &    1994 &  -0.15 &   0.25 &   11 \\ 
   FN  Aql &    ba &    1995 &   1.52 &  --- &    1 \\ 
   FN  Aql &    gg &    1995 &  -0.61 &   0.21 &   13 \\ 
   FN  Aql &    ba &    1996 &   0.75 &   0.29 &   13 \\ 
   FN  Aql &    ba &    1997 &   0.45 &   0.22 &   17 \\ 
 & & & & & \\
V1162  Aql &    gg &    1991 &   0.56 &   0.18 &    7 \\ 
V1162  Aql &    gg &    1992 &   0.51 &   0.18 &   21 \\ 
V1162  Aql &    gg &    1993 &   0.34 &   0.55 &   11 \\ 
V1162  Aql &    gg &    1994 &  -0.94 &   0.13 &    9 \\ 
V1162  Aql &    gg &    1995 &  -0.50 &   0.44 &   14 \\ 
 & & & & & \\
   RT  Aur &    gg &    1987 &   0.91 &   0.68 &    2 \\ 
   RT  Aur &    gg &    1991 &  -0.88 &   0.34 &    7 \\ 
   RT  Aur &    gg &    1994 &  -0.51 &   0.53 &    4 \\ 
   RT  Aur &    gg &    1995 &  -0.52 &   0.36 &    3 \\ 
   RT  Aur &    kk &    1996 &   0.54 &   0.15 &    7 \\ 
   RT  Aur &    gg &    1996 &   0.09 &   0.11 &   16 \\ 
   RT  Aur &    kk &    1997 &   0.64 &   0.26 &    6 \\ 
   RT  Aur &    gg &    1997 &  -0.99 &   1.00 &    3 \\ 
   RT  Aur &    gg &    1998 &  -0.19 &   0.16 &   11 \\ 
 & & & & & \\
   RX  Aur &    gg &    1987 &   0.05 &   0.77 &    3 \\ 
   RX  Aur &    gg &    1988 &   0.04 &  --- &    1 \\ 
   RX  Aur &    gg &    1989 &   1.52 &  --- &    1 \\ 
   RX  Aur &    gg &    1990 &   0.00 &   0.69 &   18 \\ 
   RX  Aur &    gg &    1991 &   0.09 &   0.26 &   18 \\ 
   RX  Aur &    im &    1994 &   0.38 &   0.29 &    4 \\ 
   RX  Aur &    gg &    1994 &   0.42 &   1.90 &    3 \\ 
   RX  Aur &    im &    1995 &  -0.06 &   0.18 &    8 \\ 
   RX  Aur &    gg &    1995 &   0.33 &  --- &    1 \\ 
   RX  Aur &    gg &    1996 &   1.40 &   0.74 &    9 \\ 
   RX  Aur &    im &    1996 &  -0.47 &   0.11 &    3 \\ 
   RX  Aur &    im &    1997 &  -0.09 &   0.28 &    3 \\ 
   RX  Aur &    gg &    1997 &   0.02 &   0.87 &    3 \\ 
   RX  Aur &    im &    1998 &   0.09 &   0.18 &    7 \\ 
   RX  Aur &    gg &    1998 &   1.59 &   0.61 &    6 \\ 
 & & & & & \\
   CK  Cam &    gg &    1996 &   0.27 &   0.33 &   10 \\ 
   CK  Cam &    kk &    1996 &   0.79 &   0.55 &    8 \\ 
   CK  Cam &    kk &    1997 &   0.31 &   0.32 &    5 \\ 
   CK  Cam &    gg &    1997 &  -0.54 &   0.19 &    6 \\ 
   CK  Cam &    gg &    1998 &   0.06 &   0.37 &    9 \\ 
 & & & & & \\
   SU  Cas &    s4 &    1985 &  -0.83 &   0.57 &    3 \\ 
   SU  Cas &    s4 &    1986 &  -1.31 &   0.53 &   16 \\ 
   SU  Cas &    gg &    1987 &   0.88 &   0.51 &    2 \\ 
   SU  Cas &    b0 &    1987 &  -0.10 &   0.21 &   19 \\ 
   SU  Cas &    s4 &    1987 &  -0.85 &   0.38 &   28 \\ 
   SU  Cas &    s4 &    1988 &  -0.94 &   1.49 &    6 \\ 
   SU  Cas &    gg &    1991 &   0.05 &   0.18 &   15 \\ 
   SU  Cas &    gg &    1992 &   0.18 &   0.30 &    3 \\ 
   SU  Cas &    gg &    1993 &  -0.73 &   0.50 &    2 \\ 
   SU  Cas &    gg &    1994 &  -0.28 &   0.26 &   12 \\ 
   SU  Cas &    gg &    1995 &  -0.02 &   0.28 &    9 \\ 
   SU  Cas &    kk &    1996 &   0.75 &   0.13 &   10 \\ 
   SU  Cas &    gg &    1996 &   0.10 &   0.24 &   16 \\ 
   SU  Cas &    gg &    1997 &   0.12 &   0.47 &    8 \\ 
   SU  Cas &    kk &    1997 &   0.19 &   0.18 &    6 \\ 
   SU  Cas &    gg &    1998 &  -0.16 &   0.34 &    4 \\ 
 & & & & & \\
 V379  Cas &    gg &    1993 &  -0.99 &   0.51 &    6 \\ 
 V379  Cas &    gg &    1994 &  -0.15 &   0.40 &    8 \\ 
 V379  Cas &    gg &    1995 &  -0.75 &   0.32 &   10 \\ 
 V379  Cas &    gg &    1996 &   0.91 &   0.41 &    7 \\ 
 V379  Cas &    gg &    1997 &   0.39 &   0.58 &    8 \\ 
 V379  Cas &    gg &    1998 &  -1.59 &   3.37 &    2 \\ 
 & & & & & \\
 V636  Cas &    b9 &    1978 &  -0.10 &   0.14 &    7 \\ 
 V636  Cas &    b9 &    1979 &  -0.45 &   0.10 &   29 \\ 
 V636  Cas &    b9 &    1980 &  -0.20 &   0.14 &   19 \\ 
 V636  Cas &    b9 &    1981 &  -0.08 &   0.26 &    6 \\ 
 V636  Cas &    gg &    1991 &  -0.06 &   0.11 &   22 \\ 
 V636  Cas &    gg &    1993 &  -0.65 &   0.42 &    8 \\ 
 V636  Cas &    gg &    1994 &   0.05 &   0.25 &    7 \\ 
 V636  Cas &    gg &    1995 &  -0.33 &   0.48 &   10 \\ 
 & & & & & \\
  Del  Cep &    b9 &    1980 &   0.00 &   0.04 &   91 \\ 
  Del  Cep &    s4 &    1986 &   2.39 &   0.37 &   20 \\ 
  Del  Cep &    ba &    1996 &   2.62 &   0.33 &   10 \\ 
  Del  Cep &    kk &    1996 &   2.75 &   0.15 &    9 \\ 
  Del  Cep &    kk &    1997 &   0.85 &   1.06 &    9 \\ 
  Del  Cep &    ba &    1997 &   0.34 &   0.93 &   19 \\ 
 & & & & & \\
   IR  Cep &    gg &    1990 &   0.46 &   0.48 &   16 \\ 
   IR  Cep &    gg &    1991 &   0.29 &   0.34 &   27 \\ 
   IR  Cep &    gg &    1993 &   0.52 &   0.56 &   10 \\ 
   IR  Cep &    gg &    1994 &  -0.20 &   0.60 &   12 \\ 
   IR  Cep &    gg &    1995 &  -1.00 &   0.54 &   18 \\ 
   IR  Cep &    gg &    1996 &  -0.08 &   0.36 &   15 \\ 
   IR  Cep &    gg &    1997 &   0.45 &   0.68 &    8 \\ 
 & & & & & \\
    X  Cyg &    b9 &    1978 &   1.13 &   0.20 &   51 \\ 
    X  Cyg &    b9 &    1979 &   0.58 &   1.05 &    3 \\ 
    X  Cyg &    b9 &    1980 &  -0.82 &   0.23 &   56 \\ 
    X  Cyg &    s4 &    1986 &   0.32 &   0.24 &   52 \\ 
    X  Cyg &    s4 &    1987 &   0.10 &   0.33 &   18 \\ 
    X  Cyg &    b9 &    1987 &   0.03 &   0.19 &   16 \\ 
    X  Cyg &    s4 &    1988 &  -0.17 &   0.71 &    5 \\ 
    X  Cyg &    b9 &    1988 &  -0.30 &   0.50 &    6 \\ 
    X  Cyg &    gg &    1991 &   0.32 &   0.21 &   21 \\ 
    X  Cyg &    gg &    1992 &   0.15 &   0.18 &   15 \\ 
    X  Cyg &    gg &    1993 &   0.31 &   0.32 &   16 \\ 
    X  Cyg &    gg &    1994 &  -0.06 &   0.17 &   14 \\ 
    X  Cyg &    gg &    1995 &  -0.26 &   0.18 &   20 \\ 
    X  Cyg &    kk &    1996 &   1.27 &   0.30 &    9 \\ 
    X  Cyg &    ba &    1996 &   1.08 &   0.30 &   12 \\ 
    X  Cyg &    kk &    1997 &   0.69 &   0.22 &    7 \\ 
    X  Cyg &    ba &    1997 &   0.41 &   0.24 &   20 \\ 
 & & & & & \\
   CD  Cyg &    im &    1980 &   0.57 &   0.22 &    8 \\ 
   CD  Cyg &    im &    1981 &  -0.14 &   0.28 &    8 \\ 
   CD  Cyg &    im &    1982 &  -0.02 &   0.39 &    4 \\ 
   CD  Cyg &    im &    1983 &  -0.79 &   0.07 &    2 \\ 
   CD  Cyg &    im &    1984 &   0.18 &   0.01 &    2 \\ 
   CD  Cyg &    im &    1986 &  -0.14 &   0.31 &    6 \\ 
   CD  Cyg &    gg &    1991 &   0.82 &   0.10 &    2 \\ 
   CD  Cyg &    gg &    1992 &  -0.79 &   0.30 &   19 \\ 
   CD  Cyg &    gg &    1993 &  -1.59 &   0.29 &   11 \\ 
   CD  Cyg &    gg &    1994 &  -1.71 &   0.35 &    9 \\ 
   CD  Cyg &    gg &    1995 &  -1.76 &   0.35 &   12 \\ 
 & & & & & \\
   DT  Cyg &    b0 &    1977 &   1.10 &   0.61 &    6 \\ 
   DT  Cyg &    b0 &    1978 &  -0.37 &   0.18 &   28 \\ 
   DT  Cyg &    b0 &    1979 &  -0.87 &   0.19 &    3 \\ 
   DT  Cyg &    b0 &    1980 &  -0.99 &   0.08 &   19 \\ 
   DT  Cyg &    gg &    1996 &  -0.34 &   0.09 &    3 \\ 
   DT  Cyg &    kk &    1996 &   0.23 &   0.11 &   10 \\ 
   DT  Cyg &    kk &    1997 &   0.67 &   0.15 &    6 \\ 
   DT  Cyg &    gg &    1997 &   0.05 &   0.11 &   21 \\ 
 & & & & & \\
V1726  Cyg &    gg &    1993 &  -0.47 &   0.32 &   16 \\ 
V1726  Cyg &    gg &    1994 &  -0.22 &   0.16 &   16 \\ 
V1726  Cyg &    gg &    1995 &  -0.29 &   0.15 &   15 \\ 
V1726  Cyg &    gg &    1996 &   0.54 &   0.26 &   19 \\ 
V1726  Cyg &    gg &    1997 &   0.70 &   0.16 &   17 \\ 
 & & & & & \\
 Zeta  Gem &    b9 &    1978 &   0.33 &   0.15 &   29 \\ 
 Zeta  Gem &    b9 &    1979 &   0.14 &   0.28 &    7 \\ 
 Zeta  Gem &    b9 &    1980 &   0.02 &   0.18 &   15 \\ 
 Zeta  Gem &    b9 &    1981 &   0.03 &   0.16 &   10 \\ 
 Zeta  Gem &    b9 &    1982 &  -0.69 &   0.39 &    4 \\ 
 Zeta  Gem &    gg &    1987 &   0.69 &   0.20 &    7 \\ 
 Zeta  Gem &    gg &    1991 &   0.39 &   0.12 &   16 \\ 
 Zeta  Gem &    gg &    1994 &  -0.81 &   0.47 &    5 \\ 
 Zeta  Gem &    gg &    1995 &  -0.18 &   0.07 &    2 \\ 
 Zeta  Gem &    kk &    1996 &   0.87 &   0.17 &    6 \\ 
 Zeta  Gem &    gg &    1996 &  -0.26 &   0.15 &   17 \\ 
 Zeta  Gem &    gg &    1997 &  -0.24 &   0.79 &    4 \\ 
 Zeta  Gem &    kk &    1997 &   0.82 &   0.27 &    5 \\ 
 Zeta  Gem &    gg &    1998 &  -0.12 &   0.38 &   11 \\ 
 & & & & & \\
    W  Gem &    im &    1979 &  -0.45 &   0.37 &   11 \\ 
    W  Gem &    im &    1980 &   0.24 &   0.44 &   11 \\ 
    W  Gem &    im &    1981 &   2.41 &   1.05 &    5 \\ 
    W  Gem &    im &    1982 &   0.20 &   0.51 &    6 \\ 
    W  Gem &    im &    1985 &  -0.98 &   0.78 &    4 \\ 
    W  Gem &    im &    1986 &  -0.66 &   0.84 &    2 \\ 
    W  Gem &    im &    1987 &  -0.46 &   0.87 &    2 \\ 
    W  Gem &    gg &    1987 &   1.27 &   0.93 &    2 \\ 
    W  Gem &    im &    1988 &   2.07 &   0.84 &    2 \\ 
    W  Gem &    im &    1989 &  -0.81 &   0.08 &    2 \\ 
    W  Gem &    gg &    1991 &   0.02 &   0.33 &    8 \\ 
    W  Gem &    gg &    1994 &  -1.93 &   1.24 &    2 \\ 
    W  Gem &    gg &    1995 &  -1.55 &   1.06 &    2 \\ 
    W  Gem &    gg &    1996 &  -0.19 &   0.35 &   13 \\ 
    W  Gem &    gg &    1997 &   0.59 &   1.28 &    2 \\ 
    W  Gem &    gg &    1998 &   0.14 &   0.33 &   10 \\ 
 & & & & & \\
    V  Lac &    gg &    1989 &   5.33 &   5.05 &    3 \\ 
    V  Lac &    gg &    1990 &   0.53 &   0.27 &    9 \\ 
    V  Lac &    gg &    1991 &   1.17 &   0.46 &    9 \\ 
    V  Lac &    gg &    1993 &  -0.78 &   0.59 &    6 \\ 
    V  Lac &    gg &    1994 &  -2.40 &   1.31 &    2 \\ 
    V  Lac &    gg &    1995 &  -0.01 &   0.44 &   10 \\ 
    V  Lac &    gg &    1996 &   1.10 &   0.31 &    6 \\ 
 & & & & & \\
    X  Lac &    b9 &    1987 &  -1.25 &   1.61 &   14 \\ 
    X  Lac &    b9 &    1988 &  -0.03 &   1.97 &    8 \\ 
    X  Lac &    gg &    1991 &   1.10 &   0.34 &   10 \\ 
    X  Lac &    gg &    1992 &   0.50 &   0.27 &   22 \\ 
    X  Lac &    gg &    1993 &  -0.90 &   0.35 &    8 \\ 
    X  Lac &    gg &    1994 &  -1.13 &   0.47 &    5 \\ 
    X  Lac &    ba &    1995 &   0.59 &  --- &    1 \\ 
    X  Lac &    gg &    1995 &  -0.87 &   0.20 &   11 \\ 
    X  Lac &    ba &    1996 &   0.63 &   0.40 &    9 \\ 
    X  Lac &    gg &    1996 &   0.35 &   0.15 &    6 \\ 
    X  Lac &    ba &    1997 &   0.33 &   0.22 &   17 \\ 
 & & & & & \\
   RR  Lac &    im &    1981 &   0.52 &   0.16 &    3 \\ 
   RR  Lac &    im &    1986 &  -0.34 &   0.31 &    8 \\ 
   RR  Lac &    b0 &    1987 &  -0.64 &   0.29 &   18 \\ 
   RR  Lac &    im &    1987 &  -0.20 &   0.38 &    2 \\ 
   RR  Lac &    b0 &    1988 &  -1.21 &   0.44 &    9 \\ 
   RR  Lac &    im &    1988 &  -0.31 &   0.77 &    5 \\ 
   RR  Lac &    im &    1989 &  -1.44 &   0.66 &    5 \\ 
   RR  Lac &    gg &    1991 &   0.71 &   0.20 &   12 \\ 
   RR  Lac &    im &    1992 &  -2.39 &   0.69 &    2 \\ 
   RR  Lac &    gg &    1992 &   0.63 &   0.26 &   25 \\ 
   RR  Lac &    gg &    1993 &  -0.72 &   0.42 &   10 \\ 
   RR  Lac &    im &    1994 &  -0.23 &   0.98 &    2 \\ 
   RR  Lac &    gg &    1995 &  -1.84 &   0.45 &   10 \\ 
   RR  Lac &    gg &    1996 &   0.88 &   0.47 &    5 \\ 
 & & & & & \\
   BG  Lac &    im &    1989 &  -0.59 &   0.17 &   12 \\ 
   BG  Lac &    im &    1991 &  -0.62 &   0.17 &    4 \\ 
   BG  Lac &    im &    1992 &  -0.92 &   0.20 &    3 \\ 
   BG  Lac &    im &    1993 &  -0.81 &   0.37 &    5 \\ 
   BG  Lac &    im &    1994 &  -0.48 &   0.40 &    5 \\ 
   BG  Lac &    ba &    1996 &   0.19 &   0.43 &   12 \\ 
   BG  Lac &    im &    1996 &  -0.81 &   0.01 &    2 \\ 
   BG  Lac &    ba &    1997 &  -0.73 &   0.41 &   17 \\ 
 & & & & & \\
   SV  Mon &    im &    1979 &   0.46 &   0.91 &    3 \\ 
   SV  Mon &    im &    1980 &  -0.66 &   0.95 &   13 \\ 
   SV  Mon &    im &    1981 &  -0.36 &   2.74 &    2 \\ 
   SV  Mon &    im &    1982 &   1.15 &   0.16 &    3 \\ 
   SV  Mon &    im &    1984 &   0.91 &   0.12 &    3 \\ 
   SV  Mon &    im &    1986 &  -0.07 &   0.94 &    4 \\ 
   SV  Mon &    im &    1989 &  -3.27 &   1.08 &    8 \\ 
   SV  Mon &    gg &    1996 &   0.93 &   0.77 &    6 \\ 
   SV  Mon &    gg &    1998 &   1.17 &   0.51 &    9 \\ 
 & & & & & \\
    Y  Oph &    gg &    1996 &  -0.07 &   0.12 &   32 \\ 
    Y  Oph &    gg &    1997 &   0.11 &   0.16 &   13 \\ 
 & & & & & \\
   RS  Ori &    im &    1979 &   0.21 &   0.74 &    3 \\ 
   RS  Ori &    im &    1980 &  -0.04 &   0.24 &    9 \\ 
   RS  Ori &    im &    1981 &  -0.55 &   0.60 &    3 \\ 
   RS  Ori &    im &    1982 &   0.12 &   0.46 &    3 \\ 
   RS  Ori &    im &    1983 &  -0.11 &   0.17 &    5 \\ 
   RS  Ori &    im &    1985 &   0.17 &   0.46 &    3 \\ 
   RS  Ori &    im &    1986 &   0.78 &   0.48 &    2 \\ 
   RS  Ori &    im &    1987 &   0.43 &   0.11 &    3 \\ 
   RS  Ori &    im &    1989 &  -0.43 &   0.23 &    4 \\ 
   RS  Ori &    gg &    1991 &  -1.27 &   0.51 &    8 \\ 
   RS  Ori &    im &    1991 &   0.04 &   1.32 &    2 \\ 
   RS  Ori &    gg &    1995 &  -1.25 &   0.67 &    2 \\ 
   RS  Ori &    gg &    1996 &  -0.03 &   0.58 &   14 \\ 
   RS  Ori &    gg &    1998 &   1.40 &   0.59 &    7 \\ 
 & & & & & \\
 V440  Per &    b9 &    1978 &  -0.51 &   0.08 &   14 \\ 
 V440  Per &    b9 &    1979 &  -0.38 &   0.11 &   27 \\ 
 V440  Per &    b9 &    1980 &  -0.41 &   0.15 &   16 \\ 
 V440  Per &    b9 &    1981 &  -0.01 &   0.25 &    6 \\ 
 V440  Per &    gg &    1991 &   0.01 &   0.18 &   15 \\ 
 V440  Per &    gg &    1993 &  -0.69 &   0.74 &    3 \\ 
 V440  Per &    gg &    1994 &  -0.37 &   0.25 &   12 \\ 
 V440  Per &    gg &    1995 &   0.01 &   0.18 &   13 \\ 
 V440  Per &    gg &    1996 &   0.22 &   0.14 &   15 \\ 
 V440  Per &    gg &    1997 &   0.34 &   0.35 &   10 \\ 
 & & & & & \\
    U  Sgr &    b0 &    1979 &   0.70 &   0.19 &    5 \\ 
    U  Sgr &    b0 &    1982 &  -0.03 &   0.18 &    2 \\ 
    U  Sgr &    b0 &    1984 &  -0.33 &   0.65 &    4 \\ 
    U  Sgr &    b0 &    1985 &  -0.43 &   0.18 &   31 \\ 
    U  Sgr &    s4 &    1986 &  -0.05 &   0.21 &   30 \\ 
    U  Sgr &    s4 &    1987 &   0.49 &   0.24 &   21 \\ 
    U  Sgr &    s4 &    1988 &  -0.16 &   2.59 &    3 \\ 
    U  Sgr &    gg &    1990 &   0.51 &   0.21 &   12 \\ 
    U  Sgr &    b0 &    1991 &   0.06 &   0.18 &    5 \\ 
    U  Sgr &    gg &    1991 &   0.40 &   0.11 &   31 \\ 
    U  Sgr &    gg &    1992 &   0.28 &   0.19 &   17 \\ 
    U  Sgr &    gg &    1993 &   0.18 &   0.25 &   31 \\ 
    U  Sgr &    gg &    1994 &  -0.44 &   0.21 &   37 \\ 
    U  Sgr &    gg &    1995 &  -0.69 &   0.16 &   18 \\ 
    U  Sgr &    b0 &    1996 &  -0.40 &   0.38 &   18 \\ 
    U  Sgr &    b0 &    1997 &   0.40 &   0.53 &   14 \\ 
 & & & & & \\
   WZ  Sgr &    gg &    1994 &   0.17 &   0.17 &   26 \\ 
   WZ  Sgr &    gg &    1995 &  -1.53 &   0.66 &   19 \\ 
   WZ  Sgr &    b0 &    1996 &  -1.25 &   0.39 &   21 \\ 
   WZ  Sgr &    gg &    1996 &  -0.42 &   0.40 &   17 \\ 
   WZ  Sgr &    b0 &    1997 &  -1.02 &   0.39 &   14 \\ 
   WZ  Sgr &    gg &    1997 &  -0.43 &   0.34 &   18 \\ 
 & & & & & \\
   BB  Sgr &    gg &    1994 &  -0.21 &   0.12 &   27 \\ 
   BB  Sgr &    gg &    1995 &  -0.37 &   0.18 &   16 \\ 
   BB  Sgr &    gg &    1996 &   0.51 &   0.15 &   17 \\ 
   BB  Sgr &    gg &    1997 &   0.40 &   0.18 &   11 \\ 
 & & & & & \\
   ST  Tau &    im &    1978 &   0.12 &   0.25 &    8 \\ 
   ST  Tau &    im &    1979 &   0.65 &   0.45 &   14 \\ 
   ST  Tau &    im &    1980 &   0.69 &   0.74 &    7 \\ 
   ST  Tau &    b9 &    1981 &   0.59 &   0.62 &    6 \\ 
   ST  Tau &    im &    1981 &   2.49 &   1.06 &    3 \\ 
   ST  Tau &    b9 &    1982 &   0.29 &   0.35 &   24 \\ 
   ST  Tau &    gg &    1996 &   1.62 &   0.53 &    5 \\ 
   ST  Tau &    gg &    1997 &   0.19 &   0.38 &    6 \\ 
   ST  Tau &    gg &    1998 &   0.95 &   0.51 &    6 \\ 
 & & & & & \\
   SZ  Tau &    b0 &    1981 &  -0.07 &   0.61 &    8 \\ 
   SZ  Tau &    b0 &    1982 &  -0.14 &   0.31 &   16 \\ 
   SZ  Tau &    gg &    1987 &   0.27 &   1.13 &    2 \\ 
   SZ  Tau &    gg &    1991 &  -0.33 &   0.25 &   17 \\ 
   SZ  Tau &    gg &    1994 &   0.13 &   0.74 &    5 \\ 
   SZ  Tau &    gg &    1995 &  -0.90 &   0.45 &    7 \\ 
   SZ  Tau &    gg &    1996 &   0.26 &   0.39 &   15 \\ 
   SZ  Tau &    kk &    1996 &   1.12 &   0.43 &    7 \\ 
   SZ  Tau &    kk &    1997 &   0.95 &   0.38 &    7 \\ 
   SZ  Tau &    gg &    1997 &   0.33 &   0.38 &    7 \\ 
   SZ  Tau &    gg &    1998 &   0.37 &   1.12 &    8 \\ 
 & & & & & \\
   EU  Tau &    b0 &    1981 &  -0.97 &   1.96 &    5 \\ 
   EU  Tau &    b0 &    1982 &   0.18 &   1.02 &   20 \\ 
   EU  Tau &    b0 &    1987 &  -0.99 &   0.67 &   16 \\ 
   EU  Tau &    gg &    1991 &  -0.81 &   0.31 &   13 \\ 
   EU  Tau &    gg &    1995 &  -0.47 &   0.07 &    2 \\ 
   EU  Tau &    gg &    1996 &   0.63 &   0.37 &   16 \\ 
   EU  Tau &    gg &    1997 &  -0.32 &   0.47 &    5 \\ 
   EU  Tau &    gg &    1998 &   0.17 &   0.45 &    5 \\ 
 & & & & & \\
    T  Vul &    b0 &    1977 &   0.89 &   0.93 &    4 \\ 
    T  Vul &    b0 &    1978 &   0.01 &   0.09 &   25 \\ 
    T  Vul &    b0 &    1979 &   0.41 &   0.31 &    3 \\ 
    T  Vul &    b0 &    1980 &  -0.13 &   0.06 &   61 \\ 
    T  Vul &    b0 &    1987 &   0.59 &   0.19 &   13 \\ 
    T  Vul &    b0 &    1988 &  -0.13 &   0.32 &   10 \\ 
    T  Vul &    ba &    1996 &   1.94 &   0.64 &   11 \\ 
    T  Vul &    kk &    1996 &   2.39 &   0.31 &    8 \\ 
    T  Vul &    kk &    1997 &   2.40 &   0.31 &    7 \\ 
    T  Vul &    ba &    1997 &   1.47 &   0.54 &   20 \\ 
 & & & & & \\
    X  Vul &    b9 &    1987 &  -0.93 &   0.34 &   12 \\ 
    X  Vul &    b9 &    1988 &  -1.64 &   0.58 &    9 \\ 
    X  Vul &    gg &    1991 &   0.51 &   0.38 &   17 \\ 
    X  Vul &    gg &    1992 &  -0.07 &   0.29 &   20 \\ 
    X  Vul &    gg &    1993 &   0.04 &   0.43 &   14 \\ 
    X  Vul &    gg &    1994 &  -0.77 &   0.55 &    8 \\ 
    X  Vul &    gg &    1995 &   0.17 &   0.38 &   12 \\ 
    X  Vul &    gg &    1996 &   0.14 &   0.30 &    7 \\ 
 & & & & & \\

\enddata

\vskip .1truein


\vskip .1truein


\noindent

\end{deluxetable}


\begin{thebibliography}{}

\bibitem[\protect\citeauthoryear{abt et al.}{1990}]{agl90}
Abt, H. A., Gomez, A. E., and Levy, S. G. 1990, ApJS, 74, 551.

\bibitem[\protect\citeauthoryear{alcock et al.}{1995}]{alc95}
Alcock et al. 1995, AJ, 109, 1653

\bibitem[\protect\citeauthoryear{alcock et al.}{2003}]{alc03}
Alcock, C., Alves, D. R., Becker, A. et al. 2003, ApJ, 598, 597


\bibitem[\protect\citeauthoryear{aldoretta et al.}{2014}]{ald04}
Aldoretta, D. J., Caballero-Nieves, S. M., Gies, D. R. et al. 2015,
AJ, 149, 26 


\bibitem[\protect\citeauthoryear{anderson}{2014}]{and14}
Anderson, R. I. 2014, A\&Ap, 566, 10

\bibitem[\protect\citeauthoryear{anderson}{2015}]{and15}
Anderson, R. I., Sahlmann, J., Holl, B., Eyer, L., Palaversa, L.,
Mowlavi, N., S\"uveges, M., and Roelens, M. 
2015, ApJ, in press (arXiv:1503.04116v1)

\bibitem[\protect\citeauthoryear{baranne et al.}{1979}]{bar79}
Baranne, A., Mayor, M. and Poncet, J. L. 1979, Vistas in Astronomy,
23, 279

\bibitem[\protect\citeauthoryear{barnes et al.}{2005}]{bar05}
Barnes III T.G., Jeffery E.J., Montemayor T.J., Skillen I. 2005,
    ApJS, 156, 227

\bibitem[\protect\citeauthoryear{benedict et al.}{2007}]{ben07}
Benedict, G. F., McArthur, B. E., Feast, M. W. et al. 2007, AJ, 133, 1810

\bibitem[\protect\citeauthoryear{berdnikov}{1994}]{ber94}
Berdnikov, L. N. 1994, AstL, 20, 232

\bibitem[\protect\citeauthoryear{berdnikov et al.}{1997}]{ber97}
Berdnikov, L. N., Ignatova, V. V., Pastukhova, E. N., and Turner, D. G
1997, AstL, 23, 177
	
\bibitem[\protect\citeauthoryear{berdnikov et al.}{2007}]{ber07}
Berdnikov, L. N., Dambis, A. K., and Vozyakova, O. V. 2000, A\&ApS,
143, 211
	
\bibitem[\protect\citeauthoryear{bersier et al.}{1994}]{bers94}
Bersier, D., Burki, G., Mayor, M. and Duquennoy, A 1994, A\&ApS, 108, 25

\bibitem[\protect\citeauthoryear{bersier}{2002}]{bers02}
 Bersier D. 2002  ApJS, 140, 465

\bibitem[\protect\citeauthoryear{brown verschueren}{1997}]{brv97}
Brown, A. G. A. and Verschueren, W. 1997, A\&A, 319, 811

\bibitem[\protect\citeauthoryear{caballero et al.}{2014}]{cab14}
Caballero-Nieves, S. M., Nelan, E. P., Gies, D. R. et al. 2014,
AJ, 147, 40

\bibitem[\protect\citeauthoryear{caldwell et al.}{2001}]{cal01}
Caldwell, J. A. R., Coulson, I. M., Dean, J. F.,and Berdnikov,
L. N. 2001 JAD, 7, No. 4

\bibitem[\protect\citeauthoryear{chini et al.}{2012}]{chi12}
Chini, R., Hoffmeister, V. H., Nasseri, A., Stahl, O., and Zinnecker, H. 
2012, MNRAS, 424, 1925

\bibitem[\protect\citeauthoryear{caldwell coulson}{1985}]{cac85}
Coulson, I. M., and Caldwell, J. A. R. 1985, SAAO Circ, 9, 5 

\bibitem[\protect\citeauthoryear{caldwell coulson gieren}{1985}]{ccg85}
Coulson, I.,M., Caldwell, J. A. R. and Gieren, W. P. 1985, ApjS, 57, 595


\bibitem[\protect\citeauthoryear{derosa et al.}{2014}]{der14}
De Rosa, R. J., Patience, J., Wilson, P. A. et al. 2014, MNRAS, 437, 1216 

\bibitem[\protect\citeauthoryear{duchene and krauss}{2013}]{duk13}
Duchene, G. and Krauss, A. 2013, ARA\&A, 51, 269

\bibitem[\protect\citeauthoryear{duquennoy mayor}{1991}]{dum91}
Duquennoy, A. and Mayor, M. 1991, A\&A, 248, 485

\bibitem[\protect\citeauthoryear{evans et al.}{2011}]{eva11}
Evans, N. R., DeGioia-Eastwood, K., Gagne, M., Townsley, L., Broos, P.,
Wolk, S., Naze, Y., Corcoran, M., Oskinova, L., Moffat, A. F. J.,
Wang, J. and Walborn, N. R. 2011a ApJS, 194, 13    Tr 16

\bibitem[\protect\citeauthoryear{evans et al. 2015}{2015}]{eva15}
Evans, N. R., Szabo, R., Derekas, A. et al. 2015, MNRAS, accepted

\bibitem[\protect\citeauthoryear{evans massa proffitt}{2009}]{emp09}
Evans N.R., Massa D. and Proffitt C.  2009, AJ, 137, 3700

\bibitem[\protect\citeauthoryear{evans et al.}{2013}]{eva13}
Evans, N. R., Bond, H. E., Schaefer, G. H., Mason, B. D., Karovska, M., and Tingle, E.
2013, ApJ, 146, 93

\bibitem[\protect\citeauthoryear{evans}{1992}]{eva92}
Evans, N. R. 1992, ApJ, 384, 220

\bibitem[\protect\citeauthoryear{evans}{1995}]{eva95}
Evans, N. R. 1995, ApJ, 445, 393

\bibitem[\protect\citeauthoryear{evans et al.}{1999}]{eva99}
Evans, N. R., Carpenter, K., Robinson, R., Massa, D., Wahlgren, G. M., 
Vinko, J., and Szabados, L. 1999, ApJ, 524, 379

\bibitem[\protect\citeauthoryear{evans et al.}{2008}]{eva08}
Evans, N. R., Schaefer, G. H., Bond, H. E. et al. 2008, ApJ, 136, 1137
1995, ApJ, 445, 393

\bibitem[\protect\citeauthoryear{fernie et al.}{1995}]{fer95}
Fernie, J.D., Beattie, B., Evans, N.R., and Seager, S. 1995, IBVS No. 4148 
Fernie et al. http://www.astro.utoronto.ca/DDO/research/cepheids/table\_physical.html


\bibitem[\protect\citeauthoryear{freedman et al.}{1994}]{fre94}
Freedman, W., L., Hughes, S. M., Madore, B. F. et al. 1994, ApJ, 427, 628. 

\bibitem[\protect\citeauthoryear{garmany conti massey}{1980}]{gar80}
   Garmany, C. D., Conti, P. S., and Massey, P.  1980, ApJ, 242, 1063


\bibitem[\protect\citeauthoryear{gieren}{1981}]{gie81}
Gieren, W. P. 1981, ApJS, 46, 287 

\bibitem[\protect\citeauthoryear{gorynya et al.}{1992}]{gor92}
Gorynya, N. A.; Irsmambetova, T. R.; Rastorgouev, A. S.; Samus, N. N.
 1992, SvAL, 18, 316	

\bibitem[\protect\citeauthoryear{gorynya et al.}{1996}]{gor96}
Gorynya, N. A., Rastorguev, A. S., and Samus, N. N. 1996, AstL, 22, 175

\bibitem[\protect\citeauthoryear{gorynya et al.}{1998}]{gor98}
Gorynya, N. A.; Samus', N. N.; Sachkov, M. E.; Rastorguev, A. S.;
Glushkova, E. V.; Antipin, S. V.  1998, AstL, 24, 815

\bibitem[\protect\citeauthoryear{groenewegen}{2013}]{gro13}
Groenewegen, M. A. T. 2013, A \& Ap, 550, A70

\bibitem[\protect\citeauthoryear{groenewegen oudmaijer}{2000}]{gro00}
Groenewegen, M. A. T. and Oudmaijer, R. D.  2000, A\&A, 356, 849   


\bibitem[\protect\citeauthoryear{imbert}{1999}]{imb99}
  Imbert M. 1999, A\&ApS, 140, 79

\bibitem[\protect\citeauthoryear{joy}{1937}]{joy37}
Joy, A. H. 1937, ApJ, 86, 363 

\bibitem[\protect\citeauthoryear{kienzle et al.}{1999}]{kie99}
Kienzle, F., Moskalik, P., Bersier, D., and Pont, F. 1999, A\&Ap, 341, 818


\bibitem[\protect\citeauthoryear{kimiki kobulnicky}{2012}]{kik12}
Kimiki, D. C. and Kobulnicky, H. A. 2012, ApJ, 751, 15

\bibitem[\protect\citeauthoryear{kobulnicky et al.}{2014}]{kob14}
Kobulnicky, H. A., Kiminki, D. C., Lundquist, M. J. et al. 2014, 
ApJS, 213, 34

\bibitem[\protect\citeauthoryear{kiss vinko}{2000}]{kiv00}
Kiss, L. and Vinko, J. 2000, MNRAS, 314, 420

\bibitem[\protect\citeauthoryear{klagyivik szabados}{2009}]{kls09}
Klagyivik, P. and Szabados, L. 2009, A \& Ap, 504, 959

\bibitem[\protect\citeauthoryear{kouwnehoven et al.}{2007}]{kou07}
Kouwenhoven, M. B. N., Brown, A. G. A., Portegies Zwart, S. F., and 
Kaper, L. 2007, A\&A, 474, 77 

\bibitem[\protect\citeauthoryear{kratter matzner}{2006}]{krm06}
Kratter, K. M. and Matzner, C. D. 2006, MNRAS, 373, 1563

\bibitem[\protect\citeauthoryear{levatp et al.}{1987}]{lev87}
Levato, H., Malaroda, S., Morrell, N., and Solivella, G. 1987, ApJS,
64, 487

\bibitem[\protect\citeauthoryear{neilson et al.}{2015}]{nei15}
Neilson, H. R., Schneider, F. R. N., Izzard, R. G., Evans, N. R. 
and Langer, N. 2015, A\&A, 574, A2

\bibitem[\protect\citeauthoryear{markley}{1995}]{mar95}
Markley, F. L. 1995, CeMDA, 63, 101

\bibitem[\protect\citeauthoryear{mason et al.}{2009}]{mas09}
Mason, B. D., Hartkopf, W. I., Gies, D. R., Henry, T. J., and
Helsel, J. W. 2009, AJ, 137, 3358

\bibitem[\protect\citeauthoryear{neilson et al.}{2012}]{nei12}
Neilson, H. R., Langer, N., Engle, S. G., Guinan, E., and Izzard, R. 2012, ApJ, 760, L18

\bibitem[\protect\citeauthoryear{perez granger}{2007}]{peg07}
P\'erez, Fernando and Granger, Brian E. 2007, Comp Sci Eng, 9, 21

\bibitem[\protect\citeauthoryear{petterson cottrell albrow}{2004}]{pca04}
Petterson, O. K. L., Cottrell, P. L., and Albrow, M. D. 2004, MNRAS, 
   350, 95

\bibitem[\protect\citeauthoryear{prada moroni et al.}{2012}]{pra12}
Prada Moroni, P. G., Gennaro, M., and Bono, G. et al. 2012, ApJ, 749, 108

\bibitem[\protect\citeauthoryear{raghavan et al.}{2010}]{rag10}
Raghavan, D., McAlister, H. A., Henry, T. J., Latham, D. W., Marcy,
G. W., Mason, B. D., Gies, D. R., White, R. J., ten Brummelaar, T. A. 
2010, ApJS, 190, 1

\bibitem[\protect\citeauthoryear{roberts lehar dreher}{1987}]{rld87}
Roberts, D. H., Lehar, J., and Dreher, J. W. 1987, AJ, 93, 968

\bibitem[\protect\citeauthoryear{sana evans}{2011}]{sae11}
Sana, H. and Evans, C. J. 2011, IAU Symp 272, eds C. Neiner, G. Wade, G. 
Meynet, and G. Peters.  

\bibitem[\protect\citeauthoryear{sana et al.}{2012}]{san12}
Sana, H.,  de Mink, S. E.,  de Koter, A.,  Langer, N.,  Evans, C. J.,
Gieles, M.,  Gosset, E.,  Izzard, R. G.,  Le Bouquin, J.-B.,
Schneider, F. R. N. 2012, Sci, 337, 444

\bibitem[\protect\citeauthoryear{shane}{1958}]{sha58}
Shane, W. W. 1958, ApJ, 127, 573 

\bibitem[\protect\citeauthoryear{shatsky tokovinin}{2002}]{sht02}
Shatsky, N. and Tokovinin, A. 2002, A\&Ap, 382, 92

\bibitem[\protect\citeauthoryear{storm et al.}{2004}]{sto04}
Storm, J., Carney, B. W., Gieren, W. P., Fouque, P., Latham, D. W.,
and  Fry, A. M. 2004, A\&Ap, 415, 531

\bibitem[\protect\citeauthoryear{storm et al.}{2011}]{sto11}
Storm, J., Gieren, W., Fouqu\'e, P., Barnes, T. G., Pietrzy\'nski, G.,
Nardetto, N., Weber, M., Granzer, T., and Strassmeier, K. G. 2011,
A\&Ap, 534, A94

\bibitem[\protect\citeauthoryear{sugars evans}{1996}]{sue96}
Sugars, B. J. A. and Evans, N. R. 1996, AJ, 112, 1670

\bibitem[\protect\citeauthoryear{szabados}{1983}]{sza83}
Szabados, L. 1983, ApSS, 96, 185

\bibitem[\protect\citeauthoryear{szabados}{2003}]{sza03}
Szabados, L. 2003 ASP Conf 298, 237

\bibitem[\protect\citeauthoryear{szabados et al.}{2012}]{sza12}
Szabados, L., Derekas, A., Kiss, C., and Klagyivik, P. 2012 MNRAS, 426, 3154 

\bibitem[\protect\citeauthoryear{szabados et al.}{2014}]{sza14}
Szabados, L., Cseh, B., Kovacs, J. et al. (11 authors) 2014, MNRAS, 442, 3155

\bibitem[\protect\citeauthoryear{templeton karovska}{2009}]{tek09}
Templeton, M. R., and Karovska, M. 2009, ApJ, 691, 1470

\bibitem[\protect\citeauthoryear{tokovinin}{1987}]{tok87}
Tokovinin, A. A. 1987, Sov Ast, 31, 98

\bibitem[\protect\citeauthoryear{tokovinin}{2014}]{tok14}
Tokovinin, A. 2014,  AJ, 147, 87

\bibitem[\protect\citeauthoryear{wallerstein}{1979}]{wal79}
Wallerstein, G. 1979, PASP, 91, 770

\bibitem[\protect\citeauthoryear{wolff}{1978}]{wol78}
Wolff, S. C. 1978, ApJ, 222, 556

-------------------











\end{thebibliography}
\end{document}